\documentclass{article}[12 pt]
\usepackage{authblk} 
\usepackage{graphicx} 
\usepackage{amsmath,amssymb}
\usepackage{natbib}
\usepackage{threeparttable} 
\usepackage[capposition=top]{floatrow} 
\usepackage{setspace}
\usepackage{silence}
\usepackage{xr-hyper}
\usepackage{xr} 
\usepackage{hyperref}
\usepackage[dvipsnames]{xcolor}
\usepackage{longtable}
\usepackage{array}
\usepackage{subcaption} 

\usepackage{tikz}
\usepackage{float}
\usetikzlibrary{shapes.geometric, arrows.meta, decorations,decorations.markings}
\tikzstyle{standard} = [rectangle, rounded corners, minimum width=2cm, minimum height=1cm,text centered, draw=black]
\tikzstyle{arrow} = [thick,-latex]

\newcommand{\single}{\baselineskip 15pt}
\DeclareMathOperator{\logit}{logit}

\DeclareMathOperator{\argmin}{arg min}

\makeatletter
\newcommand*{\addFileDependency}[1]{
  \typeout{(#1)}
  \@addtofilelist{#1}
  \IfFileExists{#1}{}{\typeout{No file #1.}}
}
\makeatother

\newif\ifarXiv
\arXivtrue 

\newcommand*{\myexternaldocument}[1]{%
    \externaldocument{#1}%
    \addFileDependency{#1.tex}%
    \addFileDependency{#1.aux}%
}
\myexternaldocument{SupplementaryMaterials}

\title{Assessing treatment effects in observational data with missing confounders: A comparative study of practical doubly-robust and traditional missing data methods}
\author[1,2,*]{Brian~D. Williamson}
\author[1]{Chloe Krakauer}
\author[1]{Eric Johnson}
\author[3]{Susan Gruber}
\author[4]{Bryan~E. Shepherd}
\author[5]{Mark~J. van der Laan}
\author[6]{Thomas Lumley}
\author[7]{Hana Lee}
\author[8]{José J. Hernández-Muñoz}
\author[7]{Fengyu Zhao}
\author[8]{Sarah~K. Dutcher}
\author[9]{Rishi Desai}
\author[1]{Gregory~E. Simon}
\author[1,2]{Susan~M. Shortreed}
\author[1,2]{Jennifer~C. Nelson}
\author[1,2]{Pamela~A. Shaw}
\affil[1]{Biostatistics Division, Kaiser Permanente Washington Health Research Institute, Seattle, WA, USA}
\affil[2]{Department of Biostatistics, University of Washington, Seattle, WA, USA}
\affil[3]{TL Revolution, LLC, Cambridge, MA, USA}
\affil[4]{Department of Biostatistics, Vanderbilt University, Nashville, Tennessee, USA}
\affil[5]{Department of Biostatistics, School of Public Health, University of California at Berkeley, Berkeley, CA, USA}
\affil[6]{Department of Statistics, University of Auckland, Auckland, New Zealand}
\affil[7]{Office of Biostatistics, Center for Drug Evaluation and Research, US Food and Drug Administration, Silver Spring, MD, USA}
\affil[8]{Office of Surveillance and Epidemiology, Center for Drug Evaluation and Research, US Food and Drug Administration, Silver Spring, MD, USA}
\affil[9]{Division of Pharmacoepidemiology and Pharmacoeconomics, Department of Medicine, Brigham and Women’s Hospital, Harvard Medical School, Boston, MA, USA}
\affil[*]{Corresponding Author: Biostatistics Division, Kaiser Permanente Washington Health Research Institute, 1730 Minor Ave Ste 1360, Seattle, WA 98101. Email: brian.d.williamson@kp.org}

\date{\today}
\begin{document}
\onehalfspacing
\maketitle
\section*{Funding}
The author(s) disclosed receipt of the following financial support for the research, authorship, and/or publication of this article: This project was supported by Task Order 75F40123F19006 under Master Agreement 75F40119D10037 from the US Food and Drug Administration (FDA) and  the U.S. National Institutes of Health (NIH) grant R37-AI131771.

\section*{Acknowledgments}
The contents are those of the authors and do not necessarily represent the official views of, nor endorsement, by FDA/HHS, National Institutes of Health, or the U.S. Government.

\section*{Declaration of competing interest} None.

\section*{Ethical approval and informed consent}
This study does not meet the criteria for human subject research as defined by Kaiser Permanente Washington Health Research Institute policies, Health and Human Services (HHS) regulations set forth in 45 CFR 46, and FDA regulations set forth in 21 CFR 56. The study involves public health surveillance activity as defined by HHS regulations at 45 CFR 46.102(l)(2).

\section*{Data availability}
The datasets generated and analyzed during this study are not publicly available because they contain detailed information from the electronic health records in the health systems participating in this study and are governed by HIPAA. Data are however available from the authors upon reasonable request, with permission of all health systems involved and fully executed data use agreement.

\newpage
\begin{abstract}
In pharmacoepidemiology, safety and effectiveness are frequently evaluated using readily available administrative and electronic health records data. In these settings, detailed confounder data are often not available in all data sources and therefore missing on a subset of individuals. Multiple imputation (MI) and inverse-probability weighting (IPW) are go-to analytical methods to handle missing data and are dominant in the biomedical literature. Doubly-robust methods, which are consistent under fewer assumptions, can be more efficient with respect to mean-squared error. We discuss two practical-to-implement doubly-robust estimators, generalized raking and inverse probability-weighted targeted maximum likelihood estimation (TMLE), which are both currently under-utilized in biomedical studies. We compare their performance to IPW and MI in a detailed numerical study for a variety of synthetic data-generating and missingness scenarios, including scenarios with rare outcomes and a high missingness proportion. Further, we consider plasmode simulation studies that emulate the complex data structure of a large electronic health records cohort in order to compare anti-depressant therapies in a rare-outcome setting where a key confounder is prone to more than 50\% missingness. We provide guidance on selecting a missing data analysis approach, based on which methods excelled with respect to the bias-variance trade-off across the different scenarios studied.
\end{abstract}

\doublespacing

\section{Introduction}

Medical claims data have become a mainstay in evaluating the safety and effectiveness of medications post-approval. They are available on broad populations but lack detailed clinical information regarding patient health status, disease severity, and other clinical factors. This lack of data can be a challenge when evaluating the effects of a medication on health outcomes in observational studies due to the inability to control for key confounders. Electronic health records (EHR) data have more detailed clinical data than do medical claims, but in many settings EHR data may only be available on a subset. For example, in the US Food and Drug Administration (FDA) Sentinel Initiative \citep{platt2018fda} system that assesses the safety of approved medical products, national health insurance plans capture administrative claims data on 500 million patients, while large integrated healthcare delivery systems additionally provide clinical data from the EHR (e.g., laboratory test results) on a subset of approximately 75 million patients. Clinical laboratory tests, patient-reported outcomes, and vital signs are routinely available in EHRs and may be able to improve confounder control \citep{flory17}. 
Many approaches exist in the statistical literature to handle missing or mismeasured data; however, these methods have been under-evaluated in studies that combine administrative claims and EHR data, which are characterized by rare outcomes and high proportions of missing confounder data. In particular, trade-offs involving bias and increased variability must be considered in these settings. 

Model-based multiple imputation (MI) is a traditional approach for handling missing data and can be used for two-phase designs, where full information is only available on a subset. Inference using model-based approaches, such as MI, relies on distributional assumptions about the data generating process. Several flexible imputation methods that have been observed to perform well in a variety of settings, including multiple imputation through chained equations, random forest algorithms and gradient boosting \citep{vanbuuren2018flexible,weberpals2024principled,luo2022evaluating}. However, MI can suffer from bias from model misspecification and from invalid inference when imputation models are not compatible with the outcome model \citep{rubin1976inference,bartlett15,vanbuuren2018flexible}.  MI approaches in the setting of non-linear outcome models are particularly prone to non-compatibility, with the usual Rubin’s variance estimate leading to overly wide confidence intervals, something not widely appreciated in applied settings \citep{bartlett15,giganti2020accounting}. 

A second class of methods for studies with missing data are design-based approaches, where inference relies on the sampling rather than assumptions about the data-generating process. In the context of missing data, the classic design-based approach is the Horvitz-Thompson inverse probability weighted (IPW) estimator: a model is developed for the missingness mechanism in order to weight the completely observed data and appropriately draw inference on the target population \citep{sarndal2003model}. Calibration of weights (also known as generalized raking (GR)) is a standard estimation approach in the survey literature that uses data available on the entire analysis cohort to adjust the IPW sampling weights in a manner that improves the efficiency of the IPW estimator \citep{deville1992calibration,breslow2009improved,breslow2009using}.  Calibration estimators are much less commonly used in the biomedical literature. Survey calibration estimators can offer a substantial improvement over the IPW estimator in an ideal implementation, yielding the efficient doubly-robust augmented inverse probability weighted (AIPW) estimator \citep{breslow2009improved,lumley2011connections}. However, general AIPW estimators can result in poor efficiency gains in observational settings due to an inadequate model of the missingness mechanism or a poor choice of the calibration variables. One nice feature of calibration-based AIPW (i.e., GR) is that, unlike the more general class of AIPW estimators, a poor choice of calibration variables may not improve efficiency, but it is also not expected to worsen efficiency relative to the IPW estimator \citep{lumley2011connections}. Bias for doubly-robust estimators can also occur in observational settings, where it may be difficult to correctly specify either of the missingness or outcome models.

The rise of machine learning has yielded new tools for estimation and possible bias reduction, including those applicable to the missing-data or two-phase design setting. Targeted maximum likelihood estimation (TMLE) targets the parameter of interest in a way that reduces bias and variance for that parameter and is typically implemented with a super learner (i.e. a data-adaptive ensemble learner) for key quantities, e.g., the outcome regression model or treatment propensity score, \citep{vanderlaan2006targeted,vanderlaan2007super}. The resulting targeted maximum likelihood estimator (which we will also refer to as a TMLE) is sometimes accompanied by an increased variance due to relaxed model assumptions, but is asymptotically efficient among nonparametric estimators when models for both the target and nuisance parameters are correctly specified (\cite{gruber2012tmle}). The TMLE approach for two-stage sampling designs incorporates machine learning while maintaining locally efficient doubly robust properties \citep{rose2011targeted}, and thus is a modern and flexible way to address missing data, while still allowing for standard inference. Despite being introduced over a decade ago, these estimators have not been widely adopted.

When choosing a method to handle missing data, bias and variance are direct trade-offs as one weighs more flexible estimation approaches (such as ensemble learning) and simpler, less variable, parametric approaches that rely on model assumptions. Inverse-weighting approaches have traditionally been dismissed as inferior to multiple imputation due to being too variable; however, there has been a growing appreciation that through the use of auxiliary variables to calibrate the weights, the GR approach can achieve the maximally efficient AIPW estimator \citep{lumley2011connections,breslow2009improved,breslow2009using}. In observational studies reliant on EHR, where there is often a wealth of data available on individuals and where associations and missingness mechanisms are likely complex, flexible machine learning approaches are attractive; however, there is little work to directly compare the relatively new TMLE approaches with GR or even traditional approaches like MI, which can also be configured to be quite flexible, for handling missing data. 

In a series of numerical studies, we compare the MI, GR, and TMLE approaches for handling partially observed data on confounders. Our motivating example arises from an integrated health care system with data from insurance claims, pharmacy dispensations, and patient visits for mental health care. We aim to evaluate the risk of self-harm or hospitalization with a mental health diagnosis following initiation of psychotherapy or medication for treatment of depression. In our setting, key confounders such as depression severity and suicidal ideation measured using the Patient Health Questionnaire \citep[PHQ;][]{kroenke2010patient} are available in the EHR only for a subset of individuals. In clinical practice, these highly prognostic variables may exhibit differential missingness patterns between different medication exposure groups, a common challenge in pharmacoepidemiological settings \citep{simon2016PHQ9SI}. Several numerical experiments were conducted to compare methods with fully-synthetic and plasmode-based simulated data that vary the complexity of the outcome and missing data mechanisms, as well as outcome proportion and amount of missing data. We focus on comparing methods practical to implement in standard software, summarizing the relative performance of the methods, including bias, relative efficiency, and nominal coverage. We conclude with guidance for selecting an analysis approach, based on which methods excelled with respect to the bias-variance trade-off across different settings. We provide the code used to implement the methods studied here, all of which are implemented in existing R packages hosted on CRAN, at \url{https://github.com/PamelaShaw/Missing-Confounders-Methods} \citep{shaw2024comparison}.

\section{Methods}
\subsection{Data structure and notation}
We consider a general data structure, where each observation consists of an outcome, exposure of interest, potential confounding variables for the outcome-exposure association that are subject to missingness, and a set of other features (e.g., additional confounders and other important variables related to the missingness or treatment propensity models) that are fully observed. 
As is common in our motivating setting, we consider a binary outcome $Y$ and binary treatment exposure $X$. 
We denote the potential confounding variables that are subject to missingness by $W \in \mathbb{R}^q$. Fully measured covariates are denoted by $Z \in \mathbb{R}^s$, 
and the exposure or treatment of interest by $X \in \{0, 1\}$, with $(X,W,Z) \in \mathbb{R}^p$.
Consistent with the survey literature, we refer to variables observed on the entire cohort $(X,Z,Y)$ as phase I variables, and those that are only available for a subset ($W$) as phase II variables \citep{breslow2009improved}. Finally, we use $R = (R_1, \ldots, R_q) \in \{0,1\}^{q}$ to denote the phase II subset indicator, where a value of 1 indicates complete data.  In the ideal setting, all variables are fully observed with the data unit $(Y,X,Z,W) \sim \mathbb{P}_0$.
In the (assumed) observed setting, the data unit is $O := (Y, X, Z, R, RW) \sim P_0$.

\subsection{Estimands}
An important first step in any analysis is to define the target of estimation, or estimand. There are several possible estimands of interest in a setting with a binary treatment exposure and binary outcome.

First, we define treatment-specific mean outcome values. Using potential (or counterfactual) outcomes notation \citep{rubin1974estimating}, where $Y(x)$ refers to the potential outcome when exposure is equal to $x$, we can write  
\begin{align*}
\mu_1 =& \ E\{Y(1)\} \text{ and } \mu_0 = E\{Y(0)\},
\end{align*}
where $E$ is shorthand for the expected value under the data-generating distribution. Under the assumptions of consistency, no unmeasured confounding, and positivity \citep[Supplementary Materials Section~\ref{sec:causal_assumptions} and, e.g.,][]{cole2009consistency,vanderlaan2011targeted}, we can identify these parameters with the observed data, writing
\begin{align*}
\mu_1 =& \ E\{E(Y \mid X = 1, W, Z)\} \\
\mu_0 =& \ E\{E(Y \mid X = 0, W, Z)\}. 
\end{align*}
Based on these parameters, we can define the marginal risk difference (mRD), relative risk (mRR), and odds ratio (mOR):
\begin{align*}
  mRD =& \ \mu_1 - \mu_0 \\
  mRR =& \ \frac{\mu_1}{\mu_0} \\
  mOR =& \ \frac{\mu_1 / (1 - \mu_1)}{\mu_0 / (1 - \mu_0)}.
\end{align*}
 These parameters are \textit{marginal} because they are averages over the distribution of the covariates. 
 Often, one of the mRD, mRR, or mOR is of interest in causal inference, because (under the assumptions listed above) these estimands measure the causal effect of the exposure in the population under study, a contrast in the mean outcome value when everyone in the population was exposed versus unexposed.

A fourth estimand of interest is the odds ratio conditional on covariates (we will often use the shorthand \textit{conditional odds ratio} or cOR). This is the natural parameter from a logistic regression model, and is the target of interest in many applied settings. This is the natural parameter from a logistic regression model, and is the target of interest in many applied settings. Suppose that we fit the following logistic regression model to the ideal data (with no missingness): $\logit P(Y = 1 \mid X = x, Z = z, W = w) = \beta_0 + \beta_1 x + \beta_2 z + \beta_3 w$. In this model, $\exp(\beta_1)$ is the cOR comparing $X = 1$ to $X = 0$ conditional on Z and W; we will consider this to be the cOR of interest. It is \textit{conditional} because the interpretation of $\beta_1$ is the log odds for a fixed level of $Z$ and $W$. While we focus on the cOR, the conditional risk difference (cRD) and conditional relative risk (cRR) could easily be estimated using a linear or Poisson working regression model, respectively. 

We do not need to believe that the logistic regression model holds for the cOR to be defined. Suppose instead that we specify the above model as the target \textit{working} model, i.e., the model of interest to be fit to the entire cohort had the data been complete. Then the cOR is the projection of the true underlying data onto the space defined by the working logistic regression model. 

This distinction allows us to define two levels of estimand. We define the \textit{oracle} estimand as a function of only the data-generating distribution (typically unknown); the mRD, mRR, and mOR as defined above are all examples of oracle estimands. If the outcomes are generated from the same model as the working logistic regression model, then the cOR is also an oracle estimand. In contrast, we define the \textit{census} estimand as a function of both the data-generating distribution and a working regression model. We can define a census version of each estimand listed above, using the working regression model in place of the true conditional probability of outcome given the covariates. For the marginal parameters defined above, the census version replaces the true treatment-specific mean outcome values with expectations of the outcome probabilities under the working model. The census estimand represents the natural target of estimation for certain methods, as we will highlight below, and performance for estimating the census estimand is often considered when comparing estimators in missing-data contexts. Importantly, if the working regression model corresponds to the data-generating model, then the oracle and census estimands are identical. If the goal of the analysis is to understand the true treatment effect, then the oracle estimand should be the target of estimation and a method whose natural target is the census estimand will be biased unless the census and oracle estimands are identical. Conversely, if the goal of the analysis is to estimate the association in a posited multivariable regression model, then the census estimand is the target of inference. 

\subsection{Missing data analysis approaches}\label{sec:estimators}
Our goal is to estimate and make inference on the estimands defined above, using $n$ independent and identically distributed observations $O_1, \ldots, O_n$.  We will compare several estimators that handle the missing confounder data, including model-based estimators (including MI), design-based estimators (IPW and GR), and nonparametric estimators (TMLE). Our goal is to compare approaches that are readily implemented in current software, including the approaches common in practice. To simplify the notation, for the remainder of this manuscript we consider a setting where there is a single indicator $R$ of whether all of $W$ is observed. To this end, we define the \textit{missing-data model} $\pi(y, x, z) = P(R = 1 \mid Y = y, X = x, Z = z)$, the \textit{propensity score model} $g(z, w) = P(X = 1 \mid Z = z, W = w)$, and the \textit{outcome model} $Q(x, z, w) = E(Y \mid X = x, Z = z, W = w)$. For simplicity, we will often drop the arguments for these functions and refer to them as $\pi$, $g$, and $Q$.

We consider MI by chained equations \citep[MICE;][]{van2007multiple,raghunathan2001multivariate} as the primary MI approach, due to its popularity in applied settings and its relative success compared to other imputation approaches in the setting of missing data in the confounders \citep{weberpals2024principled}. Briefly, MICE requires choosing a number of imputations $M$, a number of iterations $K$ (controls the convergence of each imputation run), and an imputation method for each variable with missing data. Rather than relying on a joint distribution for imputations, at each iteration, MICE imputes each $\{W_j\}_{j=1}^q$ conditional on the other variables in a specified sequence (i.e. chain) of equations \citep{austin2021missing}. 
After the $K$ iterations, this results in $M$ imputed datasets $O_{n,m} = (Y, X, Z, W_m)$. We implemented MICE using the \texttt{R} package \texttt{mice} \citep{micepkg}, choosing 20 imputations, a maximum of 25 iterations, and the default imputation method for each variable \citep{micepkg}. We also considered multiple imputation with random forest (MI-RF), which also did well in \citet{weberpals2024principled}, and with XGBoost \citep{chen2015xgboost} (MI-XGB), anticipating potential advantages using this flexible method for our most complex simulation settings. MI-RF was implemented using the \texttt{mice} package and default settings. MI-XGB was implemented using the \texttt{mixgb} package \citep{mixgbpkg}, with five-fold cross-validation to select the optimal number of boosting rounds. In all cases, after performing MI, we fit a logistic regression model to estimate the outcome probability on each imputed dataset $O_{n,m}$ and pooled the results, using Rubin's rules for inference \citep{rubin1987multiple}.

We consider two design-based estimators: IPW and GR. Both estimators are based on weighted complete-case estimating equations. For both, we first fit a logistic regression model to estimate the probability of observing data; the estimated weights $\hat{\pi}_{i, \text{IPW}}^{-1} = \widehat{P}(R = 1 \mid Y_i, X_i, Z_i)^{-1}$ resulting from this logistic regression model fit are the final weights for IPW. Survey calibration estimators use auxiliary variables available on the entire cohort to adjust the design weights, here estimated by $\hat{\pi}_{i, \text{IPW}}^{-1}$, in a way that leverages information in the auxiliary data regarding the target of estimation. When the target of inference is one or more regression coefficients, the random vector of the expected value of the efficient influence functions \citep[EIFs,][]{bickel1993efficient} for the regression coefficients given the observed phase I data (in our setting, given the variables derived from claims data), denoted $h_i$, is the ideal auxiliary variable for calibration which estimates the design-efficient AIPW estimator \citep{breslow2009improved,lumley2011connections}. 
Specifically, for GR, we obtain calibrated weights $\hat{\pi}_{i, \text{GR}}^{-1} = a_i\hat{\pi}^{-1}_{i,\text{IPW}}$, where $a$ is defined by 
\begin{align*}
    a = \argmin_{a' \in \mathbb{R}^n} & \sum_{i=1}^n R_i d(a_i'\hat{\pi}_{i, \text{IPW}}^{-1}, \hat{\pi}_{i, \text{IPW}}^{-1}) \\
    \text{subject to } & \sum_{i=1}^n \hat{h}_i = \sum_{i=1}^n R_i a_i' \hat{h}_i \hat{\pi}_{i, \text{IPW}}^{-1},
\end{align*}
where $d$ is a distance measure. We use $d(a, b) = a\ln(a/b) - a + b$ for non-negative calibrated weights \citep{chen2022optimal}. The $\hat{h}_i$ for regression parameters in a generalized linear model or proportional hazards regression model can be easily obtained from standard software \citep[see, e.g.,][]{oh2021raking, boe2024practical}, as demonstrated by our vignette \citep{shaw2024comparison}. To obtain the EIF estimates $\hat{h}_i$, we use MICE with $M = 10$ imputations: for each imputation $m$, we first obtain imputed values of $W$ for all study participants; second, we fit the target outcome model using imputed rather than observed values for the phase II variables, which is the key to constructing  auxiliary variables that are functions only of phase I data and nuisance parameters \citep{shepherd2023multiwave}; and finally we obtain the EIFs for all regression parameters based on the fitted model. The final $\hat{h}_i$ are the averages of the EIFs over the $M$ imputed datasets. Note that, unlike MI, GR is a doubly-robust approach and thus does not need the imputation model to be correct for consistent estimation, so long as the outcome model is correct (which requires confounding to be adjusted for appropriately). A correct imputation model will confer better efficiency.  We use the R package \texttt{survey} \citep{lumley2004survey,surveypkg} to implement calibration for the GR estimator. Finally, for both IPW and GR, we use weighted logistic regression to fit the outcome model.

The MI, IPW, and GR strategies all naturally target the census cOR, because they are based on the working parametric outcome regression model. They also provide an estimator of the outcome probability, which we can use to estimate marginal quantities. Suppose that we have an estimator $Q_n(x, z, w) = \widehat{P}(Y = 1 \mid X = x, Z = z, W = w)$ obtained using logistic regression. Suppose further that we have a set of weights $\hat{\pi}_i^{-1}$; for MI, $\hat{\pi}_i^{-1} = 1$ for $i = 1, \ldots, n$, while for IPW and GR $\hat{\pi}_i^{-1}$ are  inverse probability weights (calibrated weights for GR). To estimate $\mu_1$ (the mean outcome value under treatment), we can first use standard software to predict using the fitted logistic regression model after setting $X = 1$ for all observations, yielding $Q_{n}(1, Z_i, W_i) = \widehat{P}(Y = 1 \mid X_i = 1, Z_i, W_i)$; we then take an average over the empirical distribution of the covariates:
\begin{align*}
\mu_{1,n} = \frac{1}{n}\sum_{i=1}^n R_i\hat{\pi}^{-1}_iQ_n(1, Z_i, W_i).
\end{align*}
This procedure requires that $\hat{\pi}$ is a consistent estimator of $\pi$ and that $Q_n$ is a consistent estimator of $Q$. We do not implement further augmentation \citep{breslow2009improved} because our aim was to compare commonly-used approaches (without augmentation) with augmented procedures like GR.
The same procedure applied after setting $X = 0$ for all observations yields an estimator $\mu_{0,n}$ of $\mu_0$. These estimators can be plugged in to yield estimators of the mRD, mRR, and mOR: 
\begin{align*}
    mRD_n = & \ \mu_{1,n} - \mu_{0,n} \\
    mRR_n = & \ \frac{\mu_{1,n}}{\mu_{0,n}} \\
    mOR_n = & \ \frac{\mu_{1,n}/(1 - \mu_{1,n})}{\mu_{0,n}/(1 - \mu_{0,n})}.
\end{align*}
The delta method \citep[see, e.g.,][]{bickel1993efficient} can be used to obtain a variance estimator. For MICE, this process is repeated for the $M$ imputed datasets; the resulting point and variance estimators are combined using Rubin's rules \citep{rubin1987multiple}. We use the R package \texttt{marginaleffects} to perform the delta method \citep{marginaleffectspkg}. We illustrate in a vignette how to implement marginal estimates and confidence intervals for each of IPW, GR, and MI in R \citep{shaw2024comparison}. If a linear or Poisson regression model is fit instead (e.g., if there is interest in the cRD or cRR), similar steps can be taken to estimate $P(Y = 1 | X_i = 1, Z_i, W_i)$ using the fitted regression model. The marginalization proceeds exactly as outlined above.

Finally, we consider a TMLE approach. In our missing-data setting, we use the inverse probability of coarsening weighted (IPCW)-TMLE \citep{rose2011targeted}. As with the IPW estimators described above, IPCW-TMLE is a weighted complete-case estimator. Following common guidelines \citep[see, e.g.,][]{phillips2023practical}, we  use an ensemble of candidate learners (prediction functions) to estimate the probability of missing data, ranging from simple (e.g., generalized linear models) to more complex (e.g., random forests \citep{breiman2001rf}). Specifically, we use the convex ensemble super learner, which is the convex combination of the individual algorithms with weights chosen to minimize a cross-validated loss function \citep{vanderlaan2007super}; we let $\pi_n$ denote the final estimator of $\pi$. Let  $Q_n$ be an initial estimator of $Q$, and $g_n$ of $g$ (possibly also obtained using a super learner). To obtain the IPCW-TMLE, we fluctuate $Q_n$ in a direction defined by the EIFs for the treatment-specific means; the direction is given by
\begin{align*}
    H_1(x,z,w) =& \ I(x = 1) \text{ and }
    H_0(x,z,w) = I(x = 0).
\end{align*}
We obtain the fluctuation using logistic regression with outcome $Y$, covariates $[H_1(X_i,Z_i,W_i),H_0(X_i, \allowbreak Z_i,W_i)]$, offset $Q_n(X_i,Z_i,W_i )$, and weights 
\begin{align*}
\left\{\frac{I(X_i = 1)}{g_n(Z_i, W_i)} + \frac{I(X_i = 0)}{1 - g_n(Z_i, W_i)}\right\}\frac{R_i}{\pi_n(X_i, Z_i)}.
\end{align*}
This yields parameter estimates $(\epsilon_{1,n}, \epsilon_{0,n})$, and our fluctuated estimator of $Q$ is
\begin{align*}
    \logit \{Q_n^*(x,z,w)\} =& \ \logit \{Q_n(x,z,w)\} + \epsilon_{1,n}H_1(x,z,w) + \epsilon_{0,n}H_0(x,z,w).
\end{align*}
The final IPCW-TMLE of the mRD is
\begin{align*}
    \frac{1}{n}\sum_{i=1}^n \frac{R_i}{\pi_n(X_i, Z_i)}Q_n^*(1, Z_i, W_i) - \frac{1}{n}\sum_{i=1}^n \frac{R_i}{\pi_n(X_i, Z_i)}Q_n^*(0, Z_i, W_i).
\end{align*}
Similar calculations can be performed for the mRR and mOR. 
We use the R package \texttt{tmle} \citep{gruber2012tmle,twoStageDesignTMLEpkg} to implement the IPCW-TMLE. 

To estimate the cOR using TMLE, recall that the cOR is defined in terms of a set of variables (here, $Z$ and $W$). This cOR can be expressed as the parameter in a working model known as a marginal structural model (MSM) \citep{robins2000marginal,gruber2012tmle}. The MSM is a parametric model specifying a relationship between the outcome and certain covariates. In our case, to facilitate comparisons between the TMLE and the MI and design-based estimators described above, a natural MSM is exactly the working logistic regression model used in those procedures. We specify the MSM
\begin{align*}
    \logit P(Y = 1 \mid X = x, Z = z, W = w) =& \ \beta_0 + \beta_1 x + \beta_2 z + \beta_3 w;
\end{align*}
the cOR $\exp(\beta_1)$ can be interpreted as the projection of the true causal effect parameter onto the working model \citep{gruber2012tmle}. Importantly, if we used a different MSM from the logistic regression model defined above for our other estimators, the resulting cORs would not be the same and results would not be directly comparable. 

Understanding the target of estimation is an important consideration when selecting an estimation procedure. Estimators that are targeted towards the census cOR include MI, the design-based estimators, and the IPCW-TMLE for the cOR (by construction). In contrast, the IPCW-TMLEs for the mRD, mRR, and mOR are all targeted towards the oracle estimand. MI relies heavily on the working outcome regression model and will generally be biased when this is misspecified or if the data are missing not-at-random; MI will be robust to complex missing-at-random patterns if a sufficiently flexible algorithm (e.g., MICE) is used. IPW can be biased when the missingness mechanism is misspecified. GR is subject to bias when both the missingness mechanism and outcome regression model are misspecified. The IPCW-TMLEs described above will generally be biased when the missing-data model is misspecified or when both the outcome regression model and propensity score model are misspecified. See Table~\ref{tab:required_models} for a summary of estimators and requirements for consistency.         

\section{Numerical study with synthetic data} \label{synthetic}

\subsection{Simulation methods}

\subsubsection{Overview}
We considered several data-generating mechanism (DGM) scenarios to extensively examine the relative performance of the estimators described in Section~\ref{sec:estimators}. We vary the missing-data model, the outcome model, the outcome probability (12\% and 5\%), and the proportion of missing data (40\% and 80\%). Across scenarios, we fix the parametric analysis models  (e.g. working logistic regression outcome models). We were specifically interested in relatively high levels of missing data, which  occur in our motivating setting.  We considered two missing-at-random (MAR) scenarios: simple and complex, in which the simple scenario missingness mechanism coincides with the simple working missing-data model used in IPW and the complex does not. Similarly, we consider a simple and complex outcome DGM, where only the simple DGM is aligned with the assumed working analysis model. Finally, we considered a set of missing-not-at-random (MNAR) scenarios. For all scenarios, we fixed the population size at N=10,000. The outcome probabilities were chosen to be similar to that in the plasmode data example. The non-null treatment effect was selected so that it was moderate, but still detectable with generally at least 80\% power for the more efficient methods, even in the scenario of high missingness and lower outcome probability. Below, we provide further details of our simulation specifications.

\subsubsection{Base case}
We consider as a \textit{base case} a simple logistic missing-data model, a simple generalized linear outcome model, approximately 12\% outcome probability, and 40\% missing data, where all estimators were expected to perform well. We first generated data 
{$(X_\text{latent}, Z_1, Z_2, W_1, W_2, \allowbreak U_1, U_2) \sim N(0, \Sigma)$,}  
where $X_\text{latent}$ is a latent continuous variable associated with treatment $X$; $Z_1$ and $Z_2$ are always-observed confounders; $W_1$ and $W_2$ are confounders subject to missingness; and $U_1$ and $U_2$ are never-observed possible confounders. $\Sigma$ is defined such that all variables have variance 1, each element of $(Z_2,W_2,U_2)$ has correlation 0.4 with $X_\text{latent}$, and otherwise pairwise correlations in $\Sigma$ are 0.2.
Next, we dichotomized the exposure by setting 
$X_i = 1$ if $X_{i, \text{latent}}$ was less than the 40th percentile of $X_\text{latent}$. 
Third, we generated outcomes according to 
\begin{align*}
 \logit P(Y = 1 \mid X = x, W = w, Z = z) =& \ -2.4 + \ln(1.5)x + \ln(1.5)w_1 \\
 & -\ln(1.75)w_2 + \ln(1.5)z_1 - \ln(1.3)z_2.
\end{align*}
This model was designed to have a moderate treatment effect and clear confounding by both $Z$ and $W$, and to be simple enough for parametric approaches to estimate.
We then generated missing-data indicators according to 
\begin{align*}
 \logit P(R = 1 \mid X = x, Z = z, Y = y) =& \ -0.67 + \ln(2.5)x + \ln(1.5)z_1 + \ln(1.5)z_2 + \ln(2.5)y.
\end{align*}
As with the outcome model, this model was designed to be estimable by simple parametric approaches. We refer to this missingness model as MAR, since all variables that determine the missingness are observed. This model is MAR with respect to the sampling (IPW) model; however, with respect to the complete case outcome model, it is not MAR (and will not lead to valid inference) since the missingness depends on the outcome. This distinction shows the ambiguity inherent in the MAR terminology \citep{seaman2013meant}.  See Section~\ref{sec:synthetic-data-gen} (Figure~\ref{fig:mar-diagram}) for a causal diagram describing this scenario.
Finally, we set $W_{1,\text{obs}} = W_1$ and $W_{2,\text{obs}} = W_2$ if $R = 1$ and missing otherwise. This model ensures that approximately 40\% of participants are missing data on $(W_1, W_2)$.

\subsubsection{Complex outcome and MAR DGM scenarios}
The more complex DGM outcome model was 
\begin{align*}
 \logit P(Y = 1 \mid X = x, W = w, Z = z) =& \ -3 + \ln(1.5)x - 0.6 w_1 + 0.5 w_2 \\
 &+ 0.1 I(z_1 < -0.5) + 0.8 I(z_1 > 2) - 0.4 I(z_2 < -1) \\
 &+ 1 w_1w_2 + 3 w_2I(z_2 < -1) + 1 w_1 I(z_1 > 2).
\end{align*}
The more complex DGM missing-data model was 
\begin{align*}
\logit P(R = 1 \mid X = x, Z = z, Y = y) =& \ -0.9 + x -2I(z_2 < -1) + 2I(z_1 > 1) \\
&-0.9 I(z_1 < -0.5) + 3I(z_1 > 1)x + 0.2y -3I(z_2 < -1)y.
\end{align*}
Both of these models include nonlinear terms and interactions that the analyst does not model, and were designed to highlight that even relatively minor complexities can result in biased results if ignored. 

\subsubsection{MNAR DGM scenarios}
We considered two MNAR scenarios, similar to \cite{weberpals2024principled}. In the first (MNAR-value, Figure~\ref{fig:mnar-value-diagram}), $R_i$ depends on the value of the missing covariate:
\begin{align*}
  \logit P(R = 1 \mid X = x, Z = z, W = w, Y = y) =& \ -0.67 + \ln(2.5)x + \ln(1.5)z_1 + \ln(1.5)z_2 \\
  &+ \ln(2.5)y + \ln(2.5)w_1 + \ln(2.5)w_2.
\end{align*}
In the second (MNAR-unobserved, Figure~\ref{fig:mnar-unobserved-diagram}), $R_i$ depends on an unobserved variable $U$:
\begin{align*}
  \logit P(R = 1 \mid X = x, Z = z, U = u, Y = y) =& \ -0.97 + \ln(2.5)x + \ln(1.5)z_1 \\
  &+ \ln(1.5)z_2 + \ln(2.5)u_2 + \ln(2.5)y.
\end{align*}
Outcome model misspecification in the MNAR-value scenarios followed the same form as above. For the MNAR-unobserved scenarios, we considered the outcome model
\begin{align*}
    \logit P(Y = 1 \mid X = x, W = w, Z = z, U = u) =& \ -2.4 + \ln(1.5)x + \ln(1.5)w_1 \\
 & -\ln(1.75)w_2 + \ln(1.5)z_1 - \ln(1.3)z_2 - \ln(1.75)u_2,
\end{align*}
which is the same as the simple outcome model above but for dependence on $U$, the unobserved variable. This additional dependence on $U$ introduces bias in the complete-case outcome model under the MNAR-unobserved mechanism \citep{lee2023assumptions}.

\subsubsection{Varying outcome and missingness probabilities}
Finally, we created scenarios with a rarer outcome (5\% probability) or higher missingness proportion (80\%) by modifying the appropriate intercept in the regression models specified above. In Supplementary Materials Section~\ref{sec:synthetic-data-gen} (Tables~\ref{tab:sim_x_scenarios}--\ref{tab:sim_y_scenarios}), we provide more details regarding the specific models used to generate data for all scenarios.

\subsubsection{Estimators} \label{subsubsec:estimators}

For each DGM Scenario, we considered a broad set of estimators (Table~\ref{tab:estimators}). We considered a benchmark model based on the (latent) ideal data $(Y, X, Z, W)$ and with the correctly-specified outcome regression model for the target estimand. The benchmark model for the census estimand (BNMK-C) is always the working regression model. The benchmark model for the oracle estimand (BNMK-O) is the data-generating model; when the working regression model is correctly specified (e.g., in the base case), the benchmark models are identical. We expect the benchmark to have the best performance for estimating any given parameter. The remaining estimators use the observed data $(Y, X, Z, R, RW)$ that are subject to missing values in $W$. We consider two estimators that commonly appear in practice that do not account for missing data: the complete-case (CC) estimator that drops observations with missing data and fits the working logistic regression model, and the confounded (CNFD) estimator that drops $W$ and fits a modified working logistic regression model. This latter model does not account for the measured confounders $W$, and thus we expect effect estimates to be biased.  We consider the IPW, GR, and MI estimators MICE and MI-RF in all scenarios. We also explored MI-XGB in a subset of base case scenarios, but due to poor performance in the plasmode simulations did not fully consider (further details in Section 4). Finally, we consider two implementations of the IPCW-TMLE. The first (T-M) uses a super learner to estimate the missing-data model and uses logistic regression models for the outcome regression and propensity score; the second (T-MTO) uses a super learner for all three nuisance functions, and is the way that TMLE is typically applied in practice. T-M was designed to explore the benefit of flexibly modeling the missing-data probability alone, providing a natural comparator to IPW, MI, and GR. In all cases, we use a 10-fold cross-validated ensemble super learner consisting of logistic regression, boosted trees, and random forests (described more fully in Tables~\ref{tab:sl_tmle_m} and \ref{tab:sl_tmle_mto}). We further considered two modifications to the TMLE procedures: 1) augmenting the dataset with a clever covariate \citep{rose2011targeted}, the fitted probability from the initial (confounded) outcome regression model dropping $W$ and 2) specifying a modified super learner library for rare outcome settings (Supplementary Materials Section~\ref{sec:modified-tmle}).



\subsubsection{Performance metrics}
For each scenario, we generated 2500 random independent datasets from the data-generating mechanism. For each dataset, we fit the estimators as described above, obtaining estimates of the cOR, mOR, mRD, and mRR and analytic standard errors (ASEs). We assessed performance by computing the mean and median bias (estimate - estimand); median percent bias; empirical standard error (ESE) of the point estimates; robust root mean squared error (rRMSE), defined as $\sqrt{\text{median bias}^2 + \text{MAD}^2}$, where the median absolute deviation is $\text{MAD}(x) = \text{median}\{\lvert x - \text{median}(x) \rvert\}$; the nominal coverage, defined as the proportion of the 2500 replications where the estimand was contained within the nominal 95\% confidence interval using the ASE and appropriate normal distribution quantiles; and the oracle coverage, defined similarly to the nominal coverage but using the ESE rather than the ASE. In Section~\ref{sec:supp-tables-synthetic}, we report the proportion of replications where each estimator returned a result (100\% in all synthetic scenarios shown here).

\subsection{Simulation results}

\subsubsection{Base case (12\% outcome probability, 40\% MAR missing)}
\begin{figure}[!htb]
\caption[Missing at random, Census estimand, 12\% outcome probability, 40\% missingness]{\textbf{Synthetic Data MAR Simulation: Census clogOR}. Comparing estimators of the census clogOR estimand for the \textbf{base case} with 40\% missingness and 12\% outcome probability, where the data generating models were simple or complex. \textbf{Top graph}: Percent Bias (median, IQR, min and max of converged simulations); \textbf{Middle graph}: Robust RMSE (rRMSE), using median bias and MAD; \textbf{Bottom graphs}: Nominal and oracle coverage, respectively, with blue confidence bands at $ .95 \pm 1.96 \sqrt{\frac{.05\cdot .95}{2500}}$. True clogOR values are 0.405 and 0.371 for simple and complex outcome models, respectively.}

\includegraphics[scale=0.65]{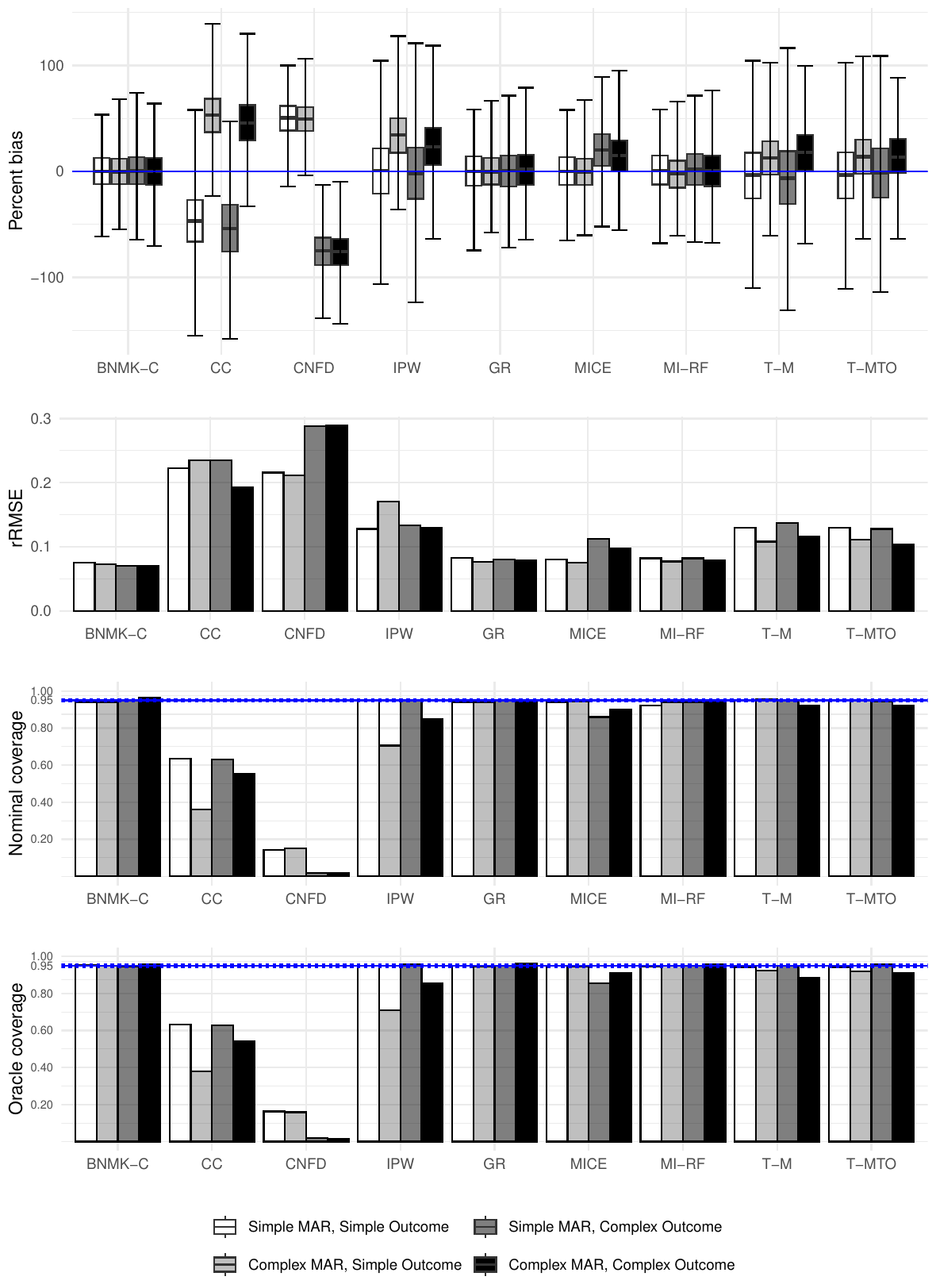}
\label{fig:clogOR_mar_census_highout_lowmissing}
\end{figure}

\begin{figure}[!htb]
\caption[Missing at random, oracle estimand, 12\% outcome probability, 40\% missingness]{\textbf{Synthetic Data MAR Simulation: Oracle clogOR}. Comparing estimators of the oracle clogOR estimand for the \textbf{base case} with 40\% missingness and 12\% outcome probability, where the data generating models were simple or complex. \textbf{Top graph}: Percent Bias (median, IQR, min and max of converged simulations); \textbf{Middle graph}: Robust RMSE (rRMSE), using median bias and MAD; \textbf{Bottom graphs}: Nominal and oracle coverage, respectively, with blue confidence bands at $ .95 \pm 1.96 \sqrt{\frac{.05\cdot .95}{2500}}$. The true clogOR value is 0.405 for both simple and complex models.}

\includegraphics[scale=0.65]{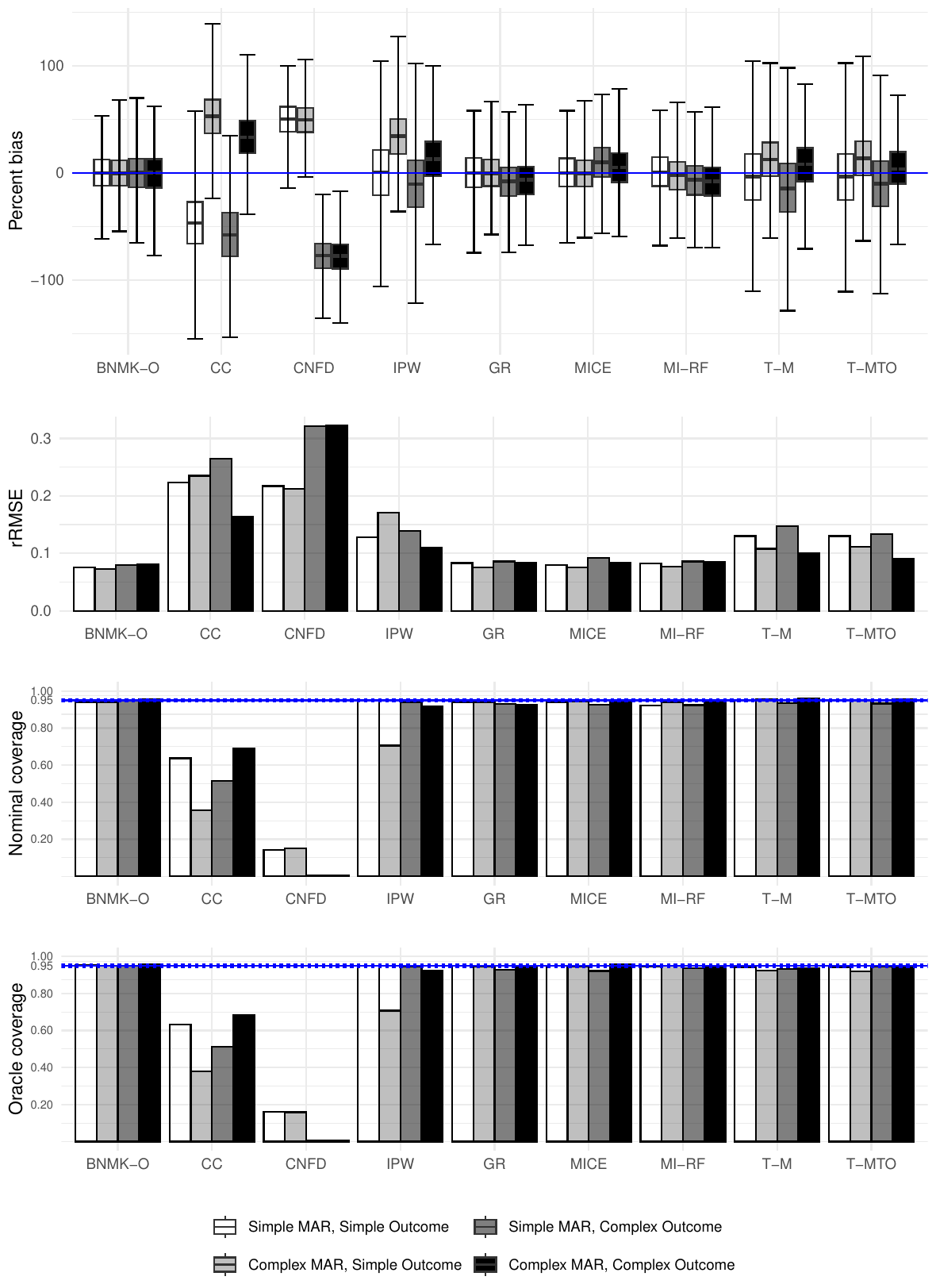}
\label{fig:clogOR_mar_oracle_highout_lowmissing}
\end{figure}

Figure \ref{fig:clogOR_mar_census_highout_lowmissing} (Figure \ref{fig:clogOR_mar_oracle_highout_lowmissing}) shows the estimator relative performance for the cOR census (oracle) estimand for the base case simulations (detailed results in Tables~\ref{tab:x1_m1.1_y1.1_cOR_census}--\ref{tab:x1_m2.2_y4.1_cOR_oracle}).
In the simple outcome and MAR scenario, the oracle and census estimands are the same and all methods, except the CNFD and CC approaches, maintained 95\% nominal coverage.  The GR estimator is consistent for the census estimand in all MAR cases because the analytic working model is the census model and so, as expected, maintained correct coverage for this parameter for these settings; unexpectedly, GR also maintained close to the correct coverage for both census and oracle estimands when the analytic working models both were misspecified (i.e. under the complex outcome and MAR scenario). GR also maintained the best efficiency for the census estimand and generally within about 5\% of the best efficiency for the oracle estimand (i.e. lowest rRMSE) relative to the other methods for all 4 combinations of simple versus complex data generating scenarios. Conversely, MICE suffered from bias and inefficiency when the true outcome model was complex (i.e., the working model was misspecified); this bias was worse for the census estimand. The MI procedures, even in the complex outcome scenario, were still more efficient than CC, IPW and the TMLE estimators; however, which imputation procedure did the best varied by setting. Generally the MI estimates were close when the outcome DGM was simple, with an apparent very slight advantage to MICE. When the outcome scenario was complex MI-RF had lower rRMSE than MICE for both the census and oracle estimands; MI-XGB tended to have lower rRMSE for the oracle cOR estimand while MI-RF had lower rRMSE for the census estimand (results for MI-XGB are only displayed in Tables~\ref{tab:x1_m1.1_y1.1_cOR_census}--\ref{tab:x1_m1.1_y4.1_cOR_census}, \ref{tab:x1_m1.1_y1.1_cOR_oracle}--\ref{tab:x1_m1.1_y4.1_cOR_oracle}). The TMLE estimators were generally more efficient than IPW and CC, but not as efficient as MI and GR. Further, TMLE estimators, which were implemented as weighted two-phase estimators, suffered from some bias when the missing data model was more complex, with oracle coverage only around 92\%; however, IPW fared much worse for the complex MAR model, with oracle (and nominal) coverage dropping to about 70\%.  TMLE-M was less efficient than TMLE-MTO for the complex outcome scenario. The CNFD method demonstrated severe bias, with nominal coverage under 20\%, across all scenarios. 

Figures~\ref{fig:mRD_mar_census_highout_lowmissing} and \ref{fig:mRD_mar_oracle_highout_lowmissing} show the mRD census and oracle estimands, respectively.  The relative performance for this estimand was similar to that seen for the cOR, with GR generally performing the best or within 5\% of the best method, which was sometimes an MI method for the simple outcome scenarios. For mRD,  MICE was generally the most efficient of the imputation methods for the simple outcome scenarios and MI-RF was otherwise the most efficient imputation method. One notable difference for the marginal estimand is that TMLE-MTO was the only estimator to maintain 95\% coverage for the oracle mRD when both the outcome and MAR scenarios were complex. Figures~\ref{fig:mlogRR_mar_census_highout_lowmissing} and \ref{fig:mlogRR_mar_oracle_highout_lowmissing} show results for the mRR census and oracle estimands, respectively; relative performance for mRR in this setting was similar to that for mRD. 

\subsubsection{Scenarios with varying outcome and missing data probabilities}

In the case of 12\% outcome probability and 80\% missingness, the methods' relative performance  was largely the same as the base case. Figures~\ref{fig:clogOR_mar_census_highout_highmissing}--\ref{fig:mlogRR_mar_oracle_highout_highmissing} show results for the census and oracle estimands for cOR, mRD and mRR for this setting. For the cOR, GR and MI were consistently the top two approaches in terms of relative efficiency and lowest rRMSE. Overall, GR was better at achieving 95\% coverage across all 4 scenarios (simple versus complex missingness and outcome models) for the cOR. While imputation methods approached closer to 10\% efficiency gains relative to GR when the outcome was simple, each imputation method suffered from bias: MI-RF showed less than 90\% coverage even when the outcome was simple, whereas MICE showed bias and less than the nominal coverage in the complex outcome scenario. In general, MI-RF had lower rRMSE and higher coverage than MICE, except for the simple outcome and MAR scenario. MI-RF also tended to overestimate standard errors, resulting in coverage greater than 95\% in some cases. The TMLE methods did not always improve on the efficiency of IPW, except in cases where there was a complex MAR mechanism. Similar relative performance was seen for the mRD, as in the base case; again, TMLE-MTO was the only method to maintain the nominal 95\% coverage for the oracle estimand under the complex DGM for both the outcome and missingness.

In the extreme case of 80\% missing data and 5\% outcome probability, there were approximately 50 expected cases amongst those with complete data for both the simple and complex outcome DGM. In this setting, the most efficient methods (e.g. MICE in setting of the simple outcome model) had approximately 80\% power or higher to detect the treatment effect cOR = .405 and mRD = .01, whereas the power for the least efficient methods (IPW and TMLE) could be as low as  10\%. Figures~\ref{fig:clogOR_mar_census_lowout_highmissing}--\ref{fig:mlogRR_mar_oracle_lowout_highmissing} show the results for  cOR, mRD, and mRR.  The relative performance remained the same. GR still generally maintained the best overall performance, having close to the nominal coverage for the census estimand and oracle cOR and mRD across all scenarios. TMLE continued to do well for the oracle mRD when both the missingness and outcome scenarios were complex, with TMLE-M the only estimator maintaining the 95\% nominal coverage, and TMLE-MTO coming close to it, as well as the having smallest rRMSE after GR. MICE had stronger gains in efficiency relative to the base case, reaching $\sim 15$\% for the simple outcome scenarios, but these gains disappeared for the complex outcome DGM scenarios. GR sometimes suffered from slight over coverage, MI-RF exhibited bias in most scenarios, and the TMLE methods were prone to underestimating the SE, resulting in lower nominal than oracle coverage. This underestimation was more exacerbated for the mRD than for the mRR.

\subsubsection{MNAR Scenarios}
\begin{figure}[!htb]
\caption[Missing not at random, census truth, high outcome, low missing]{\textbf{Synthetic Data MNAR Simulation: Census clogOR}. Comparing estimators of the census estimand with \textbf{40\% confounder missingness} and \textbf{12\% outcome probability}. \textbf{Top graph}: \%Bias (median, IQR, min and max of converged simulations); \textbf{Middle graph}: Robust RMSE (rRMSE), using median bias and MAD; \textbf{Bottom graphs}: Nominal and oracle coverage, respectively, with blue confidence bands at $ .95 \pm 1.96 \sqrt{\frac{.05\cdot .95}{2500}}$. True clogOR values are 0.405 for simple outcome models and 0.404 and 0.371 for complex outcome models for MNAR unobserved and MNAR value, respectively.}

\includegraphics[scale=0.65]{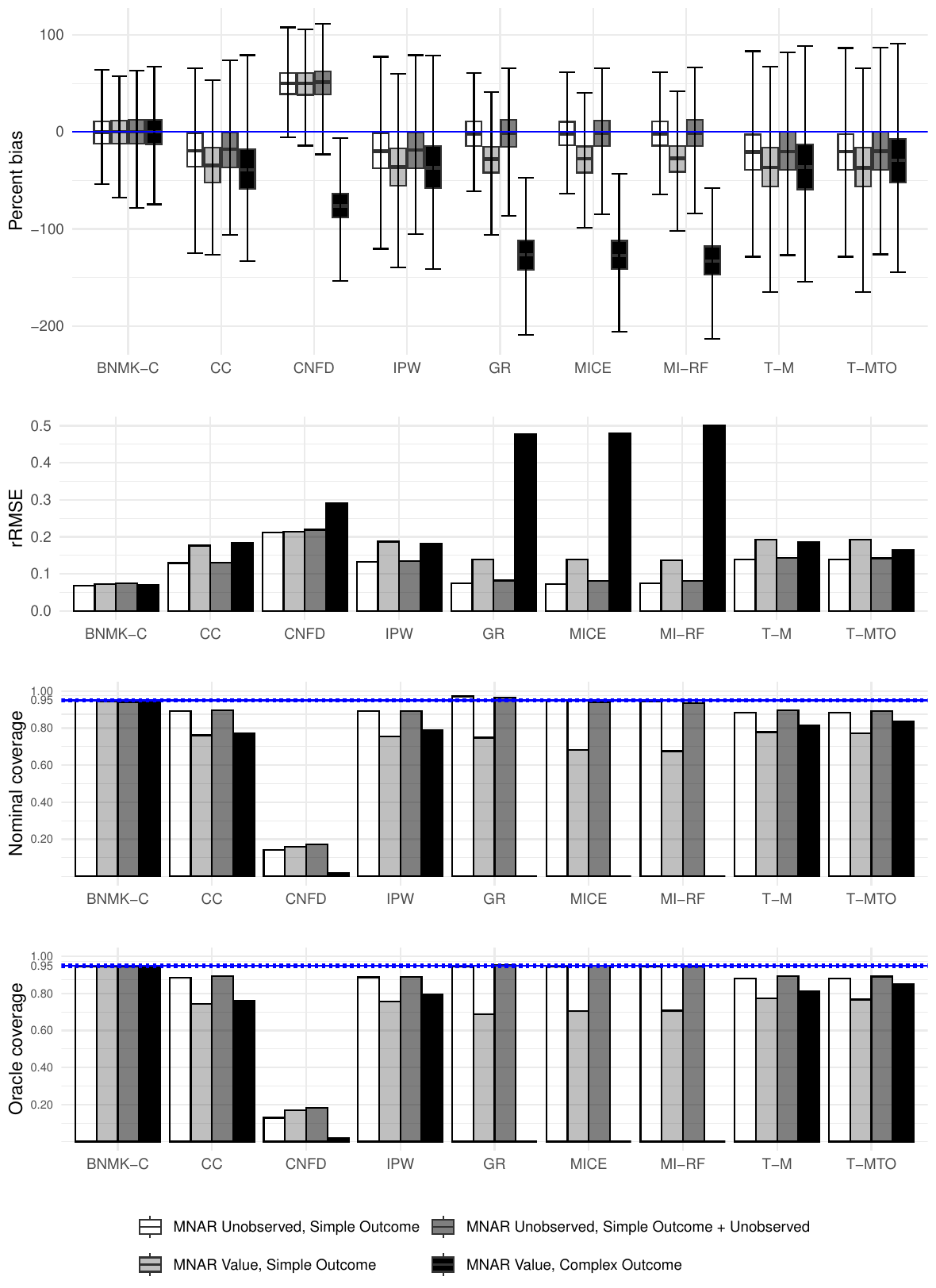}
\label{fig:clogOR_mnar_census_highout_lowmissing}
\end{figure}

\begin{figure}[!htb]
\caption[Missing not at random, oracle truth, high outcome, low missing]{\textbf{Synthetic Data MNAR Simulation: Oracle clogOR}. Comparing estimators of the oracle estimand with \textbf{40\% missingness} and \textbf{12\% outcome probability}. \textbf{Top graph}: \%Bias (median, IQR, min and max of converged simulations); \textbf{Middle graph}: Robust RMSE (rRMSE), using median bias and MAD; \textbf{Bottom graphs}: Nominal and oracle coverage, respectively, with blue confidence bands at $ .95 \pm 1.96 \sqrt{\frac{.05\cdot .95}{2500}}$. The correct clogOR value is 0.405 across all scenarios.}

\includegraphics[scale=0.65]{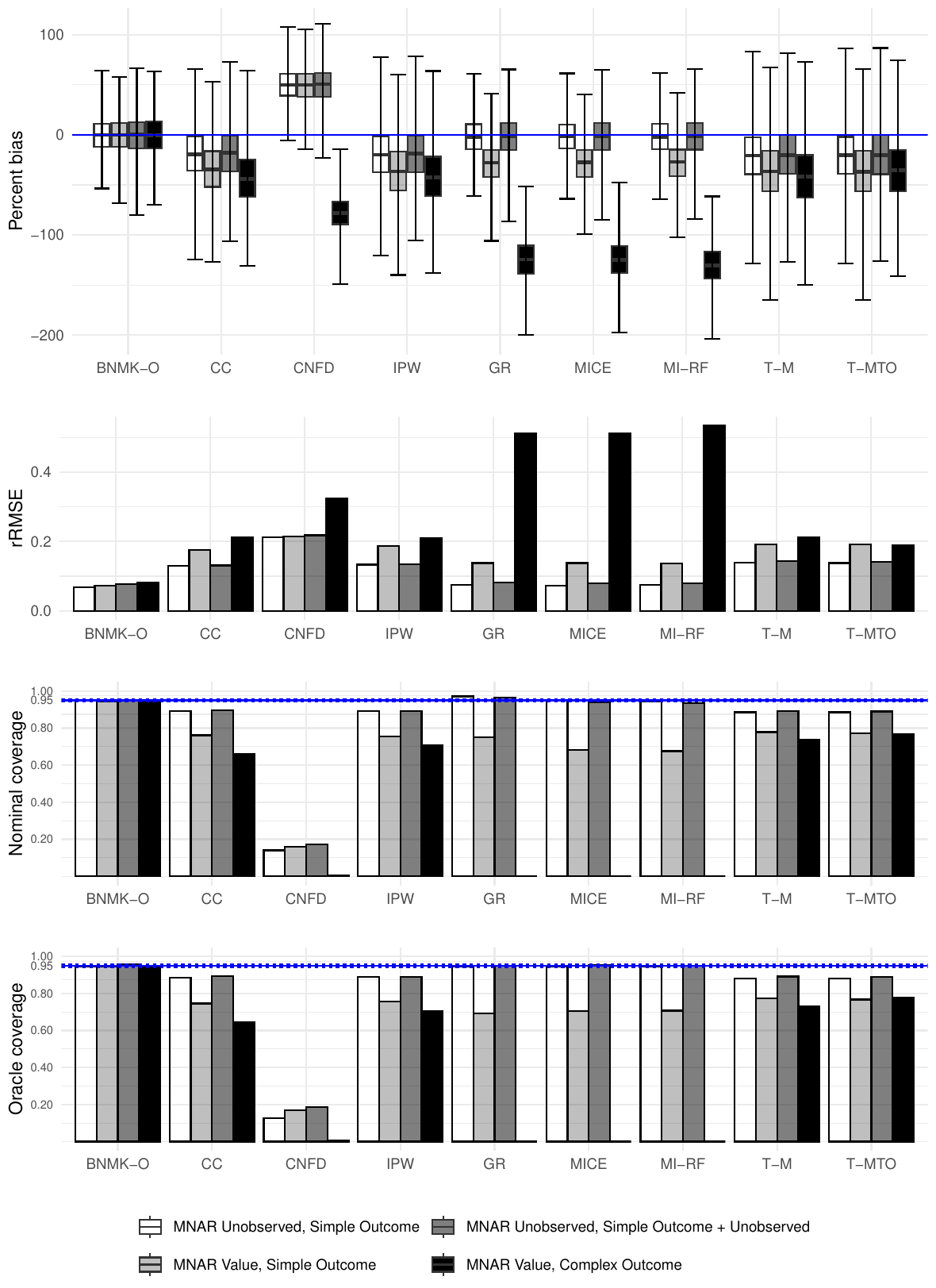}
\label{fig:clogOR_mnar_oracle_highout_lowmissing}
\end{figure}

Figures \ref{fig:clogOR_mnar_census_highout_lowmissing} and \ref{fig:clogOR_mnar_oracle_highout_lowmissing}  (Figures~\ref{fig:mRD_mnar_census_highout_lowmissing}--\ref{fig:mlogRR_mnar_oracle_highout_lowmissing}, Tables~\ref{tab:x1_m2.5_y1.1_cOR_census}--\ref{tab:x1_m2.6_y4.1_cOR_oracle}) show the MNAR scenarios for 40\% missingness and 12\% outcome probability for cOR (mRD, mRR) census and oracle estimands, respectively. Overall, all non-benchmark methods suffered from bias and less than the 95\% nominal coverage in the MNAR scenarios. Bias and coverage is worse for the MNAR-value scenarios and patterns of performance being a bit more complex across the different DGMs. For the MNAR-value scenarios, IPW and TMLE methods tended to maintain the highest coverage probability, but there were exceptions. In the simple outcome setting, GR and MI had the best coverage for the mRD. For lower outcome proportion and higher missingness, the nominal coverage of the GR estimator was highest and close to the 95\% level in part due to inflated SE estimation, particularly for cOR and mRR. Conversely, the nominal coverage of the TMLE approaches worsened relative to the oracle coverage, and were lower than the nominal 95\%, due to an underestimated SE in these scenarios. In the MNAR-value and complex outcome DGM, IPW and TMLE approaches were also the most efficient estimators. The CC estimator, while biased, also had amongst the lowest rRMSE in several MNAR-value scenarios. The bias for the CC estimator is largely dictated through dependence of the missingness indicator R on the outcome Y \citep{daniel2012using}, with weaker dependence leading to less bias. 

With MNAR-unobserved, TMLE and IPW methods had more bias and less efficiency relative to MICE and GR. In this setting, the relative performance of MICE and GR methods was similar to that for the MAR scenarios. For the high missingness and lower outcome probability scenarios, GR again suffered from variance inflation, and at times over coverage, and TMLE approaches suffered from underestimated SE (Figures~\ref{fig:cOR_mnar_census_highout_highmissing}--\ref{fig:mlogRR_mnar_oracle_lowout_highmissing}).


\section{Estimating the effect of anti-depressant therapy initiation on health outcomes using plasmode simulation}

\subsection{Study cohort}

We extracted electronic health records data for 112,770 individuals aged 13 years and older initiating antidepressant medication or psychotherapy for treatment of depression at Kaiser Permanente Washington (KPWA) from January 1, 2008 to December 31, 2018. Severity of depressive symptoms is a key prognostic variable, correlated with future depressive symptoms, self-harm, and mental health hospitalization. Thus, important potential confounders that should be controlled for in treatment comparisons come from the patient health questionnaire 9-item depression questionnaire (PHQ-9) \citep{kroenke2001} which is routinely collected and documented in the EHR. This FDA Sentinel Initiative project does not meet the criteria for human subject research as defined by Kaiser Permanente Washington Health Research Institute
policies, Health and Human Services, and the FDA (\citeauthor{commonrule,OHRP}). The study involves public health surveillance activity defined by HHS regulation 45 CFR 46.102(l)(2).

In distributed-data systems like the FDA Sentinel Initiative \citep{platt2018fda}, some data-contributing sites may only have ready access to insurance claims information; in these settings, depressive symptom severity would not be routinely available from these sites. In our sample, the PHQ-9 is available on only 50,337 individuals (45\%). We use the sum of first 8 items to summarize depressive symptoms and the 9th item as a measure of suicidal ideation~\citep{simon2016PHQ9SI}. Other variables we use as covariates are: sex, age at time of treatment initiation, Charlson comorbidity index score \citep{simard2018validation}, an anxiety diagnosis in the past year, alcohol use disorder in the past year, recorded self-harm in the prior 6 months, and hospitalization with a mental health diagnosis in the prior 6 months. See Supplementary Materials Section~\ref{sec:plasmode-gen} for further details. Table~\ref{tab:plasmodetab1} describes the study cohort overall and by treatment initiation status.
  
In a plasmode simulation study, we considered two outcomes with different frequency in our population: 1) a composite outcome of self-harm (fatal or non-fatal) or hospitalization with a mental health diagnosis in the 5 years following treatment initiation and 2) self-harm (fatal or non-fatal) in the 365 days following treatment initiation, with an outcome proportion in the complete-data cohort of 10.3\% (N=5193) and  0.7\% (N=358), respectively. We refer to these scenarios as the ``common'' and ``rare'' outcome scenarios.

\subsection{Plasmode simulation methods}

We seek to compare the same estimation methods described in Section \ref{subsubsec:estimators} for four sets of plasmode simulation studies, where we vary the rarity of the outcome and the complexity of the underlying DGMs. We used two different DGMs that were derived by fitting different models to the KPWA cohort: 1) a parametric generalized linear model (GLM) (i.e., logistic regression) with a few interactions and 2) a tree-based approach allowing for complex interactions. These two DGMs allow us to investigate the performance of estimators under a simple and more complex DGM, where the latter was chosen to be have a functional form notably different than the typical linear parametric working model. In both DGMs, we used the same approach (i.e., GLM or trees) to estimate the propensity score, outcome regression, and missing-data models. We only consider the BNMK-O estimator for the GLM DGM because there is no single oracle cOR of interest in the tree-based DGM.

We began by estimating the missing-data model using the entire sample of 112,770 individuals; in this model, the outcome was an indicator of whether the PHQ was observed or not. Next, we estimated the treatment propensity score and outcome regression models using the complete-data cohort (N=50,337). Importantly, using this complete-data cohort is convenient for plasmode simulation, but precludes making general conclusions about the larger population. Table~\ref{tab:plasmodetab2} and Figures~\ref{fig:treatTree}--\ref{fig:missTree} describe the final parametric and tree-based models, respectively. With these models fixed, we generate plasmode data as follows. We use bootstrap sampling with replacement from the complete-data cohort (i.e., individuals with fully observed PHQ-9) to generate complete covariate data for each person for 1000 plasmode datasets of size N=50,337, thereby maintaining the complex data structure in the KPWA cohort. For each DGM approach, we use the fitted models to generate treatment given the plasmode sample of  covariates, the outcome given the treatment and covariates, and then the missing data indicators for the PHQ-9 given the treatment, covariates and outcome for the 1000 plasmode datasets. Importantly, the data are MAR because the probability of missing data depends only on observed variables. In all cases, the working models that we use for analysis are misspecified, because the true data-generating models contain interactions that are not modeled (Table~\ref{tab:plasmodetab2}). The plasmode sampling for the simulations drew only from individuals with PHQ-9 measured (i.e., the complete-data cohort of size 50,337) and we use the generated missing data indicators to ``hide'' PHQ information for selected individuals.

Similar performance metrics as for the synthetic data simulations were calculated for each estimator across the plasmode simulations. Because the value of the DGM treatment effect estimands derived from the KPWA were close to null, bias instead of percent bias was calculated. The estimators returned results in $> 98$\% of plasmode datasets (Section~\ref{sec:supp-tables-plasmode}).

\subsection{Plasmode simulation results}

\begin{figure}[!htb]
\caption[mRD, census truth, bias]{\textbf{Plasmode Data Simulation: Census mRD}. Estimator relative performance for the census mRD estimand, across GLM and tree-based data generating models for the rare and common outcomes.  \textbf{Top graph}: \textbf{\textcolor{black}{Bias}} (median, IQR, min and max of converged simulations); \textbf{Middle graph}: Robust RMSE (rRMSE), using median bias and MAD; \textbf{Bottom graphs}: Nominal and oracle coverage, respectively, with blue confidence bands at $ .95 \pm 1.96 \sqrt{\frac{.05\cdot .95}{1000}}$. True values of mRD from lightest to darkest scenarios are 7.6E-4, 1.2E-4, -1.6E-2, and -6.2E-3.}

\includegraphics[scale=0.65]{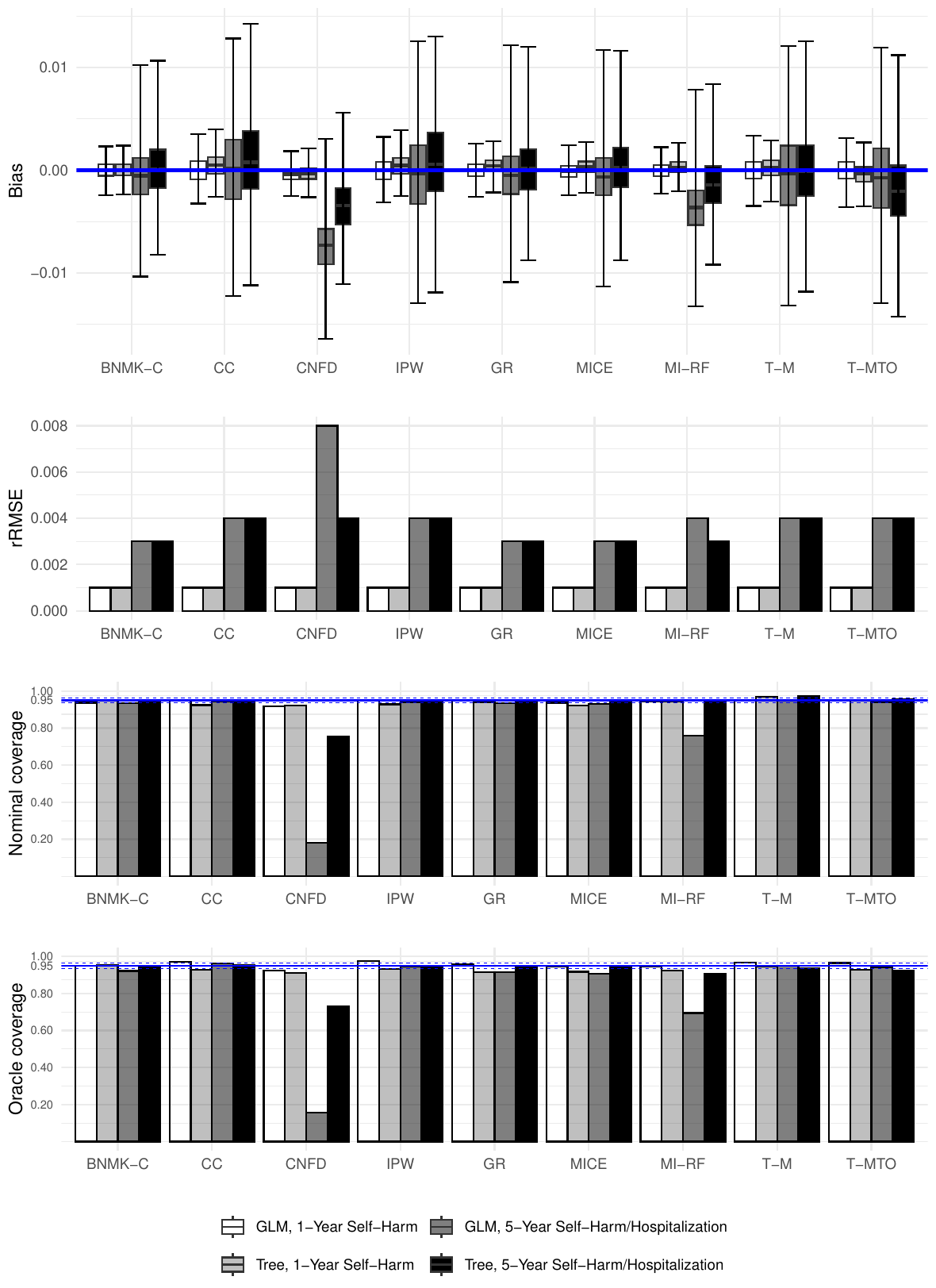}
\label{fig:plasmode_mRD_census}
\end{figure}

\begin{figure}[!htb]
\caption[mRD, oracle truth, regular bias]{\textbf{Plasmode Data Simulation: Oracle mRD}. Estimator relative performance for the oracle mRD estimand, across GLM and tree-based data generating models for the rare and common outcomes.  \textbf{Top graph}: Bias (median, IQR, min and max of converged simulations); \textbf{Middle graph}: Robust RMSE (rRMSE), using median bias and MAD; \textbf{Bottom graphs}: Nominal and oracle coverage, respectively, with blue confidence bands at $ .95 \pm 1.96 \sqrt{\frac{.05\cdot .95}{1000}}$. True values of mRD from lightest to darkest scenarios are 7.5E-4, -2.4E-4, -1.7E-2, and -7.6E-3.}

\includegraphics[scale=0.65]{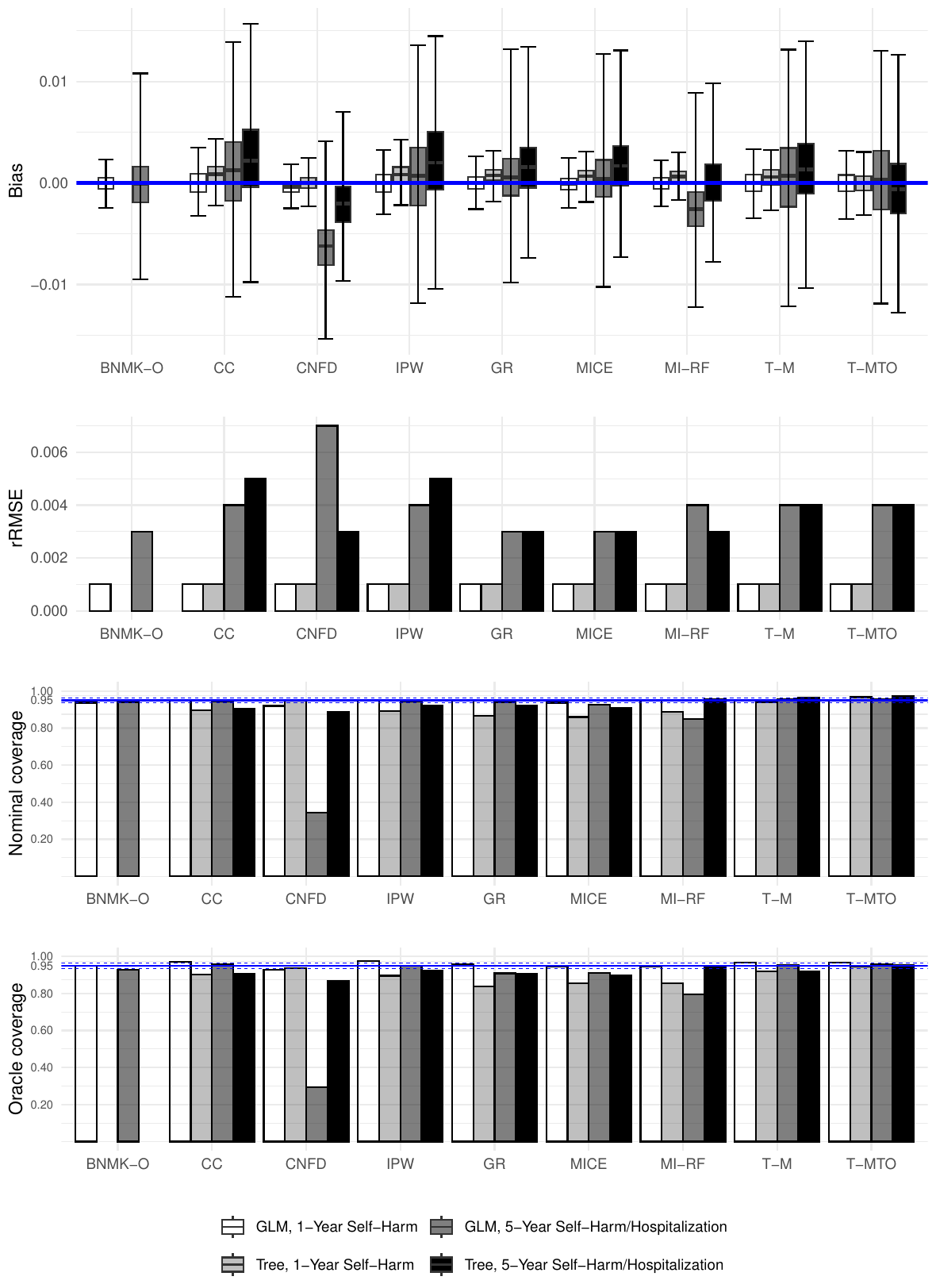}
\label{fig:plasmode_mRD_oracle}
\end{figure}

Figure \ref{fig:plasmode_mRD_census} (Figure \ref{fig:plasmode_mRD_oracle}) shows the relative performance of the estimation methods for the mRD census (oracle) estimand, with Figures~\ref{fig:plasmode_clogOR_census}--\ref{fig:plasmode_mlogRR_oracle} showing the results for cOR and mRR. For the common outcome, IPW, MICE, GR and the two TMLE approaches generally maintained low bias and 95\% empirical coverage for all estimands. TMLE maintained closer to 95\% coverage for the tree-based scenarios. The CNFD approach (which omits the PHQ variables from the regression model) was appreciably biased across both the GLM and tree DGM scenarios. GR and MICE performed similarly (rRMSE within 5\% of each other) and generally maintained the best efficiency with an approximately 20--40\% efficiency advantage over the CC, IPW and TMLE methods; better efficiency gains were seen for the GLM DGM. MI-RF had mixed performance. For the cOR and marginal estimands under the GLM DGM scenario, MI-RF had some bias, with higher RMSE relative to MICE, and poor coverage; whereas for the marginal estimands under the tree DGM, MI-RF achieved the 95\% coverage and best relative efficiency.  MI-XGB suffered from bias and poor coverage (Tables~\ref{tab:xNA_mNA_yglm_SH_365day_mRD_census}--\ref{tab:xNA_mNA_ytree_SH_HOSP_1826day_mRD_oracle}); one difference between the synthetic and plasmode data scenarios is the covariates prone to missingness are continuous for the former and a mix of continuous and discrete for the latter. 
In Supplementary Materials Section~\ref{sec:XGB-troubles}), we describe work done to improve the performance of MI-XGB; however, since our goal was to evaluate the estimation approaches as they are implemented in current software, this investigation was limited and, ultimately, MI-XGB did not perform as well as MICE.

For the rare outcome, the relative performance of the estimators was similar for the GLM DGM but not for the tree-based DGM scenario. In the tree-based rare outcome scenario, MICE and GR had less than 95\% coverage (85 -- 92\%) for the cOR and marginal estimands. Interestingly, MI-RF did not maintain high coverage in the tree-based DGM scenarios. The TMLE-MTO maintained the best overall performance across all estimands with respect to maintaining close to the nominal 95\% coverage, with slight efficiency gains over TMLE-M. For the more complex tree DGM and rare outcome, the RMSE for TMLE-MTO was the smallest for the mRD. Given the very poor performance for the common outcome, MI-XGB was not implemented for this setting.


\section{Discussion}

Medications are approved for use in the population after a benefit-risk profile is established in randomized clinical trials, which are highly controlled settings. Often, the safety and comparative effectiveness of available medications must be further studied in post-approval settings, where studies commonly rely on imperfect EHR and claims data. Choosing an appropriate and efficient method to address missing data in studies reliant on these imperfect observational data, especially in settings of rare outcomes, is important in order to leverage all available patient data and to draw reliable conclusions regarding safety and effectiveness. An important component of this is incorporating domain knowledge into both the outcome regression model and missing-data model: careful specification can result in reduced bias and variance. Further, methods of analyses typically must be well-established algorithms implemented in widely available software for acceptability and accessibility to regulators and clinical researchers.

Through extensive synthetic and plasmode simulations, we investigated the performance of several standard approaches to addressing missing data --- including IPW and MI  --- and two more robust methods, GR and IPCW-TMLE. We evaluated all methods with respect to both a census estimand, defined by a working outcome regression model; and an oracle estimand, defined purely as a summary of the population distribution. 

Our numerical experiments highlight the importance of first choosing a target estimand and then determining an estimation procedure. As expected, in cases where the working outcome regression model coincided with the data-generating outcome model and a simple MAR missing data model (the ``base case'' simulations), all methods performed well, with bias near zero and confidence interval coverage near the nominal 95\% level. However, even in this standard case we noted several important findings. The doubly-robust GR approach, which reduces to a tailored re-weighting of complete cases, maintained nearly the same efficiency as MI, the method generally expected to be the most efficient for correctly-specified parametric models. While MI can be highly efficient, the MI method (e.g., MICE or MI-RF) that performs best can vary across scenarios, but we do not know the DGM scenario in practice; for this reason, GR is often a safer choice than MI. Perhaps less surprising, both the CNFD and CC approaches were biased and highly inefficient; and  IPW and TMLE were inefficient. Importantly, the CNFD model --- resulting from omitting the confounders with missing values --- is always biased and is not recommended; all missing data approaches that we evaluated outperformed the CNFD approach.

If the working outcome regression model is misspecified, the census and oracle estimands are not equal and the relative performance of each method depends on its target. For the census estimand,  GR performed well, often having close to the nominal 95\% coverage and the best relative efficiency, with theoretical consistency for the oracle esimand if the missingness or outcome model is correctly specified. Correct model specification is a tall order. \cite{han2021combining} showed that even in settings where the amount of outcome model misspecification was small enough to remain undetectable as sample sizes approached infinity (i.e., an asymptotically local alternative), GR can still provide efficiency advantages over MI, in terms of smaller RMSE for the census estimand. In the current work, even in cases where the missingness mechanism and outcome models were both misspecified, GR generally maintained similar or better RMSE than MI. If the oracle estimand is of interest, we found TMLE often maintained consistency but with larger variance relative to the other estimators (due to relaxed parametric assumptions), while GR had some bias but a small variance. This bias-variance tradeoff can be acceptable in some cases, for example, postmarket safety studies where detecting risk is a primary goal. Notably in both the synthetic data and plasmode simulations, when both the missingness and outcome regression models were complex, GR often still performed well. The TMLE approaches became advantageous for some of the more complex data scenarios, consistently maintaining good coverage when MI and GR estimators did not, particularly for the marginal estimands and in the rare-outcome plasmode setting. In these cases the TMLE-MTO also maintained the best efficiency for the oracle mRD.

 We observed some underestimation of standard errors for TMLE in the rare outcome setting. This is expected when using influence-curve based standard error estimators in small-sample settings. An alternative is to use the targeted bootstrap for variance estimation \citep{gruber2012tmle,coyle2018targeted}.

There are several limitations to our findings. We only considered settings with missing confounder data. 
We assumed that the confounding could be controlled through adjustment; further research is necessary in cases where the confounders are not easily modeled. Our numerical experiments, while extensive, reflect a limited set of possibilities for model misspecification. Relative performance of the methods studied here may differ in other settings.  Using both synthetic and plasmode simulations is a useful tool for understanding performance; for example, we found that MI-XGB performed poorly with categorical covariates in our plasmode simulations but performed well with continuous covariates in our synthetic simulations.

In practice, we do not know whether our models are misspecified. After carefully choosing a target estimand, one pragmatic approach is to fit both doubly-robust methods that we considered here, with one designated as the primary procedure and the other a sensitivity analysis. If the GR point estimate agrees with that of TMLE, this provides at least some encouraging evidence that the census and oracle estimands are similar. If the point estimates differ, it may be that either the missing-data model in the TMLE is misspecified or the census and oracle estimands are different. This information can aid in interpreting the final result. The (often) wider confidence interval returned by TMLE reflects weaker assumptions than GR; by comparing against the interval returned by GR, one can examine how these assumptions impact inference. 

Both GR and TMLE approaches for handling missing data are straightforward to implement in standard software. We developed detailed vignettes to further facilitate and encourage their use. Together, these two approaches provide a thorough analysis that is efficient and robust against model misspecification.

\clearpage

\begin{table}
    \begin{threeparttable}
    \begin{tabular}{p{0.6in}|p{0.65in}p{1.1in}p{1.2in}p{1.1in}}
         & & \multicolumn{3}{c}{Correct specification for consistent estimation of:} \\
       Analysis approach  & Required models for estimation & oracle parameters & census parameters in regression models & marginal census parameters\\
       \hline
        IPW & $Q$ and $\pi$ & $Q$ and $\pi$ & $\pi$ or CD-MCAR & $\pi$ or (Q and  CD-MCAR)\\
        MI & $Q$ and $f$ & $Q$ and $f$ & $f$ & $f$ \\
        GR & $Q$, $\pi$ and ($\eta$ or $\eta_Q$)  & $Q$ and ($\pi$ or $\eta$) & $\pi$ or $\eta_Q$ or CD-MCAR & ($\pi$ or $\eta_Q$) or ($Q$ and CD-MCAR)\\
        IPCW-TMLE & $Q$, $g$ and $\pi$ & ($Q$ or $g$) and $\pi$ & $\pi$ or [($Q$ or $g$) and CD-MCAR] & $\pi$ or [($Q$ or $g$) and CD-MCAR] \\ \hline
    \end{tabular}
    \caption{Model specifications for missing-data estimation procedures and requirements for consistency for census and oracle estimands, under missing-at-random and assuming that there are no unmeasured confounders. The working outcome model $Q$ is assumed to be the same as the census model. The missing-data model is $\pi$, the imputation model is $f$, the treatment/exposure propensity score model is $g$, and the augmentation term/optimal raking variable for the oracle parameter is $\eta$ (a property of the data-generating mechanism) and for the census parameter is $\eta_Q$ (which depends on both $Q$ and the data-generating mechanism).}
    \label{tab:required_models}
    \begin{tablenotes}
        \item IPW: inverse probability weighted outcome regression where weights account for missing data; CD-MCAR: covariate-dependent missing completely at random, where missing data can depend only on always-observed covariates, not outcomes \citep{seaman2013meant}; MI: multiple imputation; GR: generalized raking; IPCW: inverse probability of coarsening weighted; TMLE: targeted minimum loss-based estimation.
    \end{tablenotes}
    \end{threeparttable} 
\end{table}

\clearpage

\begin{table} \label{estimators}
    \centering
    \begin{tabular}{p{1.8 in}|lll}
         &  &  & Natural target  \\ 
         Estimator & Abbreviation & Input data & of estimation \\
         \hline
        Census Benchmark & BNMK-C  & Ideal data $(Y,X,Z,W)$ & All \\ 
        Oracle Benchmark & BNMK-O & Ideal data $(Y,X,Z,W)$ & All \\
        Complete-case & CC & $(RY, RX, RZ, R, RW)$ & cOR \\
        Confounded & CNFD & $(Y,X,Z)$ & cOR \\
        Inverse probability weighting & IPW & $(Y, X, Z, R, RW)$ & cOR \\
        Generalized raking & GR & $(Y, X, Z, R, RW)$ & cOR \\
        MI via chained equations & MICE & $(Y, X, Z, R, RW)$ & cOR \\
        MI with random forests & MI-RF & $(Y, X, Z, R, RW)$ & cOR \\
        IPCW-TMLE with simple outcome and propensity score model & T-M & $(Y, X, Z, R, RW)$ & mRD, mRR, mOR\\
        IPCW-TMLE & T-MTO & $(Y, X, Z, R, RW)$ & mRD, mRR, mOR\\ \hline
    \end{tabular}
    \caption{Estimators considered across all simulation scenarios.*}
    \label{tab:estimators}
    \footnotesize{$^*$MI: multiple imputation; IPCW, inverse probability of coarsening weighted; TMLE, targeted minimum loss-based estimation} 
\end{table}

\clearpage

\ifarXiv

\renewcommand{\thesection}{S\arabic{section}}
\renewcommand{\thetable}{S\arabic{table}}  
\renewcommand{\thefigure}{S\arabic{figure}}
\renewcommand{\figurename}{Supplemental Figure} 
\renewcommand{\tablename}{Supplemental Table} 
\renewcommand{\theequation}{S\arabic{equation}}
\setcounter{table}{0}
\setcounter{figure}{0}
\setcounter{tocdepth}{2}
\setcounter{section}{0}
\setcounter{equation}{0}

\newif\ifnotarXiv
\notarXivfalse

\ifarXiv
\section*{SUPPLEMENTARY MATERIALS}
\fi

\ifnotarXiv
\usepackage{authblk}
\usepackage{graphicx} 
\usepackage{subcaption} 

\usepackage{amsmath,amssymb}
\usepackage{natbib}
\usepackage{longtable}
\usepackage{array}
\usepackage[margin=0.75in]{geometry}
\usepackage[capposition=top]{floatrow} 
\usepackage[colorlinks=true,linkcolor=cyan,urlcolor=cyan]{hyperref} 
\hypersetup{
    colorlinks,
    citecolor=black,
    filecolor=black,
    linkcolor=black,
    urlcolor=black
}
\usepackage{silence} 
\usepackage[dvipsnames]{xcolor}

\usepackage{tikz}
\usepackage{float}
\usetikzlibrary{shapes.geometric, arrows.meta, decorations,decorations.markings}
\tikzstyle{standard} = [rectangle, rounded corners, minimum width=2cm, minimum height=1cm,text centered, draw=black]
\tikzstyle{arrow} = [thick,-latex]

\WarningFilter*{latex}{Text page \thepage\space contains only floats}

\WarningFilter*{latex}{Float too large for page by}

\newcommand{\single}{\baselineskip 15pt}
\DeclareMathOperator{\logit}{logit}
\documentclass{article}

\title{Supplementary Materials for ``Assessing treatment effects in observational data with missing confounders: A comparative study of practical doubly-robust and traditional missing data methods''}

\author[1,2]{Brian~D. Williamson}
\author[1]{Chloe Krakauer}
\author[1]{Eric Johnson}
\author[3]{Susan Gruber}
\author[4]{Bryan~E. Shepherd}
\author[5]{Mark~J. van der Laan}
\author[6]{Thomas Lumley}
\author[7]{Hana Lee}
\author[8]{José J. Hernández-Muñoz}
\author[7]{Fengyu Zhao}
\author[8]{Sarah~K. Dutcher}
\author[9]{Rishi Desai}
\author[1]{Gregory~E. Simon}
\author[1,2]{Susan~M. Shortreed}
\author[1,2]{Jennifer C. Nelson}
\author[1,2]{Pamela A. Shaw}

\affil[1]{Biostatistics Division, Kaiser Permanente Washington Health Research Institute, Seattle, WA, USA}
\affil[2]{Department of Biostatistics, University of Washington, Seattle, WA, USA}
\affil[3]{TL Revolution, LLC, Cambridge, MA, USA}
\affil[4]{Department of Biostatistics, Vanderbilt University, Nashville, Tennessee, USA}
\affil[5]{Department of Biostatistics, School of Public Health, University of California at Berkeley, Berkeley, CA, USA}
\affil[6]{Department of Statistics, University of Auckland, Auckland, New Zealand}
\affil[7]{Office of Biostatistics, Center for Drug Evaluation and Research, US Food and Drug Administration, Silver Spring, MD, USA}
\affil[8]{Office of Surveillance and Epidemiology, Center for Drug Evaluation and Research, US Food and Drug Administration, Silver Spring, MD, USA}
\affil[9]{Division of Pharmacoepidemiology and Pharmacoeconomics, Department of Medicine, Brigham and Women’s Hospital, Harvard Medical School, Boston, MA, USA}


\date{\today}

\setcounter{tocdepth}{3}

\begin{document}
\single

\maketitle

\tableofcontents
\newpage

\fi

\section{Supplemental Methods}

\subsection{Data structure and notation}

We use the same notation as in the main manuscript. We consider a binary outcome $Y$ and binary treatment exposure $X$. The confounders subject to missingness are $W \in \mathbb{R}^q$, while the always-observed confounders are $Z \in \mathbb{R}^s$. $R = (R_1, \ldots, R_q) \in \{0,1\}^{q}$ is an indicator of complete data. In the (assumed) observed setting, the data unit is $O := (Y, X, Z, R, RW) \sim P_0$.

\subsection{Causal Inference Assumptions}\label{sec:causal_assumptions}

Since we do not observe both potential outcomes for each participant in the study, additional assumptions are necessary to estimate the average treatment effect using our data. These assumptions allow us to \textit{identify} the causal estimand, relating the causal estimand to a statistical estimand \citep{vanderlaan2011targeted}. 

The \textit{consistency} assumption states that the observed outcome is the potential outcome that would be observed if we set the exposure to its observed level; in other words, that if $X = x$, then $Y(x) = Y$. 

The \textit{randomization} or \textit{no unmeasured confounders} assumption states that the potential outcome is independent of the exposure, given the measured covariates; in other words, that $Y(x) \perp X \mid (Z,W)$. 

The \textit{positivity} assumption states that there is a positive probability of exposure level $x$ within all possible strata of the covariates; in other words, $P(X = x \mid Z = z, W = w) > 0$ for all $(z,w)$.

\newpage

\section{Synthetic-data Simulation Study}\label{sec:synthetic-data-gen}


\subsection{Causal Diagrams}\label{sec:causal-diagrams}

The following graphs show the causal relationships between the (Y,X,\textbf{Z},\textbf{W},R) for the different missing data scenarios in the synthetic data simulations.

\begin{figure}[H]
\begin{tikzpicture}[node distance=2cm]
    \node (x) [standard] {$X$};
    \node (y) [standard, right of=x, xshift=8cm] {$Y$};
    \node (r) [standard, below of=x, yshift=-1cm] {$R$};
    \node (z) [standard, right of=r, xshift=4cm] {$Z$};
    \node (w) [standard, right of=r, xshift=6.5cm, fill=black!15] {$W$};

    \draw[-{Latex[length=3mm]}] (x) -- (y);
    \draw[-{Latex[length=3mm]}] (z) -- (y);
    \draw[-{Latex[length=3mm]}] (w) -- (y);
    \draw[-{Latex[length=3mm]}] (z) -- (r);
    \draw[-{Latex[length=3mm]}] (x) -- (r);
    \draw[-{Latex[length=3mm]}] (y) -- (r);
    \draw[-{Latex[length=3mm]}] ([yshift=-5pt] z.west) .. controls ([yshift=-4.1cm] r) and ([xshift=-4.1cm] r) .. ([yshift=-10pt] x.west);
    \draw[-{Latex[length=3mm]}] (w.south) .. controls ([yshift=-5.1cm] r) and ([xshift=-5.1cm] r) .. ([yshift=5pt] x.west);
\end{tikzpicture}
\vspace{-2cm}
\caption{Causal diagram for the simple and complex missing at random (MAR) scenarios. Only $\textbf{W}$ (shaded in grey) is subject to missingness.}
\label{fig:mar-diagram}
\end{figure}

\begin{figure}[H]
\begin{tikzpicture}[node distance=2cm]
    \node (x) [standard] {$X$};
    \node (y) [standard, right of=x, xshift=8cm] {$Y$};
    \node (r) [standard, below of=x, yshift=-1cm] {$R$};
    \node (z) [standard, right of=r, xshift=4cm] {$Z$};
    \node (w) [standard, right of=r, xshift=6.5cm, fill=black!15] {$W$};

    \draw[-{Latex[length=3mm]}] (x) -- (y);
    \draw[-{Latex[length=3mm]}] (z) -- (y);
    \draw[-{Latex[length=3mm]}] (w) -- (y);
    \draw[-{Latex[length=3mm]}] (z) -- (r);
    \draw[-{Latex[length=3mm]}] (x) -- (r);
    \draw[-{Latex[length=3mm]}] (y) -- (r);
    \draw[-{Latex[length=3mm]}] ([yshift=-5pt] z.west) .. controls ([yshift=-4.1cm] r) and ([xshift=-4.1cm] r) .. ([yshift=-10pt] x.west);
    \draw[-{Latex[length=3mm]}] (w.south) .. controls ([yshift=-5.1cm] r) and ([xshift=-5.1cm] r) .. ([yshift=5pt] x.west);
    \draw[-{Latex[length=3mm]}] (w.south) .. controls ([yshift=-2cm] z) and ([xshift=1cm,yshift=-2cm] r) .. (r.south);
\end{tikzpicture}
\vspace{-2cm}
\caption{Causal diagram for data missing not at random scenario, where missingness depends in part on the value of $\textbf{W}$ that is subject to missingness (MNAR-value).}
\label{fig:mnar-value-diagram}
\end{figure}

\begin{figure}[H]
\begin{tikzpicture}[node distance=2cm]
    \node (x) [standard] {$X$};
    \node (y) [standard, right of=x, xshift=8cm] {$Y$};
    \node (r) [standard, below of=x, yshift=-1cm] {$R$};
    \node (z) [standard, right of=r, xshift=4cm] {$Z$};
    \node (w) [standard, right of=r, xshift=6.5cm, fill=black!15] {$W$};
    \node (u) [standard, below of=r, fill=black!30] {$U$};

    \draw[-{Latex[length=3mm]}] (x) -- (y);
    \draw[-{Latex[length=3mm]}] (z) -- (y);
    \draw[-{Latex[length=3mm]}] (w) -- (y);
    \draw[-{Latex[length=3mm]}] (z) -- (r);
    \draw[-{Latex[length=3mm]}] (x) -- (r);
    \draw[-{Latex[length=3mm]}] (y) -- (r);
    \draw[-{Latex[length=3mm]}] (u) -- (r);
    \draw[-{Latex[length=3mm]}] ([yshift=-5pt] z.west) .. controls ([yshift=-4.1cm] u) and ([xshift=-4.1cm] u) .. ([yshift=-10pt] x.west);
    \draw[-{Latex[length=3mm]}] (w.south) .. controls ([yshift=-5.1cm] u) and ([xshift=-5.1cm] u) .. ([yshift=5pt] x.west);
    \draw[ultra thick, dashed, -{Latex[length=3mm]}] (u.east) .. controls ([yshift=-3cm] w) and ([xshift=3cm] w) .. (y.south);

\end{tikzpicture}
\vspace{-2cm}
\caption{Causal diagram for data missing not at random, where missingness in part depends on an unobserved variable $U$ (MNAR-unobserved); the dashed arrow indicates the relationship depends on the DGM scenario (solid for the MNAR-unobserved scenario, and no arrow appears otherwise.}
\label{fig:mnar-unobserved-diagram}
\end{figure}
\clearpage\newpage

\subsection{Covariate Distribution}

We generate covariates according to the following model:
\begin{align*}
(X_\text{latent}, Z_s, Z_w, W_s, W_w, U_s, U_w, A_s, A_w) \sim N(0, \Sigma),
\end{align*}
where $(W_w, W_s)$ are the latent potential confounders (will only be observed on a subset), $(Z_w, Z_s)$ are covariates observed on everyone, $(U_w, U_s)$ are unobserved variables, $(A_w, A_s)$ are auxiliary variables, and $X_\text{latent}$ is a latent treatment assignment variable. Though we generated auxiliary variables $A$, we did not use them in any simulation.

For X Scenario 1, 
$$\Sigma = \begin{bmatrix}
    1 & \rho_1 & \rho_2 & \rho_1 & \rho_2 & \rho_1 & \rho_2 & \rho_1 & \rho_2 \\
    \rho_1 & 1 & \rho_3 & \rho_3 & \rho_3 & \rho_3 & \rho_3 & \rho_3 & \rho_3\\
    \rho_2 & \rho_3 & 1 & \rho_3 & \rho_3 & \rho_3 & \rho_3 & \rho_3 & \rho_3\\
    \rho_1 & \rho_3 & \rho_3 & 1 & \rho_3 & \rho_3 & \rho_3 & \rho_3 & \rho_3\\
    \rho_2 & \rho_3 & \rho_3 & \rho_3 & 1 & \rho_3 & \rho_3 & \rho_3 & \rho_3\\
    \rho_1 & \rho_3 & \rho_3 & \rho_3 & \rho_3 & 1 & \rho_3 & \rho_3 & \rho_3\\
    \rho_2 & \rho_3 & \rho_3 & \rho_3 & \rho_3 & \rho_3 & 1 & \rho_3 & \rho_3\\
    \rho_1 & \rho_3 & \rho_3 & \rho_3 & \rho_3 & \rho_3 & \rho_3 & 1 & \rho_3\\
    \rho_2 & \rho_3 & \rho_3 & \rho_3 & \rho_3 & \rho_3 & \rho_3 & \rho_3 & 1\\
\end{bmatrix},$$
where $\rho_1 = 0.4$, $\rho_2 = 0.2$, and $\rho_3 = 0.2$.

\subsection{Treatment Assignment}

In the general (X Scenario 1), we will randomly assign treatment according to the value of $X_\text{latent}$ being below the 40th percentile of its distribution. In X Scenario 1, we have covariates $(Z_s,W_s)$ with strong correlation 0.4 with $X_\text{latent}$ and covariates $(Z_w,W_w)$ with weak correlation 0.2 with $X_\text{latent}$. Additional unobserved covariates $(U_s,U_w)$ and auxiliary covariates $(A_s,A_w)$ that have strong and weak correlation with $X_\text{latent}$ are generated. See Table~\ref{tab:sim_x_scenarios}.

\begin{table}[!ht]
    \centering
    \caption{Covariate generating scenarios.}
    \label{tab:sim_x_scenarios}
    \begin{tabular}{p{2.5cm}|p{3cm}|p{7.5cm}}
        \textbf{Scenario} & \textbf{Description} & \textbf{Detailed specification} \\
        \hline
        X Scenario 1 & Multivariate Normal covariates, simple propensity score for treatment & Generate multivariate normal covariates, including latent continuous treatment-related variable, with varying correlation; $X = 0$ if this latent variable is below the 40th quantile, and $X = 1$ otherwise. \\ 
        X Scenario 1.1 & Multivariate Normal covariates, simple propensity score for treatment & Same as scenario 1 but with correlation 0.8 between $U$ and $W$\\
    \end{tabular}
\end{table}

\subsection{Missing Data Generation}

The full set of missing-data scenarios are provided in Table~\ref{tab:sim_m_scenarios}.

In the base case (Missing scenario 1.1, the data are missing at random (MAR) and the missing-data mechanism follows a simple linear logistic model, that matches the functional form of the working analysis model: 
\begin{align*}
    \logit P(R = 1 \mid X = x, Z = z, Y = y) = \alpha_0 + \alpha_X x + \alpha_Z z + \alpha_Y y.
\end{align*}
In Scenario 1.1, the missingness mechanism does depend on the outcome (here, $\alpha_Y = \ln(2.5)$). The values of $\alpha_0$, $\alpha_X$, and $\alpha_Z$ may be varied in different scenarios. In Scenario 2, the missingness mechanism is no longer a simple MAR linear logistic model. Scenario 2.2 the missingness model involves interaction terms and discretized cut-offs of the analysis variables and dependence on the outcome, while maintaining approximately a 40\% missingness rate.  Scenario 2.4 is similar to 2.2, but with an 80\% missingness rate. Scenarios 2.5 and 2.6 are missing not at random (MNAR) scenarios. In Scenario 2.5, the missingness depends on an unobserved covariate U and for Scenario 2.6, it depends on the value of the value of the variable (W) prone to missingness. Scenarios 2.7 and 2.8 are the same as 2.5 and 2.6, respectively, such that the intercept was increased to yield 80\% missingness. Scenario 3.1 is identical to Scenario 1.1, but with an increased intercept an 80\% missingness probability.

\begin{longtable}[ht]{p{3cm}|p{2.5cm}|p{8cm}}
    \caption{Missing-data model scenarios. Unless otherwise specified, $\alpha_0 = -2/3$, $\alpha_X = \ln(1.5)$, $\alpha_Z = [\ln(2.5), \ln(2)]$.}\\
    \label{tab:sim_m_scenarios}
        \textbf{Scenario} & \textbf{Description} & \textbf{Detailed specification} \\
        \hline
        Missing Scenario 1.1 & Base case simple MAR, depends on $Y$ & $\logit P(R = 1 \mid X = x, Z = z, Y = y) = \alpha_0 + \alpha_X x + \alpha_Z^\top z + \alpha_Y y$, $\alpha_X = \ln(2.5)$, $\alpha_Z = [\ln(1.5), \ln(1.5)]$, $\alpha_Y = \ln(2.5)$\\ 
        Missing Scenario 2.2 & Complex MAR specification, depends on $Y$ & $\logit P(R = 1 \mid X = x, Z = z, Y = y) = \alpha_0 + \alpha_X x + \alpha_{Z_{1,1}} I(z_1 < -0.5) + \alpha_{Z_{1,2}} I(z_1 > 1) + \alpha_{Z_1Z_2}z_1z_2 + \alpha_{Z_2} I(z_2 < -1) + \alpha_{YZ_2}yz_2 + \alpha_{XZ_2}xz_2$, $\alpha_0 = -0.9$, $\alpha_Y = 0.2$, $\alpha_X = 1$, $\alpha_{Z_{1,1}} = -0.9$, $\alpha_{Z_{1,2}} = 2$, $\alpha_{Z_1Z_2} = 0$, $\alpha_{XZ_2} = 3$, $\alpha_{YZ_2} = -3$ \\
        Missing Scenario 2.4 & Same as 2.2 but with 80\% missing & $\logit P(R = 1 \mid X = x, Z = z, Y = y) = \alpha_0 + \alpha_X x + \alpha_{Z_{1,1}} I(z_1 < -0.5) + \alpha_{Z_{1,2}} I(z_1 > 1) + \alpha_{Z_1Z_2}z_1z_2 + \alpha_{Z_2} I(z_2 < -1) + \alpha_{YZ_2}yz_2 + \alpha_{XZ_2}xz_2$, $\alpha_0 = 1.3$, other regression parameters the same as scenario 2.2\\
         Missing Scenario 2.5 & MNAR with missingness dependent upon $U$ & $\logit P(R = 1 \mid X = x, Z = z, Y = y, U = u) = \alpha_0 + \alpha_X x + \alpha_Z^\top z + \alpha_Y y + \alpha_U^\top u$, $\alpha_0 = -0.97$, $\alpha_X = \ln(2.5)$, $\alpha_Z = [\ln(1.5), \ln(1.5)]$, $\alpha_Y = \ln(2.5)$, [$\alpha_{Us},\alpha_{Uw}] = [\ln(2.5), \ln(1)]$ \\ 
        Missing Scenario 2.6 & MNAR with missingness dependent upon $W$ & $\logit P(R = 1 \mid X = x, Z = z, Y = y, W = w) = \alpha_0 + \alpha_X x + \alpha_Z^\top z + \alpha_Y y + \alpha_W^\top w$, $\alpha_X = \ln(2.5)$, $\alpha_Z = [\ln(1.5), \ln(1.5)]$, $\alpha_Y = \ln(2.5)$, $\alpha_W = [\ln(2.5), \ln(2.5]$ \\ 
        Missing Scenario 2.7 & Same MNAR as 2.5 with 80\% missing & $\logit P(R = 1 \mid X = x, Z = z, Y = y, U = u) = \alpha_0 + \alpha_X x + \alpha_Z^\top z + \alpha_Y y + \alpha_U^\top u$, $\alpha_X = \ln(2.5)$, $\alpha_Z = [\ln(1.5), \ln(1.5)]$, $\alpha_Y = \ln(2.5)$, [$\alpha_{Us},\alpha_{Uw}] = [\ln(2.5), \ln(1)]$, $\alpha_0 = 1.28$ \\
        Missing Scenario 2.8 & Same MNAR as 2.6 with 80\% missing & $\logit P(R = 1 \mid X = x, Z = z, Y = y, W = w) = \alpha_0 + \alpha_X x + \alpha_Z^\top z + \alpha_Y y + \alpha_W^\top w$, $\alpha_X = \ln(2.5)$, $\alpha_Z = [\ln(1.5), \ln(1.5)]$, $\alpha_Y = \ln(2.5)$, $\alpha_W = [\ln(2.5), \ln(2.5]$, $\alpha_0 = 1.63$ \\
        Missing Scenario 3.1 & Same MAR as 1.1 with 80\% missing & $\logit P(R = 1 \mid X = x, Z = z, Y = y) = \alpha_0 + \alpha_X x + \alpha_Z^\top z + \alpha_Y y$, $\alpha_X = \ln(2.5)$, $\alpha_Z = [\ln(1.5), \ln(1.5)]$, $\alpha_0=1.08$, $\alpha_Y = \ln(2.5)$\\ 

\end{longtable}
\clearpage
\newpage
\subsection{Outcome Generation}

The full set of outcome generation scenarios are provided in Table~\ref{tab:sim_y_scenarios}. In the Base Case Scenarios (Y Scenario 1.1), the data are generated according to a simple linear logistic model, which in the Base Case also matches the analytical form of the working model, and in the population is not subject to missingness: 
\begin{align*}
    \logit P(Y = 1 \mid X = x, W = w, Z = z) = \beta_0 + \beta_X x + \beta_W^\top w + \beta_Z^\top z.
\end{align*}
In all synthetic data scenarios, there is a treatment effect (here, $\beta_X = \ln(1.5)$). The values of $\beta_0$, $\beta_W$, and $\beta_Z$ may be varied in different scenarios.
Scenario 2.11 adds an unobserved covariate, that does not appear in the working analytical model. Scenario 4.1 involves a complex form that has interaction effects and effects of discretized versions of the covariates, that do not match the simpler analytic working model. Scenarios 1.17, 2.17, and 4.17 are rare-outcome variations of Y Scenarios 1.1, 2.11, and 4.1, respectively (approximately 5\% incidence rather than approximately 12\% incidence).

\begin{longtable}[ht]{p{2.5cm}|p{3cm}|p{6cm}|p{1.5cm}}
    \caption{Outcome regression model scenarios. Unless otherwise specified, $\beta_0 = -2.4$, $\beta_W=[\beta_{Ww},\beta_{Ws}] = [\ln(1.5), -\ln(1.75)]$, $\beta_Z=[\beta_{Zw},\beta_{Zs}] = [\ln(1.5), -\ln(1.3)]$.} \\
    \label{tab:sim_y_scenarios}
        \textbf{Scenario} & \textbf{Description} & \textbf{Detailed specification} & \textbf{Outcome rate} (\%)\\
        \hline
        Y Scenario 1.1 & Base case: Simple outcome specification, non-null treatment effect & $\logit P(Y = 1 \mid X = x, W = w, Z = z) = \beta_0 + \beta_Xx + \beta_W^\top w + \beta_Z^\top z$, $\beta_X = \ln(1.5)$ & 12.0\\
        Y Scenario 1.17 & Same as Y1.1, intercept for rare outcome & $\logit P(Y = 1 \mid X = x, W = w, Z = z) = \beta_0 + \beta_Xx + \beta_W^\top w + \beta_Z^\top z$, $\beta_0=-3.4, \, \beta_X = \ln(1.5)$ &  5.0 \\
        Y Scenario 2.11 & Unobserved covariate added to Y1.1 & $\logit P(Y = 1 \mid X = x, W = w, Z = z,Us=u) = \beta_0 + \beta_Xx + \beta_W^\top w + \beta_Z^\top z + \beta_U^\top u$, $\beta_0 = -2.5$, $\beta_X = \ln(1.5)$, $\beta_{Us} = -\ln(1.75)$ & 12.0\\
     Y Scenario 2.17 & Same as 2.11, intercept for rare outcome & $\logit P(Y = 1 \mid X = x, W = w, Z = z,Us=u) = \beta_0 + \beta_Xx + \beta_W^\top w + \beta_Z^\top z + \beta_{Us}^\top u$, $\beta_0 = -3.56$, $\beta_X = \ln(1.5)$, $\beta_U = -\ln(1.75)$ & 5.0\\
        Y Scenario 4.1 & Complex outcome specification, non-null treatment effect & $\logit P(Y = 1 \mid X = x, W = w, Z = z) = \beta_0 + \beta_X x + \beta_W^\top w + \beta_{W,\text{int}}w_w w_s + \beta_{Z_w,1}I(z_w < -0.5) + \beta_{Z_w,2}I(z_w > 2) + \beta_{Z_s}I(z_s < -1) + \beta_{WZs,\text{inter}}w_s z_s + \beta_{WZw,\text{inter}}w_s I(z_w > 2)$, $\beta_X = \ln(1.5)$, $\beta_{W,\text{int}} = 1$, $\beta_{Z_w,1} = 0.1$, $\beta_{Z_w,2} = 0.8$, $\beta_{WZs,\text{inter}} = 3$, $\beta_{WZw,\text{inter}} = 1$  & 14.9\\
        Y Scenario 4.17 & Same as Y 4.1, intercept rare outcome & $\logit P(Y = 1 \mid X = x, W = w, Z = z) = \beta_0 + \beta_X x + \beta_W^\top w + \beta_{W,\text{int}}w_w w_s + \beta_{Z_w,1}I(z_w < -0.5) + \beta_{Z_w,2}I(z_w > 2) + \beta_{Z_s}I(z_s < -1) + \beta_{WZs,\text{inter}}w_s z_s + \beta_{WZw,\text{inter}}w_s I(z_w > 2)$, $\beta_0=-4.1, \, \beta_X = \ln(1.5)$, other regression parameters the same as scenario 4.1 & 5.3 \\
\end{longtable}
\clearpage
\newpage

\subsection{Modified TMLE Approaches}\label{sec:modified-tmle}

In addition to the TMLE-M (candidate learners described in Table~\ref{tab:sl_tmle_m}) and TMLE-MTO (candidate learners described in Table~\ref{tab:sl_tmle_mto}) approaches described in the main manuscript, we considered two modifications designed to provide better performance in complex MAR and rare-outcome settings.

The first modification, which we call the IPCW-a-TMLE (the ``a'' stands for ``augmented-data''), augments the dataset passed to the TMLE with an estimate of the confounded regression model. In other words, prior to running the TMLE, we obtain a fitted regression model using all observations but dropping the confounders prone to missingness. The fitted values from this model are then passed as a covariate to the TMLE. We hoped that this would reduce bias in complex MAR settings where the initial weights might be misspecified, because the IPCW-a-TMLE can be viewed as an approximation of an augmented IPCW-TMLE \citep{rose2011targeted}. We considered both IPCW-a-TMLE-M and IPCW-a-TMLE-MTO.

The second modification, which we call the r-IPCW-TMLE (the ``r'' stands for ``rare-outcome''), specifies a super learner library for the outcome regression model that is hypothesized to perform well in rare-outcome settings, by reducing the complexity of the candidate learners. We considered only the r-IPCW-TMLE-MTO, because the TMLE-M does not use a super learner for the outcome regression model. Rather than using the outcome regression library listed in Table~\ref{tab:sl_tmle_mto}, we use a simpler library consisting of logistic regression, logistic regression with lasso, and discrete Bayesian additive regression trees (using the default settings from the \texttt{dbarts} R package, with tuning parameter $k = 2$ controlling smoothing.)

\begin{table}[!ht]
    \centering
    \caption{Algorithm used for each nuisance function (missing-data model, propensity score, and outcome regression) in the TMLE-M procedure, along with the candidate algorithms in any super learners. Unspecified tuning parameters are set to their default values. Sample size is denoted by $n$ and number of covariates by $p$.
    \newline
    $^{*}$sequence contains 10 possible values.
    }
    \label{tab:sl_tmle_m}
    \begin{tabular}{ccl}
       Nuisance function  & Algorithm & Tuning parameters \\
       \hline
        Outcome regression & glm & --- \\
        Propensity score & glm & --- \\
        Missing-data model & super learner & \\
        & glm & --- \\
        & gradient boosted trees (\texttt{xgboost}) & maximum depth $\in \{1, 3\}$ \\
        & & shrinkage $\in \{.01, .1\}$ \\
        & & number of trees = 500 \\
        & random forests (\texttt{ranger}) & minimum node size $\in \{n / 100, \ldots, n / 10\}^{*}$ \\
        & & number of trees = 500 \\
        & & mtry = $\sqrt{p}$
    \end{tabular}
\end{table}

\begin{table}[!ht]
    \centering
    \caption{Algorithm used for each nuisance function (missing-data model, propensity score, and outcome regression) in the TMLE-MTO procedure, along with the candidate algorithms in any super learners. Sample size is denoted by $n$ and number of covariates by $p$.
    \newline
    $^{*}$sequence contains 10 possible values.
    }
    \label{tab:sl_tmle_mto}
    \begin{tabular}{ccc}
    Nuisance function  & Algorithm & Tuning parameters \\
       \hline
       Outcome regression & super learner & \\
        & glm & --- \\
        & gradient boosted trees (\texttt{xgboost}) & maximum depth $\in \{1, 3\}$ \\
        & & shrinkage $\in \{.01, .1\}$ \\
        & & number of trees = 500 \\
        & random forests (\texttt{ranger}) & minimum node size $\in \{n / 100, \ldots, n / 10\}^{*}$ \\
        & & number of trees = 500 \\
        & & mtry = $\sqrt{p}$ \\
        Propensity score & super learner & \\
        & glm & --- \\
        & gradient boosted trees (\texttt{xgboost}) & maximum depth $\in \{1, 3\}$ \\
        & & shrinkage $\in \{.01, .1\}$ \\
        & & number of trees = 500 \\
        & random forests (\texttt{ranger}) & minimum node size $\in \{n / 100, \ldots, n / 10\}^{*}$ \\
        & & number of trees = 500 \\
        & & mtry = $\sqrt{p}$ \\
        Missing-data model & super learner & \\
        & glm & --- \\
        & gradient boosted trees (\texttt{xgboost}) & maximum depth $\in \{1, 3\}$ \\
        & & shrinkage $\in \{.01, .1\}$ \\
        & & number of trees = 500 \\
        & random forests (\texttt{ranger}) & minimum node size $\in \{n / 100, \ldots, n / 10\}^{*}$ \\
        & & number of trees = 500 \\
        & & mtry = $\sqrt{p}$
    \end{tabular}
\end{table}

\clearpage
\newpage

\subsection{XGBoost Algorithm Investigation}\label{sec:XGB-troubles}

While multiple imputation of missing data using extreme gradient boosting (MI-XGB) performed well in the initial synthetic scenarios (e.g., 12\% incidence, 40\% missing, correct missingness model), it produced substantial bias in the plasmode setting (5\% composite).  The core difference in these scenarios is the form of the confounder with missing data; in the synthetic scenarios, they were continuous variables, while in the plasmode scenario, it was a categorical variable.  Furthermore, the plasmode categorical variable prone to missingness is not evenly distributed, with 66\% of its values falling in the lowest risk category.
When we investigated an imputation formed from the mixgb package, we found that 99.87\% of the missing values were imputed to the lowest risk category, causing severe bias in both plasmode scenarios.  To address this, we changed the mixgb initial.fac setting from “mode” to “sample”, thus starting the imputation process from a random sample of observed values rather than the modal value.  This somewhat reduced the bias and is what is presented in the tables.
It may be possible for extreme gradient boosting to be more accurate.  While we used the built-in cross-validation function to choose an optimal number of boosting rounds, no functions are available in the package to choose the learning rate or maximum depth of the trees.  Our goal was to evaluate these approaches as they are implemented in current software, so expanding the cross-validation parameter search was out of the scope of our investigation.

\newpage


\subsection{Synthetic-data Results Tables}\label{sec:supp-results-synthetic}

\clearpage

\subsubsection{Tables: 12\% Outcome, 40\% MAR (Base Case), cOR}
\clearpage
\newpage

\begin{table}
\centering
\caption{\label{tab:x1_m1.1_y1.1_cOR_census}\textbf{Synthetic data MAR simulation: census conditional odds ratio (cOR), 12\% outcome proportion, 40\% missing proportion}. Comparing estimators under the \textbf{simple outcome and simple MAR} scenario. The value of the estimand is 0.406. The sample size is n = 10000. Maximum observed Monte-Carlo error over the 2500 simulation replications was 0.009 for all summaries besides coverage and 0.012 for coverage. ESE = empirical standard error, ASE = asymptotic standard error, MAD = mean absolute deviation, RMSE = root mean squared error, rRMSE = robust RMSE (using median bias and MAD), Oracle coverage = coverage of a confidence interval based on the ESE, Nominal coverage = coverage of a confidence interval based on the ASE.}
\centering
\fontsize{9}{11}\selectfont
\begin{tabular}[t]{>{\raggedright\arraybackslash}p{2cm}|>{\raggedright\arraybackslash}p{1cm}|>{\raggedleft\arraybackslash}p{1cm}|>{\raggedright\arraybackslash}p{1cm}|>{\raggedright\arraybackslash}p{1cm}|>{\raggedright\arraybackslash}p{1cm}|>{\raggedright\arraybackslash}p{1cm}|>{\raggedright\arraybackslash}p{1cm}|>{\raggedleft\arraybackslash}p{1cm}|>{\raggedleft\arraybackslash}p{1cm}|>{\raggedleft\arraybackslash}p{1cm}|>{\raggedleft\arraybackslash}p{1cm}|>{}p{1cm}}
\hline
Estimator & Mean bias & Median bias & ESE & ASE & MAD & RMSE & rRMSE & Oracle coverage & Nominal coverage & Power & Prop. completed\\
\hline
Benchmark model & 0.001 & -0.001 & 0.073 & 0.07 & 0.075 & 0.07 & 0.075 & 0.950 & 0.940 & 1.000 & 100\\
\hline
Complete-case & -0.191 & -0.189 & 0.12 & 0.119 & 0.118 & 0.225 & 0.223 & 0.646 & 0.636 & 0.442 & 100\\
\hline
Confounded model & 0.203 & 0.205 & 0.069 & 0.067 & 0.07 & 0.214 & 0.216 & 0.158 & 0.143 & 1.000 & 100\\
\hline
IPW & -0.001 & 0.002 & 0.129 & 0.13 & 0.128 & 0.13 & 0.128 & 0.950 & 0.953 & 0.884 & 100\\
\hline
Raking (vanilla) & 0 & 0.000 & 0.082 & 0.078 & 0.083 & 0.078 & 0.083 & 0.950 & 0.938 & 0.999 & 100\\
\hline
MICE & 0.001 & -0.001 & 0.079 & 0.076 & 0.08 & 0.076 & 0.08 & 0.949 & 0.939 & 1.000 & 100\\
\hline
MI-XGB & -0.006 & -0.007 & 0.081 & 0.077 & 0.081 & 0.077 & 0.082 & 0.946 & 0.935 & 0.999 & 100\\
\hline
MI-RF & 0.005 & 0.004 & 0.082 & 0.074 & 0.082 & 0.074 & 0.082 & 0.945 & 0.923 & 0.999 & 100\\
\hline
IPCW-TMLE-M & -0.015 & -0.014 & 0.132 & 0.136 & 0.129 & 0.137 & 0.13 & 0.949 & 0.953 & 0.835 & 100\\
\hline
IPCW-TMLE-MTO & -0.015 & -0.014 & 0.131 & 0.134 & 0.129 & 0.135 & 0.13 & 0.946 & 0.953 & 0.842 & 100\\
\hline
\end{tabular}
\end{table}

\begin{table}
\centering
\caption{\label{tab:x1_m2.2_y1.1_cOR_census}\textbf{Synthetic data MAR simulation: census conditional odds ratio (cOR), 12\% outcome proportion, 40\% missing proportion}. Comparing estimators under the \textbf{simple outcome and complex MAR} scenario. The value of the estimand is 0.406. The sample size is n = 10000. Maximum observed Monte-Carlo error over the 2500 simulation replications was 0.009 for all summaries besides coverage and 0.012 for coverage. ESE = empirical standard error, ASE = asymptotic standard error, MAD = mean absolute deviation, RMSE = root mean squared error, rRMSE = robust RMSE (using median bias and MAD), Oracle coverage = coverage of a confidence interval based on the ESE, Nominal coverage = coverage of a confidence interval based on the ASE.}
\centering
\fontsize{9}{11}\selectfont
\begin{tabular}[t]{>{\raggedright\arraybackslash}p{2cm}|>{\raggedright\arraybackslash}p{1cm}|>{\raggedleft\arraybackslash}p{1cm}|>{\raggedright\arraybackslash}p{1cm}|>{\raggedright\arraybackslash}p{1cm}|>{\raggedright\arraybackslash}p{1cm}|>{\raggedright\arraybackslash}p{1cm}|>{\raggedright\arraybackslash}p{1cm}|>{\raggedleft\arraybackslash}p{1cm}|>{\raggedleft\arraybackslash}p{1cm}|>{\raggedleft\arraybackslash}p{1cm}|>{\raggedleft\arraybackslash}p{1cm}|>{}p{1cm}}
\hline
Estimator & Mean bias & Median bias & ESE & ASE & MAD & RMSE & rRMSE & Oracle coverage & Nominal coverage & Power & Prop. completed\\
\hline
Benchmark model & -0.001 & -0.002 & 0.072 & 0.07 & 0.073 & 0.07 & 0.073 & 0.948 & 0.942 & 1.000 & 100\\
\hline
Complete-case & 0.214 & 0.215 & 0.092 & 0.092 & 0.095 & 0.233 & 0.235 & 0.354 & 0.359 & 1.000 & 100\\
\hline
Confounded model & 0.201 & 0.201 & 0.068 & 0.067 & 0.068 & 0.212 & 0.212 & 0.165 & 0.151 & 1.000 & 100\\
\hline
IPW & 0.14 & 0.140 & 0.096 & 0.097 & 0.098 & 0.17 & 0.171 & 0.691 & 0.706 & 1.000 & 100\\
\hline
Raking (vanilla) & 0 & -0.001 & 0.077 & 0.073 & 0.076 & 0.073 & 0.076 & 0.950 & 0.939 & 1.000 & 100\\
\hline
MICE & -0.002 & -0.001 & 0.075 & 0.073 & 0.075 & 0.073 & 0.075 & 0.947 & 0.944 & 1.000 & 100\\
\hline
MI-RF & -0.01 & -0.010 & 0.076 & 0.073 & 0.077 & 0.074 & 0.077 & 0.947 & 0.939 & 1.000 & 100\\
\hline
IPCW-TMLE-M & 0.051 & 0.051 & 0.094 & 0.106 & 0.095 & 0.118 & 0.108 & 0.919 & 0.958 & 0.998 & 100\\
\hline
IPCW-TMLE-MTO & 0.055 & 0.055 & 0.094 & 0.105 & 0.096 & 0.118 & 0.111 & 0.914 & 0.950 & 0.997 & 100\\
\hline
IPCW-a-TMLE-M & 0.049 & 0.048 & 0.096 & 0.107 & 0.097 & 0.117 & 0.109 & 0.924 & 0.955 & 0.996 & 100\\
\hline
IPCW-a-TMLE-MTO & 0.053 & 0.052 & 0.096 & 0.105 & 0.098 & 0.117 & 0.111 & 0.920 & 0.946 & 0.997 & 100\\
\hline
\end{tabular}
\end{table}

\begin{table}
\centering
\caption{\label{tab:x1_m1.1_y4.1_cOR_census}\textbf{Synthetic data MAR simulation: census conditional odds ratio (cOR), 12\% outcome proportion, 40\% missing proportion}. Comparing estimators under the \textbf{complex outcome and simple MAR} scenario. The value of the estimand is 0.371. The sample size is n = 10000. Maximum observed Monte-Carlo error over the 2500 simulation replications was 0.009 for all summaries besides coverage and 0.012 for coverage. ESE = empirical standard error, ASE = asymptotic standard error, MAD = mean absolute deviation, RMSE = root mean squared error, rRMSE = robust RMSE (using median bias and MAD), Oracle coverage = coverage of a confidence interval based on the ESE, Nominal coverage = coverage of a confidence interval based on the ASE. Estimators that are mismatched with the estimand (i.e., are estimating a different parameter) are emphasized using a star.}
\centering
\fontsize{9}{11}\selectfont
\begin{tabular}[t]{>{\raggedright\arraybackslash}p{2cm}|>{\raggedright\arraybackslash}p{1cm}|>{\raggedleft\arraybackslash}p{1cm}|>{\raggedright\arraybackslash}p{1cm}|>{\raggedright\arraybackslash}p{1cm}|>{\raggedright\arraybackslash}p{1cm}|>{\raggedright\arraybackslash}p{1cm}|>{\raggedright\arraybackslash}p{1cm}|>{\raggedleft\arraybackslash}p{1cm}|>{\raggedleft\arraybackslash}p{1cm}|>{\raggedleft\arraybackslash}p{1cm}|>{\raggedleft\arraybackslash}p{1cm}|>{}p{1cm}}
\hline
Estimator & Mean bias & Median bias & ESE & ASE & MAD & RMSE & rRMSE & Oracle coverage & Nominal coverage & Power & Prop. completed\\
\hline
Benchmark model & 0.002 & 0.002 & 0.071 & 0.072 & 0.07 & 0.072 & 0.07 & 0.950 & 0.953 & 1.000 & 100\\
\hline
Complete-case & -0.199 & -0.201 & 0.123 & 0.122 & 0.122 & 0.234 & 0.235 & 0.635 & 0.630 & 0.292 & 100\\
\hline
Confounded model & -0.28 & -0.279 & 0.069 & 0.068 & 0.07 & 0.288 & 0.288 & 0.017 & 0.016 & 0.271 & 100\\
\hline
IPW & -0.006 & -0.008 & 0.13 & 0.13 & 0.133 & 0.13 & 0.133 & 0.956 & 0.954 & 0.801 & 100\\
\hline
Raking (vanilla) & 0.001 & 0.002 & 0.081 & 0.081 & 0.08 & 0.081 & 0.08 & 0.950 & 0.950 & 0.996 & 100\\
\hline
MICE & 0.075 & 0.075 & 0.083 & 0.083 & 0.083 & 0.112 & 0.112 & 0.856 & 0.860 & 1.000 & 100\\
\hline
MI-XGB & 0.043 & 0.042 & 0.08 & 0.08 & 0.08 & 0.091 & 0.09 & 0.912 & 0.917 & 0.999 & 100\\
\hline
MI-RF & 0.008 & 0.008 & 0.082 & 0.079 & 0.082 & 0.079 & 0.082 & 0.948 & 0.942 & 0.998 & 100\\
\hline
IPCW-TMLE-M & -0.022 & -0.024 & 0.134 & 0.135 & 0.135 & 0.137 & 0.137 & 0.948 & 0.948 & 0.737 & 100\\
\hline
IPCW-TMLE-MTO & -0.006 & -0.006 & 0.125 & 0.123 & 0.128 & 0.123 & 0.128 & 0.951 & 0.946 & 0.842 & 100\\
\hline
IPCW-a-TMLE-M & -0.024 & -0.028 & 0.134 & 0.135 & 0.138 & 0.137 & 0.141 & 0.949 & 0.951 & 0.728 & 100\\
\hline
IPCW-a-TMLE-MTO & -0.007 & -0.007 & 0.126 & 0.122 & 0.13 & 0.122 & 0.13 & 0.953 & 0.943 & 0.838 & 100\\
\hline
\end{tabular}
\end{table}

\begin{table}
\centering
\caption{\label{tab:x1_m2.2_y4.1_cOR_census}\textbf{Synthetic data MAR simulation: census conditional odds ratio (cOR), 12\% outcome proportion, 40\% missing proportion}. Comparing estimators under the \textbf{complex outcome and complex MAR} scenario. The value of the estimand is 0.371. The sample size is n = 10000. Maximum observed Monte-Carlo error over the 2500 simulation replications was 0.009 for all summaries besides coverage and 0.012 for coverage. ESE = empirical standard error, ASE = asymptotic standard error, MAD = mean absolute deviation, RMSE = root mean squared error, rRMSE = robust RMSE (using median bias and MAD), Oracle coverage = coverage of a confidence interval based on the ESE, Nominal coverage = coverage of a confidence interval based on the ASE. Estimators that are mismatched with the estimand (i.e., are estimating a different parameter) are emphasized using a star.}
\centering
\fontsize{9}{11}\selectfont
\begin{tabular}[t]{>{\raggedright\arraybackslash}p{2cm}|>{\raggedright\arraybackslash}p{1cm}|>{\raggedleft\arraybackslash}p{1cm}|>{\raggedright\arraybackslash}p{1cm}|>{\raggedright\arraybackslash}p{1cm}|>{\raggedright\arraybackslash}p{1cm}|>{\raggedright\arraybackslash}p{1cm}|>{\raggedright\arraybackslash}p{1cm}|>{\raggedleft\arraybackslash}p{1cm}|>{\raggedleft\arraybackslash}p{1cm}|>{\raggedleft\arraybackslash}p{1cm}|>{\raggedleft\arraybackslash}p{1cm}|>{}p{1cm}}
\hline
Estimator & Mean bias & Median bias & ESE & ASE & MAD & RMSE & rRMSE & Oracle coverage & Nominal coverage & Power & Prop. completed\\
\hline
Benchmark model & 0 & 0.000 & 0.069 & 0.072 & 0.07 & 0.072 & 0.07 & 0.953 & 0.964 & 1.000 & 100\\
\hline
Complete-case & 0.171 & 0.170 & 0.09 & 0.093 & 0.092 & 0.194 & 0.193 & 0.525 & 0.555 & 1.000 & 100\\
\hline
Confounded model & -0.282 & -0.281 & 0.068 & 0.068 & 0.067 & 0.29 & 0.289 & 0.015 & 0.015 & 0.257 & 100\\
\hline
IPW & 0.087 & 0.086 & 0.095 & 0.096 & 0.097 & 0.129 & 0.13 & 0.849 & 0.850 & 0.998 & 100\\
\hline
Raking (vanilla) & 0.007 & 0.008 & 0.075 & 0.073 & 0.079 & 0.074 & 0.079 & 0.951 & 0.946 & 0.999 & 100\\
\hline
MICE & 0.054 & 0.055 & 0.076 & 0.077 & 0.08 & 0.094 & 0.097 & 0.894 & 0.899 & 1.000 & 100\\
\hline
MI-RF & 0.003 & 0.003 & 0.075 & 0.076 & 0.079 & 0.076 & 0.079 & 0.950 & 0.953 & 0.999 & 100\\
\hline
IPCW-TMLE-M & 0.066 & 0.067 & 0.093 & 0.103 & 0.094 & 0.122 & 0.116 & 0.884 & 0.923 & 0.995 & 100\\
\hline
IPCW-TMLE-MTO & 0.051 & 0.050 & 0.088 & 0.093 & 0.09 & 0.106 & 0.103 & 0.906 & 0.923 & 0.997 & 100\\
\hline
IPCW-a-TMLE-M & 0.077 & 0.076 & 0.095 & 0.103 & 0.095 & 0.128 & 0.121 & 0.871 & 0.904 & 0.998 & 100\\
\hline
IPCW-a-TMLE-MTO & 0.057 & 0.055 & 0.09 & 0.092 & 0.092 & 0.109 & 0.107 & 0.899 & 0.910 & 0.997 & 100\\
\hline
\end{tabular}
\end{table}

\begin{table}
\centering
\caption{\label{tab:x1_m1.1_y1.1_cOR_oracle}\textbf{Synthetic data MAR simulation: oracle conditional odds ratio (cOR), 12\% outcome proportion, 40\% missing proportion}. Comparing estimators under the \textbf{simple outcome and simple MAR} scenario. The value of the estimand is 0.405. The sample size is n = 10000. Maximum observed Monte-Carlo error over the 2500 simulation replications was 0.009 for all summaries besides coverage and 0.012 for coverage. ESE = empirical standard error, ASE = asymptotic standard error, MAD = mean absolute deviation, RMSE = root mean squared error, rRMSE = robust RMSE (using median bias and MAD), Oracle coverage = coverage of a confidence interval based on the ESE, Nominal coverage = coverage of a confidence interval based on the ASE.}
\centering
\fontsize{9}{11}\selectfont
\begin{tabular}[t]{>{\raggedright\arraybackslash}p{2cm}|>{\raggedright\arraybackslash}p{1cm}|>{\raggedleft\arraybackslash}p{1cm}|>{\raggedright\arraybackslash}p{1cm}|>{\raggedright\arraybackslash}p{1cm}|>{\raggedright\arraybackslash}p{1cm}|>{\raggedright\arraybackslash}p{1cm}|>{\raggedright\arraybackslash}p{1cm}|>{\raggedleft\arraybackslash}p{1cm}|>{\raggedleft\arraybackslash}p{1cm}|>{\raggedleft\arraybackslash}p{1cm}|>{\raggedleft\arraybackslash}p{1cm}|>{}p{1cm}}
\hline
Estimator & Mean bias & Median bias & ESE & ASE & MAD & RMSE & rRMSE & Nominal coverage & Oracle coverage & Power & Prop. completed\\
\hline
Benchmark model & 0.001 & -0.001 & 0.073 & 0.07 & 0.075 & 0.07 & 0.075 & 0.940 & 0.950 & 1.000 & 100\\
\hline
Complete-case & -0.191 & -0.189 & 0.12 & 0.119 & 0.118 & 0.225 & 0.223 & 0.637 & 0.646 & 0.442 & 100\\
\hline
Confounded model & 0.204 & 0.205 & 0.069 & 0.067 & 0.07 & 0.214 & 0.217 & 0.142 & 0.158 & 1.000 & 100\\
\hline
IPW & -0.001 & 0.002 & 0.129 & 0.13 & 0.128 & 0.13 & 0.128 & 0.953 & 0.950 & 0.884 & 100\\
\hline
Raking (vanilla) & 0 & 0.000 & 0.082 & 0.078 & 0.083 & 0.078 & 0.083 & 0.938 & 0.950 & 0.999 & 100\\
\hline
MICE & 0.001 & -0.001 & 0.079 & 0.076 & 0.08 & 0.076 & 0.08 & 0.939 & 0.949 & 1.000 & 100\\
\hline
MI-XGB & -0.006 & -0.007 & 0.081 & 0.077 & 0.081 & 0.077 & 0.082 & 0.935 & 0.946 & 0.999 & 100\\
\hline
MI-RF & 0.005 & 0.004 & 0.082 & 0.074 & 0.082 & 0.074 & 0.082 & 0.923 & 0.945 & 0.999 & 100\\
\hline
IPCW-TMLE-M & -0.015 & -0.013 & 0.132 & 0.136 & 0.129 & 0.137 & 0.13 & 0.953 & 0.949 & 0.835 & 100\\
\hline
IPCW-TMLE-MTO & -0.015 & -0.013 & 0.131 & 0.134 & 0.129 & 0.135 & 0.13 & 0.953 & 0.946 & 0.842 & 100\\
\hline
\end{tabular}
\end{table}

\begin{table}
\centering
\caption{\label{tab:x1_m2.2_y1.1_cOR_oracle}\textbf{Synthetic data MAR simulation: oracle conditional odds ratio (cOR), 12\% outcome proportion, 40\% missing proportion}. Comparing estimators under the \textbf{simple outcome and complex MAR} scenario. The value of the estimand is 0.405. The sample size is n = 10000. Maximum observed Monte-Carlo error over the 2500 simulation replications was 0.009 for all summaries besides coverage and 0.012 for coverage. ESE = empirical standard error, ASE = asymptotic standard error, MAD = mean absolute deviation, RMSE = root mean squared error, rRMSE = robust RMSE (using median bias and MAD), Oracle coverage = coverage of a confidence interval based on the ESE, Nominal coverage = coverage of a confidence interval based on the ASE.}
\centering
\fontsize{9}{11}\selectfont
\begin{tabular}[t]{>{\raggedright\arraybackslash}p{2cm}|>{\raggedright\arraybackslash}p{1cm}|>{\raggedleft\arraybackslash}p{1cm}|>{\raggedright\arraybackslash}p{1cm}|>{\raggedright\arraybackslash}p{1cm}|>{\raggedright\arraybackslash}p{1cm}|>{\raggedright\arraybackslash}p{1cm}|>{\raggedright\arraybackslash}p{1cm}|>{\raggedleft\arraybackslash}p{1cm}|>{\raggedleft\arraybackslash}p{1cm}|>{\raggedleft\arraybackslash}p{1cm}|>{\raggedleft\arraybackslash}p{1cm}|>{}p{1cm}}
\hline
Estimator & Mean bias & Median bias & ESE & ASE & MAD & RMSE & rRMSE & Nominal coverage & Oracle coverage & Power & Prop. completed\\
\hline
Benchmark model & -0.001 & -0.002 & 0.072 & 0.07 & 0.073 & 0.07 & 0.073 & 0.942 & 0.948 & 1.000 & 100\\
\hline
Complete-case & 0.214 & 0.215 & 0.092 & 0.092 & 0.095 & 0.234 & 0.235 & 0.358 & 0.354 & 1.000 & 100\\
\hline
Confounded model & 0.201 & 0.201 & 0.068 & 0.067 & 0.068 & 0.212 & 0.212 & 0.150 & 0.165 & 1.000 & 100\\
\hline
IPW & 0.14 & 0.140 & 0.096 & 0.097 & 0.098 & 0.17 & 0.171 & 0.706 & 0.690 & 1.000 & 100\\
\hline
Raking (vanilla) & 0 & -0.001 & 0.077 & 0.073 & 0.076 & 0.073 & 0.076 & 0.938 & 0.951 & 1.000 & 100\\
\hline
MICE & -0.002 & -0.001 & 0.075 & 0.073 & 0.075 & 0.073 & 0.075 & 0.944 & 0.947 & 1.000 & 100\\
\hline
MI-RF & -0.01 & -0.010 & 0.076 & 0.073 & 0.077 & 0.074 & 0.077 & 0.939 & 0.948 & 1.000 & 100\\
\hline
IPCW-TMLE-M & 0.052 & 0.052 & 0.094 & 0.106 & 0.095 & 0.118 & 0.108 & 0.957 & 0.919 & 0.998 & 100\\
\hline
IPCW-TMLE-MTO & 0.055 & 0.056 & 0.094 & 0.105 & 0.096 & 0.118 & 0.111 & 0.950 & 0.914 & 0.997 & 100\\
\hline
IPCW-a-TMLE-M & 0.049 & 0.049 & 0.096 & 0.107 & 0.097 & 0.117 & 0.109 & 0.955 & 0.924 & 0.996 & 100\\
\hline
IPCW-a-TMLE-MTO & 0.053 & 0.053 & 0.096 & 0.105 & 0.098 & 0.117 & 0.111 & 0.946 & 0.920 & 0.997 & 100\\
\hline
\end{tabular}
\end{table}

\begin{table}
\centering
\caption{\label{tab:x1_m1.1_y4.1_cOR_oracle}\textbf{Synthetic data MAR simulation: oracle conditional odds ratio (cOR), 12\% outcome proportion, 40\% missing proportion}. Comparing estimators under the \textbf{complex outcome and simple MAR} scenario. The value of the estimand is 0.405. The sample size is n = 10000. Maximum observed Monte-Carlo error over the 2500 simulation replications was 0.009 for all summaries besides coverage and 0.012 for coverage. ESE = empirical standard error, ASE = asymptotic standard error, MAD = mean absolute deviation, RMSE = root mean squared error, rRMSE = robust RMSE (using median bias and MAD), Oracle coverage = coverage of a confidence interval based on the ESE, Nominal coverage = coverage of a confidence interval based on the ASE. Estimators that are mismatched with the estimand (i.e., are estimating a different parameter) are emphasized using a star.}
\centering
\fontsize{9}{11}\selectfont
\begin{tabular}[t]{>{\raggedright\arraybackslash}p{2cm}|>{\raggedright\arraybackslash}p{1cm}|>{\raggedleft\arraybackslash}p{1cm}|>{\raggedright\arraybackslash}p{1cm}|>{\raggedright\arraybackslash}p{1cm}|>{\raggedright\arraybackslash}p{1cm}|>{\raggedright\arraybackslash}p{1cm}|>{\raggedright\arraybackslash}p{1cm}|>{\raggedleft\arraybackslash}p{1cm}|>{\raggedleft\arraybackslash}p{1cm}|>{\raggedleft\arraybackslash}p{1cm}|>{\raggedleft\arraybackslash}p{1cm}|>{}p{1cm}}
\hline
Estimator & Mean bias & Median bias & ESE & ASE & MAD & RMSE & rRMSE & Nominal coverage & Oracle coverage & Power & Prop. completed\\
\hline
Benchmark model & 0.001 & 0.001 & 0.08 & 0.08 & 0.08 & 0.08 & 0.08 & 0.954 & 0.952 & 0.999 & 100\\
\hline
Complete-case${}^*$ & -0.233 & -0.235 & 0.123 & 0.122 & 0.122 & 0.263 & 0.265 & 0.514 & 0.518 & 0.292 & 100\\
\hline
Confounded model${}^*$ & -0.314 & -0.313 & 0.069 & 0.068 & 0.07 & 0.321 & 0.321 & 0.006 & 0.006 & 0.271 & 100\\
\hline
IPW${}^*$ & -0.04 & -0.042 & 0.13 & 0.13 & 0.133 & 0.136 & 0.139 & 0.940 & 0.940 & 0.801 & 100\\
\hline
Raking (vanilla)${}^*$ & -0.033 & -0.032 & 0.081 & 0.081 & 0.08 & 0.088 & 0.086 & 0.932 & 0.931 & 0.996 & 100\\
\hline
MICE${}^*$ & 0.041 & 0.041 & 0.083 & 0.083 & 0.083 & 0.092 & 0.092 & 0.926 & 0.923 & 1.000 & 100\\
\hline
MI-XGB${}^*$ & 0.008 & 0.008 & 0.08 & 0.08 & 0.08 & 0.08 & 0.08 & 0.948 & 0.947 & 0.999 & 100\\
\hline
MI-RF${}^*$ & -0.026 & -0.026 & 0.082 & 0.079 & 0.082 & 0.083 & 0.086 & 0.925 & 0.937 & 0.998 & 100\\
\hline
IPCW-TMLE-M${}^*$ & -0.056 & -0.058 & 0.134 & 0.135 & 0.135 & 0.147 & 0.147 & 0.936 & 0.931 & 0.737 & 100\\
\hline
IPCW-TMLE-MTO${}^*$ & -0.04 & -0.040 & 0.125 & 0.123 & 0.128 & 0.129 & 0.134 & 0.933 & 0.941 & 0.842 & 100\\
\hline
IPCW-a-TMLE-M${}^*$ & -0.058 & -0.062 & 0.134 & 0.135 & 0.138 & 0.147 & 0.151 & 0.931 & 0.933 & 0.728 & 100\\
\hline
IPCW-a-TMLE-MTO${}^*$ & -0.041 & -0.041 & 0.126 & 0.122 & 0.13 & 0.129 & 0.136 & 0.928 & 0.943 & 0.838 & 100\\
\hline
\end{tabular}
\end{table}

\begin{table}
\centering
\caption{\label{tab:x1_m2.2_y4.1_cOR_oracle}\textbf{Synthetic data MAR simulation: oracle conditional odds ratio (cOR), 12\% outcome proportion, 40\% missing proportion}. Comparing estimators under the \textbf{complex outcome and complex MAR} scenario. The value of the estimand is 0.405. The sample size is n = 10000. Maximum observed Monte-Carlo error over the 2500 simulation replications was 0.009 for all summaries besides coverage and 0.012 for coverage. ESE = empirical standard error, ASE = asymptotic standard error, MAD = mean absolute deviation, RMSE = root mean squared error, rRMSE = robust RMSE (using median bias and MAD), Oracle coverage = coverage of a confidence interval based on the ESE, Nominal coverage = coverage of a confidence interval based on the ASE. Estimators that are mismatched with the estimand (i.e., are estimating a different parameter) are emphasized using a star.}
\centering
\fontsize{9}{11}\selectfont
\begin{tabular}[t]{>{\raggedright\arraybackslash}p{2cm}|>{\raggedright\arraybackslash}p{1cm}|>{\raggedleft\arraybackslash}p{1cm}|>{\raggedright\arraybackslash}p{1cm}|>{\raggedright\arraybackslash}p{1cm}|>{\raggedright\arraybackslash}p{1cm}|>{\raggedright\arraybackslash}p{1cm}|>{\raggedright\arraybackslash}p{1cm}|>{\raggedleft\arraybackslash}p{1cm}|>{\raggedleft\arraybackslash}p{1cm}|>{\raggedleft\arraybackslash}p{1cm}|>{\raggedleft\arraybackslash}p{1cm}|>{}p{1cm}}
\hline
Estimator & Mean bias & Median bias & ESE & ASE & MAD & RMSE & rRMSE & Nominal coverage & Oracle coverage & Power & Prop. completed\\
\hline
Benchmark model & 0 & 0.002 & 0.079 & 0.08 & 0.081 & 0.08 & 0.081 & 0.958 & 0.954 & 0.999 & 100\\
\hline
Complete-case${}^*$ & 0.136 & 0.136 & 0.09 & 0.093 & 0.092 & 0.165 & 0.164 & 0.691 & 0.667 & 1.000 & 100\\
\hline
Confounded model${}^*$ & -0.316 & -0.315 & 0.068 & 0.068 & 0.067 & 0.323 & 0.322 & 0.006 & 0.005 & 0.257 & 100\\
\hline
IPW${}^*$ & 0.053 & 0.052 & 0.095 & 0.096 & 0.097 & 0.109 & 0.11 & 0.916 & 0.918 & 0.998 & 100\\
\hline
Raking (vanilla)${}^*$ & -0.027 & -0.026 & 0.075 & 0.073 & 0.079 & 0.078 & 0.083 & 0.926 & 0.938 & 0.999 & 100\\
\hline
MICE${}^*$ & 0.02 & 0.021 & 0.076 & 0.077 & 0.08 & 0.079 & 0.083 & 0.947 & 0.947 & 1.000 & 100\\
\hline
MI-RF${}^*$ & -0.031 & -0.031 & 0.075 & 0.076 & 0.079 & 0.083 & 0.085 & 0.937 & 0.934 & 0.999 & 100\\
\hline
IPCW-TMLE-M${}^*$ & 0.032 & 0.033 & 0.093 & 0.103 & 0.094 & 0.108 & 0.1 & 0.959 & 0.932 & 0.995 & 100\\
\hline
IPCW-TMLE-MTO${}^*$ & 0.017 & 0.016 & 0.088 & 0.093 & 0.09 & 0.094 & 0.091 & 0.955 & 0.944 & 0.997 & 100\\
\hline
IPCW-a-TMLE-M${}^*$ & 0.042 & 0.042 & 0.095 & 0.103 & 0.095 & 0.111 & 0.104 & 0.951 & 0.921 & 0.998 & 100\\
\hline
IPCW-a-TMLE-MTO${}^*$ & 0.023 & 0.021 & 0.09 & 0.092 & 0.092 & 0.095 & 0.094 & 0.946 & 0.941 & 0.997 & 100\\
\hline
\end{tabular}
\end{table}

\clearpage

\newpage

\subsubsection{Tables: 12\% Outcome, 40\% MNAR, cOR}
\newpage

\begin{table}
\centering
\caption{\label{tab:x1_m2.5_y1.1_cOR_census}\textbf{Synthetic data MNAR simulation: census conditional odds ratio (cOR), 12\% outcome proportion, 40\% missing proportion}. Comparing estimators under the \textbf{simple outcome and MNAR-unobserved} scenario. The value of the estimand is 0.406. The sample size is n = 10000. Maximum observed Monte-Carlo error over the 2500 simulation replications was 0.009 for all summaries besides coverage and 0.012 for coverage. ESE = empirical standard error, ASE = asymptotic standard error, MAD = mean absolute deviation, RMSE = root mean squared error, rRMSE = robust RMSE (using median bias and MAD), Oracle coverage = coverage of a confidence interval based on the ESE, Nominal coverage = coverage of a confidence interval based on the ASE.}
\centering
\fontsize{9}{11}\selectfont
\begin{tabular}[t]{>{\raggedright\arraybackslash}p{2cm}|>{\raggedright\arraybackslash}p{1cm}|>{\raggedleft\arraybackslash}p{1cm}|>{\raggedright\arraybackslash}p{1cm}|>{\raggedright\arraybackslash}p{1cm}|>{\raggedright\arraybackslash}p{1cm}|>{\raggedright\arraybackslash}p{1cm}|>{\raggedright\arraybackslash}p{1cm}|>{\raggedleft\arraybackslash}p{1cm}|>{\raggedleft\arraybackslash}p{1cm}|>{\raggedleft\arraybackslash}p{1cm}|>{\raggedleft\arraybackslash}p{1cm}|>{}p{1cm}}
\hline
Estimator & Mean bias & Median bias & ESE & ASE & MAD & RMSE & rRMSE & Oracle coverage & Nominal coverage & Power & Prop. completed\\
\hline
Benchmark model & -0.002 & -0.001 & 0.069 & 0.07 & 0.068 & 0.07 & 0.068 & 0.951 & 0.954 & 1.000 & 100\\
\hline
Complete-case & -0.077 & -0.079 & 0.104 & 0.105 & 0.103 & 0.131 & 0.129 & 0.889 & 0.894 & 0.886 & 100\\
\hline
Confounded model & 0.202 & 0.202 & 0.066 & 0.067 & 0.065 & 0.212 & 0.212 & 0.135 & 0.141 & 1.000 & 100\\
\hline
IPW & -0.078 & -0.080 & 0.108 & 0.11 & 0.107 & 0.135 & 0.133 & 0.893 & 0.894 & 0.857 & 100\\
\hline
Raking (vanilla) & -0.008 & -0.008 & 0.075 & 0.085 & 0.075 & 0.085 & 0.075 & 0.945 & 0.972 & 1.000 & 100\\
\hline
MICE & -0.008 & -0.007 & 0.073 & 0.074 & 0.073 & 0.075 & 0.073 & 0.946 & 0.952 & 1.000 & 100\\
\hline
MI-XGB & -0.014 & -0.014 & 0.075 & 0.077 & 0.075 & 0.079 & 0.076 & 0.946 & 0.951 & 1.000 & 100\\
\hline
MI-RF & -0.008 & -0.008 & 0.074 & 0.073 & 0.075 & 0.074 & 0.075 & 0.946 & 0.943 & 1.000 & 100\\
\hline
IPCW-TMLE-M & -0.083 & -0.084 & 0.115 & 0.116 & 0.111 & 0.143 & 0.139 & 0.895 & 0.885 & 0.801 & 100\\
\hline
IPCW-TMLE-MTO & -0.083 & -0.083 & 0.115 & 0.114 & 0.111 & 0.141 & 0.139 & 0.896 & 0.885 & 0.812 & 100\\
\hline
IPCW-a-TMLE-M & -0.083 & -0.084 & 0.115 & 0.116 & 0.111 & 0.143 & 0.139 & 0.894 & 0.884 & 0.800 & 100\\
\hline
IPCW-a-TMLE-MTO & -0.083 & -0.082 & 0.115 & 0.114 & 0.111 & 0.14 & 0.138 & 0.897 & 0.882 & 0.812 & 100\\
\hline
\end{tabular}
\end{table}

\begin{table}
\centering
\caption{\label{tab:x1_m2.6_y1.1_cOR_census}\textbf{Synthetic data MNAR simulation: census conditional odds ratio (cOR), 12\% outcome proportion, 40\% missing proportion}. Comparing estimators under the \textbf{simple outcome and MNAR-value} scenario. The value of the estimand is 0.406. The sample size is n = 10000. Maximum observed Monte-Carlo error over the 2500 simulation replications was 0.009 for all summaries besides coverage and 0.012 for coverage. ESE = empirical standard error, ASE = asymptotic standard error, MAD = mean absolute deviation, RMSE = root mean squared error, rRMSE = robust RMSE (using median bias and MAD), Oracle coverage = coverage of a confidence interval based on the ESE, Nominal coverage = coverage of a confidence interval based on the ASE.}
\centering
\fontsize{9}{11}\selectfont
\begin{tabular}[t]{>{\raggedright\arraybackslash}p{2cm}|>{\raggedright\arraybackslash}p{1cm}|>{\raggedleft\arraybackslash}p{1cm}|>{\raggedright\arraybackslash}p{1cm}|>{\raggedright\arraybackslash}p{1cm}|>{\raggedright\arraybackslash}p{1cm}|>{\raggedright\arraybackslash}p{1cm}|>{\raggedright\arraybackslash}p{1cm}|>{\raggedleft\arraybackslash}p{1cm}|>{\raggedleft\arraybackslash}p{1cm}|>{\raggedleft\arraybackslash}p{1cm}|>{\raggedleft\arraybackslash}p{1cm}|>{}p{1cm}}
\hline
Estimator & Mean bias & Median bias & ESE & ASE & MAD & RMSE & rRMSE & Oracle coverage & Nominal coverage & Power & Prop. completed\\
\hline
Benchmark model & -0.002 & 0.000 & 0.072 & 0.07 & 0.072 & 0.07 & 0.072 & 0.949 & 0.944 & 1.000 & 100\\
\hline
Complete-case & -0.14 & -0.140 & 0.109 & 0.11 & 0.107 & 0.178 & 0.176 & 0.758 & 0.762 & 0.681 & 100\\
\hline
Confounded model & 0.201 & 0.203 & 0.069 & 0.067 & 0.068 & 0.211 & 0.214 & 0.178 & 0.158 & 1.000 & 100\\
\hline
IPW & -0.147 & -0.147 & 0.115 & 0.117 & 0.116 & 0.187 & 0.187 & 0.755 & 0.755 & 0.600 & 100\\
\hline
Raking (vanilla) & -0.116 & -0.113 & 0.081 & 0.086 & 0.079 & 0.145 & 0.138 & 0.702 & 0.749 & 0.926 & 100\\
\hline
MICE & -0.115 & -0.112 & 0.08 & 0.077 & 0.081 & 0.138 & 0.138 & 0.703 & 0.681 & 0.955 & 100\\
\hline
MI-XGB & -0.116 & -0.114 & 0.081 & 0.097 & 0.082 & 0.151 & 0.14 & 0.710 & 0.820 & 0.870 & 100\\
\hline
MI-RF & -0.113 & -0.110 & 0.079 & 0.075 & 0.079 & 0.135 & 0.136 & 0.707 & 0.676 & 0.968 & 100\\
\hline
IPCW-TMLE-M & -0.146 & -0.149 & 0.124 & 0.124 & 0.122 & 0.192 & 0.192 & 0.786 & 0.779 & 0.551 & 100\\
\hline
IPCW-TMLE-MTO & -0.146 & -0.150 & 0.123 & 0.122 & 0.12 & 0.19 & 0.192 & 0.784 & 0.772 & 0.565 & 100\\
\hline
\end{tabular}
\end{table}

\begin{table}
\centering
\caption{\label{tab:x1_m2.5_y2.11_cOR_census}\textbf{Synthetic data MNAR simulation: census conditional odds ratio (cOR), 12\% outcome proportion, 40\% missing proportion}. Comparing estimators under the \textbf{simple outcome (unobserved covariate) and MNAR-unobserved} scenario. The value of the estimand is 0.404. The sample size is n = 10000. Maximum observed Monte-Carlo error over the 2500 simulation replications was 0.009 for all summaries besides coverage and 0.012 for coverage. ESE = empirical standard error, ASE = asymptotic standard error, MAD = mean absolute deviation, RMSE = root mean squared error, rRMSE = robust RMSE (using median bias and MAD), Oracle coverage = coverage of a confidence interval based on the ESE, Nominal coverage = coverage of a confidence interval based on the ASE. Estimators that are mismatched with the estimand (i.e., are estimating a different parameter) are emphasized using a star.}
\centering
\fontsize{9}{11}\selectfont
\begin{tabular}[t]{>{\raggedright\arraybackslash}p{2cm}|>{\raggedright\arraybackslash}p{1cm}|>{\raggedleft\arraybackslash}p{1cm}|>{\raggedright\arraybackslash}p{1cm}|>{\raggedright\arraybackslash}p{1cm}|>{\raggedright\arraybackslash}p{1cm}|>{\raggedright\arraybackslash}p{1cm}|>{\raggedright\arraybackslash}p{1cm}|>{\raggedleft\arraybackslash}p{1cm}|>{\raggedleft\arraybackslash}p{1cm}|>{\raggedleft\arraybackslash}p{1cm}|>{\raggedleft\arraybackslash}p{1cm}|>{}p{1cm}}
\hline
Estimator & Mean bias & Median bias & ESE & ASE & MAD & RMSE & rRMSE & Oracle coverage & Nominal coverage & Power & Prop. completed\\
\hline
Benchmark model & 0 & 0.000 & 0.075 & 0.073 & 0.074 & 0.073 & 0.074 & 0.948 & 0.942 & 0.999 & 100\\
\hline
Complete-case & -0.074 & -0.072 & 0.11 & 0.109 & 0.109 & 0.132 & 0.13 & 0.895 & 0.895 & 0.859 & 100\\
\hline
Confounded model & 0.204 & 0.207 & 0.072 & 0.069 & 0.071 & 0.215 & 0.219 & 0.188 & 0.171 & 1.000 & 100\\
\hline
IPW & -0.077 & -0.075 & 0.114 & 0.114 & 0.111 & 0.137 & 0.134 & 0.896 & 0.894 & 0.828 & 100\\
\hline
Raking (vanilla) & -0.007 & -0.006 & 0.08 & 0.088 & 0.082 & 0.088 & 0.082 & 0.949 & 0.966 & 0.996 & 100\\
\hline
MICE & -0.006 & -0.006 & 0.079 & 0.077 & 0.08 & 0.077 & 0.08 & 0.948 & 0.942 & 0.999 & 100\\
\hline
MI-RF & -0.005 & -0.004 & 0.08 & 0.076 & 0.08 & 0.076 & 0.08 & 0.948 & 0.936 & 0.998 & 100\\
\hline
IPCW-TMLE-M & -0.081 & -0.081 & 0.119 & 0.12 & 0.117 & 0.145 & 0.143 & 0.896 & 0.896 & 0.784 & 100\\
\hline
IPCW-TMLE-MTO & -0.081 & -0.081 & 0.118 & 0.118 & 0.116 & 0.143 & 0.142 & 0.895 & 0.891 & 0.790 & 100\\
\hline
IPCW-a-TMLE-M & -0.081 & -0.080 & 0.119 & 0.12 & 0.117 & 0.145 & 0.142 & 0.897 & 0.894 & 0.784 & 100\\
\hline
IPCW-a-TMLE-MTO & -0.081 & -0.079 & 0.118 & 0.118 & 0.117 & 0.143 & 0.142 & 0.897 & 0.890 & 0.796 & 100\\
\hline
\end{tabular}
\end{table}

\begin{table}
\centering
\caption{\label{tab:x1_m2.6_y4.1_cOR_census}\textbf{Synthetic data MNAR simulation: census conditional odds ratio (cOR), 12\% outcome proportion, 40\% missing proportion}. Comparing estimators under the \textbf{complex outcome and MNAR-value} scenario. The value of the estimand is 0.371. The sample size is n = 10000. Maximum observed Monte-Carlo error over the 2500 simulation replications was 0.009 for all summaries besides coverage and 0.012 for coverage. ESE = empirical standard error, ASE = asymptotic standard error, MAD = mean absolute deviation, RMSE = root mean squared error, rRMSE = robust RMSE (using median bias and MAD), Oracle coverage = coverage of a confidence interval based on the ESE, Nominal coverage = coverage of a confidence interval based on the ASE. Estimators that are mismatched with the estimand (i.e., are estimating a different parameter) are emphasized using a star.}
\centering
\fontsize{9}{11}\selectfont
\begin{tabular}[t]{>{\raggedright\arraybackslash}p{2cm}|>{\raggedright\arraybackslash}p{1cm}|>{\raggedleft\arraybackslash}p{1cm}|>{\raggedright\arraybackslash}p{1cm}|>{\raggedright\arraybackslash}p{1cm}|>{\raggedright\arraybackslash}p{1cm}|>{\raggedright\arraybackslash}p{1cm}|>{\raggedright\arraybackslash}p{1cm}|>{\raggedleft\arraybackslash}p{1cm}|>{\raggedleft\arraybackslash}p{1cm}|>{\raggedleft\arraybackslash}p{1cm}|>{\raggedleft\arraybackslash}p{1cm}|>{}p{1cm}}
\hline
Estimator & Mean bias & Median bias & ESE & ASE & MAD & RMSE & rRMSE & Oracle coverage & Nominal coverage & Power & Prop. completed\\
\hline
Benchmark model & 0 & 0.000 & 0.071 & 0.072 & 0.07 & 0.072 & 0.07 & 0.945 & 0.950 & 0.999 & 100\\
\hline
Complete-case & -0.143 & -0.144 & 0.116 & 0.115 & 0.113 & 0.184 & 0.183 & 0.777 & 0.773 & 0.501 & 100\\
\hline
Confounded model & -0.282 & -0.283 & 0.069 & 0.068 & 0.069 & 0.29 & 0.291 & 0.019 & 0.016 & 0.266 & 100\\
\hline
IPW & -0.136 & -0.138 & 0.121 & 0.121 & 0.119 & 0.181 & 0.182 & 0.804 & 0.791 & 0.496 & 100\\
\hline
Raking (vanilla) & -0.471 & -0.470 & 0.086 & 0.096 & 0.085 & 0.481 & 0.478 & 0.000 & 0.000 & 0.147 & 100\\
\hline
MICE & -0.471 & -0.473 & 0.081 & 0.081 & 0.081 & 0.478 & 0.479 & 0.000 & 0.000 & 0.233 & 100\\
\hline
MI-XGB & -0.355 & -0.355 & 0.086 & 0.091 & 0.082 & 0.367 & 0.365 & 0.015 & 0.020 & 0.043 & 100\\
\hline
MI-RF & -0.492 & -0.494 & 0.083 & 0.081 & 0.08 & 0.499 & 0.5 & 0.000 & 0.000 & 0.317 & 100\\
\hline
IPCW-TMLE-M & -0.135 & -0.134 & 0.128 & 0.129 & 0.127 & 0.187 & 0.185 & 0.820 & 0.813 & 0.456 & 100\\
\hline
IPCW-TMLE-MTO & -0.111 & -0.109 & 0.124 & 0.122 & 0.123 & 0.165 & 0.165 & 0.855 & 0.838 & 0.585 & 100\\
\hline
IPCW-a-TMLE-M & -0.135 & -0.135 & 0.128 & 0.129 & 0.128 & 0.187 & 0.186 & 0.818 & 0.813 & 0.456 & 100\\
\hline
IPCW-a-TMLE-MTO & -0.11 & -0.108 & 0.125 & 0.121 & 0.123 & 0.164 & 0.164 & 0.854 & 0.837 & 0.584 & 100\\
\hline
\end{tabular}
\end{table}

\begin{table}
\centering
\caption{\label{tab:x1_m2.5_y1.1_cOR_oracle}\textbf{Synthetic data MNAR simulation: oracle conditional odds ratio (cOR), 12\% outcome proportion, 40\% missing proportion}. Comparing estimators under the \textbf{simple outcome and MNAR-unobserved} scenario. The value of the estimand is 0.405. The sample size is n = 10000. Maximum observed Monte-Carlo error over the 2500 simulation replications was 0.009 for all summaries besides coverage and 0.012 for coverage. ESE = empirical standard error, ASE = asymptotic standard error, MAD = mean absolute deviation, RMSE = root mean squared error, rRMSE = robust RMSE (using median bias and MAD), Oracle coverage = coverage of a confidence interval based on the ESE, Nominal coverage = coverage of a confidence interval based on the ASE.}
\centering
\fontsize{9}{11}\selectfont
\begin{tabular}[t]{>{\raggedright\arraybackslash}p{2cm}|>{\raggedright\arraybackslash}p{1cm}|>{\raggedleft\arraybackslash}p{1cm}|>{\raggedright\arraybackslash}p{1cm}|>{\raggedright\arraybackslash}p{1cm}|>{\raggedright\arraybackslash}p{1cm}|>{\raggedright\arraybackslash}p{1cm}|>{\raggedright\arraybackslash}p{1cm}|>{\raggedleft\arraybackslash}p{1cm}|>{\raggedleft\arraybackslash}p{1cm}|>{\raggedleft\arraybackslash}p{1cm}|>{\raggedleft\arraybackslash}p{1cm}|>{}p{1cm}}
\hline
Estimator & Mean bias & Median bias & ESE & ASE & MAD & RMSE & rRMSE & Nominal coverage & Oracle coverage & Power & Prop. completed\\
\hline
Benchmark model & -0.002 & -0.001 & 0.069 & 0.07 & 0.068 & 0.07 & 0.068 & 0.954 & 0.950 & 1.000 & 100\\
\hline
Complete-case & -0.077 & -0.079 & 0.104 & 0.105 & 0.103 & 0.13 & 0.129 & 0.894 & 0.889 & 0.886 & 100\\
\hline
Confounded model & 0.202 & 0.202 & 0.066 & 0.067 & 0.065 & 0.213 & 0.212 & 0.140 & 0.135 & 1.000 & 100\\
\hline
IPW & -0.078 & -0.080 & 0.108 & 0.11 & 0.107 & 0.135 & 0.133 & 0.894 & 0.893 & 0.857 & 100\\
\hline
Raking (vanilla) & -0.008 & -0.008 & 0.075 & 0.085 & 0.075 & 0.085 & 0.075 & 0.972 & 0.945 & 1.000 & 100\\
\hline
MICE & -0.008 & -0.007 & 0.073 & 0.074 & 0.073 & 0.075 & 0.073 & 0.952 & 0.946 & 1.000 & 100\\
\hline
MI-XGB & -0.014 & -0.013 & 0.075 & 0.077 & 0.075 & 0.079 & 0.076 & 0.952 & 0.946 & 1.000 & 100\\
\hline
MI-RF & -0.008 & -0.008 & 0.074 & 0.073 & 0.075 & 0.074 & 0.075 & 0.943 & 0.946 & 1.000 & 100\\
\hline
IPCW-TMLE-M & -0.083 & -0.083 & 0.115 & 0.116 & 0.111 & 0.143 & 0.139 & 0.886 & 0.895 & 0.801 & 100\\
\hline
IPCW-TMLE-MTO & -0.083 & -0.082 & 0.115 & 0.114 & 0.111 & 0.141 & 0.138 & 0.886 & 0.896 & 0.812 & 100\\
\hline
IPCW-a-TMLE-M & -0.083 & -0.083 & 0.115 & 0.116 & 0.111 & 0.142 & 0.139 & 0.886 & 0.894 & 0.800 & 100\\
\hline
IPCW-a-TMLE-MTO & -0.082 & -0.081 & 0.115 & 0.114 & 0.111 & 0.14 & 0.138 & 0.882 & 0.897 & 0.812 & 100\\
\hline
\end{tabular}
\end{table}

\begin{table}
\centering
\caption{\label{tab:x1_m2.6_y1.1_cOR_oracle}\textbf{Synthetic data MNAR simulation: oracle conditional odds ratio (cOR), 12\% outcome proportion, 40\% missing proportion}. Comparing estimators under the \textbf{simple outcome and MNAR-value} scenario. The value of the estimand is 0.405. The sample size is n = 10000. Maximum observed Monte-Carlo error over the 2500 simulation replications was 0.009 for all summaries besides coverage and 0.012 for coverage. ESE = empirical standard error, ASE = asymptotic standard error, MAD = mean absolute deviation, RMSE = root mean squared error, rRMSE = robust RMSE (using median bias and MAD), Oracle coverage = coverage of a confidence interval based on the ESE, Nominal coverage = coverage of a confidence interval based on the ASE.}
\centering
\fontsize{9}{11}\selectfont
\begin{tabular}[t]{>{\raggedright\arraybackslash}p{2cm}|>{\raggedright\arraybackslash}p{1cm}|>{\raggedleft\arraybackslash}p{1cm}|>{\raggedright\arraybackslash}p{1cm}|>{\raggedright\arraybackslash}p{1cm}|>{\raggedright\arraybackslash}p{1cm}|>{\raggedright\arraybackslash}p{1cm}|>{\raggedright\arraybackslash}p{1cm}|>{\raggedleft\arraybackslash}p{1cm}|>{\raggedleft\arraybackslash}p{1cm}|>{\raggedleft\arraybackslash}p{1cm}|>{\raggedleft\arraybackslash}p{1cm}|>{}p{1cm}}
\hline
Estimator & Mean bias & Median bias & ESE & ASE & MAD & RMSE & rRMSE & Nominal coverage & Oracle coverage & Power & Prop. completed\\
\hline
Benchmark model & -0.002 & 0.000 & 0.072 & 0.07 & 0.072 & 0.07 & 0.072 & 0.944 & 0.949 & 1.000 & 100\\
\hline
Complete-case & -0.14 & -0.140 & 0.109 & 0.11 & 0.107 & 0.178 & 0.176 & 0.762 & 0.758 & 0.681 & 100\\
\hline
Confounded model & 0.201 & 0.203 & 0.069 & 0.067 & 0.068 & 0.211 & 0.214 & 0.158 & 0.178 & 1.000 & 100\\
\hline
IPW & -0.147 & -0.147 & 0.115 & 0.117 & 0.116 & 0.187 & 0.187 & 0.756 & 0.755 & 0.600 & 100\\
\hline
Raking (vanilla) & -0.116 & -0.113 & 0.081 & 0.086 & 0.079 & 0.145 & 0.138 & 0.750 & 0.703 & 0.926 & 100\\
\hline
MICE & -0.115 & -0.111 & 0.08 & 0.077 & 0.081 & 0.138 & 0.138 & 0.682 & 0.703 & 0.955 & 100\\
\hline
MI-XGB & -0.116 & -0.113 & 0.081 & 0.097 & 0.082 & 0.151 & 0.14 & 0.821 & 0.711 & 0.870 & 100\\
\hline
MI-RF & -0.113 & -0.110 & 0.079 & 0.075 & 0.079 & 0.135 & 0.136 & 0.676 & 0.708 & 0.968 & 100\\
\hline
IPCW-TMLE-M & -0.146 & -0.148 & 0.124 & 0.124 & 0.122 & 0.192 & 0.192 & 0.779 & 0.786 & 0.551 & 100\\
\hline
IPCW-TMLE-MTO & -0.146 & -0.149 & 0.123 & 0.122 & 0.12 & 0.19 & 0.192 & 0.772 & 0.785 & 0.565 & 100\\
\hline
\end{tabular}
\end{table}

\begin{table}
\centering
\caption{\label{tab:x1_m2.5_y2.11_cOR_oracle}\textbf{Synthetic data MNAR simulation: oracle conditional odds ratio (cOR), 12\% outcome proportion, 40\% missing proportion}. Comparing estimators under the \textbf{simple outcome (unobserved covariate) and MNAR-unobserved} scenario. The value of the estimand is 0.405. The sample size is n = 10000. Maximum observed Monte-Carlo error over the 2500 simulation replications was 0.009 for all summaries besides coverage and 0.012 for coverage. ESE = empirical standard error, ASE = asymptotic standard error, MAD = mean absolute deviation, RMSE = root mean squared error, rRMSE = robust RMSE (using median bias and MAD), Oracle coverage = coverage of a confidence interval based on the ESE, Nominal coverage = coverage of a confidence interval based on the ASE. Estimators that are mismatched with the estimand (i.e., are estimating a different parameter) are emphasized using a star.}
\centering
\fontsize{9}{11}\selectfont
\begin{tabular}[t]{>{\raggedright\arraybackslash}p{2cm}|>{\raggedright\arraybackslash}p{1cm}|>{\raggedleft\arraybackslash}p{1cm}|>{\raggedright\arraybackslash}p{1cm}|>{\raggedright\arraybackslash}p{1cm}|>{\raggedright\arraybackslash}p{1cm}|>{\raggedright\arraybackslash}p{1cm}|>{\raggedright\arraybackslash}p{1cm}|>{\raggedleft\arraybackslash}p{1cm}|>{\raggedleft\arraybackslash}p{1cm}|>{\raggedleft\arraybackslash}p{1cm}|>{\raggedleft\arraybackslash}p{1cm}|>{}p{1cm}}
\hline
Estimator & Mean bias & Median bias & ESE & ASE & MAD & RMSE & rRMSE & Nominal coverage & Oracle coverage & Power & Prop. completed\\
\hline
Benchmark model & -0.001 & 0.001 & 0.077 & 0.075 & 0.078 & 0.075 & 0.078 & 0.947 & 0.954 & 0.999 & 100\\
\hline
Complete-case${}^*$ & -0.075 & -0.073 & 0.11 & 0.109 & 0.109 & 0.133 & 0.131 & 0.895 & 0.895 & 0.859 & 100\\
\hline
Confounded model${}^*$ & 0.202 & 0.206 & 0.072 & 0.069 & 0.071 & 0.214 & 0.218 & 0.172 & 0.195 & 1.000 & 100\\
\hline
IPW${}^*$ & -0.078 & -0.076 & 0.114 & 0.114 & 0.111 & 0.138 & 0.135 & 0.893 & 0.896 & 0.828 & 100\\
\hline
Raking (vanilla)${}^*$ & -0.008 & -0.007 & 0.08 & 0.088 & 0.082 & 0.088 & 0.082 & 0.966 & 0.948 & 0.996 & 100\\
\hline
MICE${}^*$ & -0.007 & -0.007 & 0.079 & 0.077 & 0.08 & 0.077 & 0.08 & 0.940 & 0.949 & 0.999 & 100\\
\hline
MI-RF${}^*$ & -0.006 & -0.005 & 0.08 & 0.076 & 0.08 & 0.076 & 0.08 & 0.936 & 0.949 & 0.998 & 100\\
\hline
IPCW-TMLE-M${}^*$ & -0.082 & -0.082 & 0.119 & 0.12 & 0.117 & 0.145 & 0.143 & 0.894 & 0.895 & 0.784 & 100\\
\hline
IPCW-TMLE-MTO${}^*$ & -0.082 & -0.082 & 0.118 & 0.118 & 0.116 & 0.144 & 0.142 & 0.890 & 0.896 & 0.790 & 100\\
\hline
IPCW-a-TMLE-M${}^*$ & -0.082 & -0.081 & 0.119 & 0.12 & 0.117 & 0.145 & 0.143 & 0.894 & 0.896 & 0.784 & 100\\
\hline
IPCW-a-TMLE-MTO${}^*$ & -0.082 & -0.080 & 0.118 & 0.118 & 0.117 & 0.143 & 0.142 & 0.889 & 0.896 & 0.796 & 100\\
\hline
\end{tabular}
\end{table}

\begin{table}
\centering
\caption{\label{tab:x1_m2.6_y4.1_cOR_oracle}\textbf{Synthetic data MNAR simulation: oracle conditional odds ratio (cOR), 12\% outcome proportion, 40\% missing proportion}. Comparing estimators under the \textbf{complex outcome and MNAR-value} scenario. The value of the estimand is 0.405. The sample size is n = 10000. Maximum observed Monte-Carlo error over the 2500 simulation replications was 0.009 for all summaries besides coverage and 0.012 for coverage. ESE = empirical standard error, ASE = asymptotic standard error, MAD = mean absolute deviation, RMSE = root mean squared error, rRMSE = robust RMSE (using median bias and MAD), Oracle coverage = coverage of a confidence interval based on the ESE, Nominal coverage = coverage of a confidence interval based on the ASE. Estimators that are mismatched with the estimand (i.e., are estimating a different parameter) are emphasized using a star.}
\centering
\fontsize{9}{11}\selectfont
\begin{tabular}[t]{>{\raggedright\arraybackslash}p{2cm}|>{\raggedright\arraybackslash}p{1cm}|>{\raggedleft\arraybackslash}p{1cm}|>{\raggedright\arraybackslash}p{1cm}|>{\raggedright\arraybackslash}p{1cm}|>{\raggedright\arraybackslash}p{1cm}|>{\raggedright\arraybackslash}p{1cm}|>{\raggedright\arraybackslash}p{1cm}|>{\raggedleft\arraybackslash}p{1cm}|>{\raggedleft\arraybackslash}p{1cm}|>{\raggedleft\arraybackslash}p{1cm}|>{\raggedleft\arraybackslash}p{1cm}|>{}p{1cm}}
\hline
Estimator & Mean bias & Median bias & ESE & ASE & MAD & RMSE & rRMSE & Nominal coverage & Oracle coverage & Power & Prop. completed\\
\hline
Benchmark model & -0.001 & -0.002 & 0.081 & 0.08 & 0.081 & 0.08 & 0.081 & 0.945 & 0.947 & 0.998 & 100\\
\hline
Complete-case${}^*$ & -0.177 & -0.178 & 0.116 & 0.115 & 0.113 & 0.211 & 0.211 & 0.662 & 0.663 & 0.501 & 100\\
\hline
Confounded model${}^*$ & -0.316 & -0.317 & 0.069 & 0.068 & 0.069 & 0.323 & 0.324 & 0.004 & 0.004 & 0.266 & 100\\
\hline
IPW${}^*$ & -0.17 & -0.172 & 0.121 & 0.121 & 0.119 & 0.208 & 0.209 & 0.706 & 0.714 & 0.496 & 100\\
\hline
Raking (vanilla)${}^*$ & -0.505 & -0.504 & 0.086 & 0.096 & 0.085 & 0.514 & 0.512 & 0.000 & 0.000 & 0.147 & 100\\
\hline
MICE${}^*$ & -0.505 & -0.507 & 0.081 & 0.081 & 0.081 & 0.512 & 0.513 & 0.000 & 0.000 & 0.233 & 100\\
\hline
MI-XGB${}^*$ & -0.389 & -0.389 & 0.086 & 0.091 & 0.082 & 0.4 & 0.398 & 0.008 & 0.007 & 0.043 & 100\\
\hline
MI-RF${}^*$ & -0.527 & -0.528 & 0.083 & 0.081 & 0.08 & 0.533 & 0.534 & 0.000 & 0.000 & 0.317 & 100\\
\hline
IPCW-TMLE-M${}^*$ & -0.17 & -0.168 & 0.128 & 0.129 & 0.127 & 0.213 & 0.211 & 0.736 & 0.744 & 0.456 & 100\\
\hline
IPCW-TMLE-MTO${}^*$ & -0.145 & -0.143 & 0.124 & 0.122 & 0.123 & 0.189 & 0.189 & 0.766 & 0.784 & 0.585 & 100\\
\hline
IPCW-a-TMLE-M${}^*$ & -0.17 & -0.169 & 0.128 & 0.129 & 0.128 & 0.213 & 0.212 & 0.736 & 0.743 & 0.456 & 100\\
\hline
IPCW-a-TMLE-MTO${}^*$ & -0.144 & -0.142 & 0.125 & 0.121 & 0.123 & 0.188 & 0.188 & 0.764 & 0.786 & 0.584 & 100\\
\hline
\end{tabular}
\end{table}

\clearpage
\newpage

\subsubsection{Tables: Other Synthetic-data Scenarios}

Tables with full results from other scenarios can be found on \href{https://github.com/PamelaShaw/Missing-Confounders-Methods/tree/main/}{GitHub site \url{https://github.com/PamelaShaw/Missing-Confounders-Methods} \citep{shaw2024comparison}}.

\clearpage
\newpage


\subsection{Synthetic-Data Simulation Results Figures}\label{sec:supp-tables-synthetic}
\newpage

\subsubsection{Base Case, mRD}
\newpage

\begin{figure}[!htb]
\caption[Missing at random, census truth, high outcome, low missing]{\textbf{Synthetic Data MAR Simulation: Census mRD}. Comparing estimators of the census estimand with \textbf{40\% confounder missingness MAR} and \textbf{12\% outcome proportion}. \textbf{Top graph}: \%Bias (median, IQR, min and max of converged simulations); \textbf{Middle graph}: Robust RMSE (rRMSE), using median bias and MAD; \textbf{Bottom graphs}: Nominal and oracle coverage, respectively, with blue confidence bands at $ .95 \pm 1.96 \sqrt{\frac{.05\cdot .95}{2500}}$. True mRD values are 0.040 and 0.037 for simple and complex outcome models, respectively.}

\includegraphics[scale=0.65]{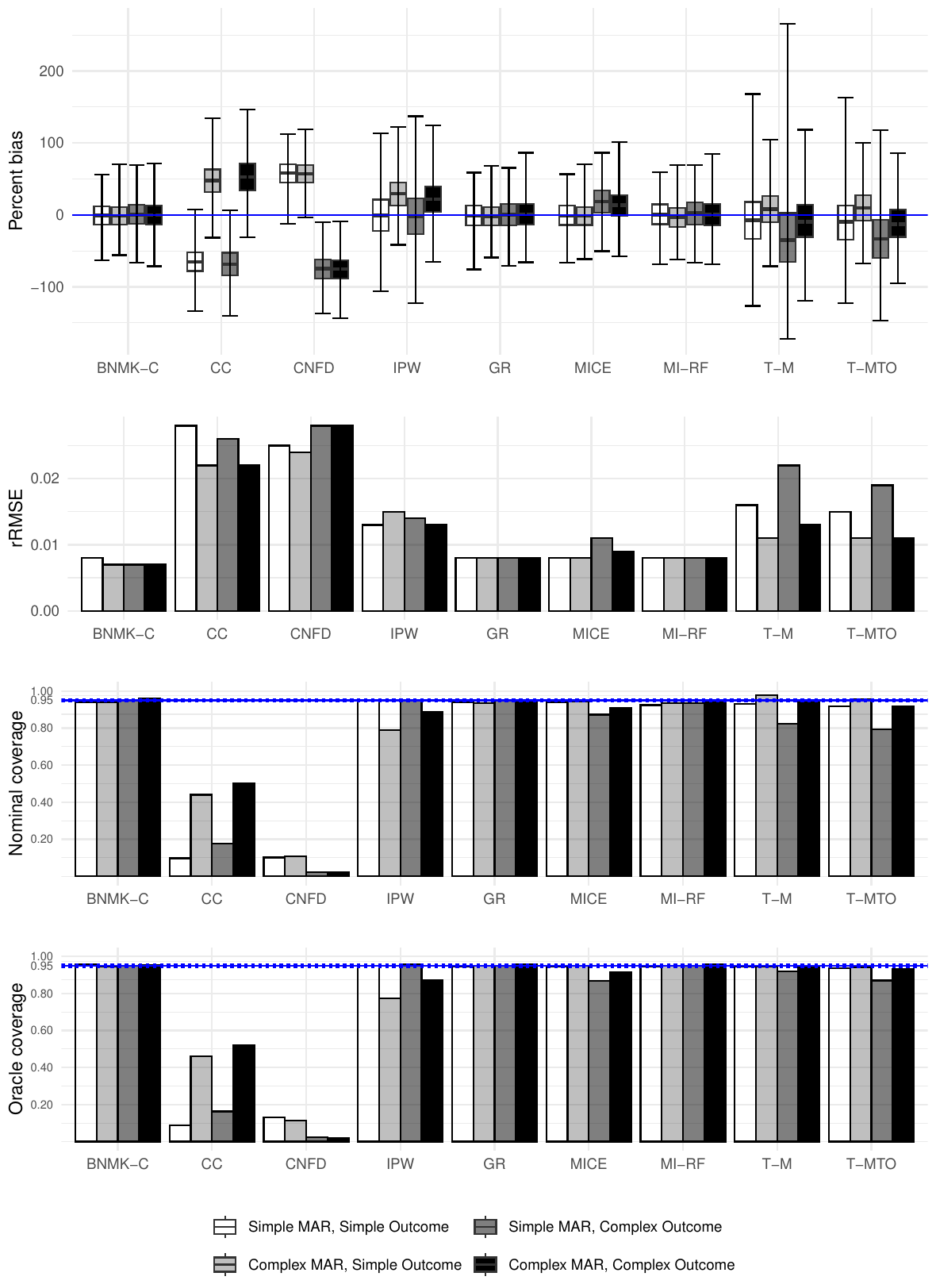}
\label{fig:mRD_mar_census_highout_lowmissing}
\end{figure}


\begin{figure}[!htb]
\caption[Missing at random, oracle truth, high outcome, low missing]{\textbf{Synthetic Data MAR Simulation: oracle mRD}. Comparing estimators of the oracle truth with \textbf{40\% missingness MAR} and \textbf{12\% incidence}. \textbf{Top graph}: \%Bias (median, IQR, min and max of converged simulations); \textbf{Middle graph}: Robust RMSE (rRMSE), using median bias and MAD; \textbf{Bottom graphs}: Nominal and oracle coverage, respectively, with blue confidence bands at $ .95 \pm 1.96 \sqrt{\frac{.05\cdot .95}{2500}}$. True mRD values are 0.040 and 0.031 for simple and complex outcome models, respectively.}

\includegraphics[scale=0.65]{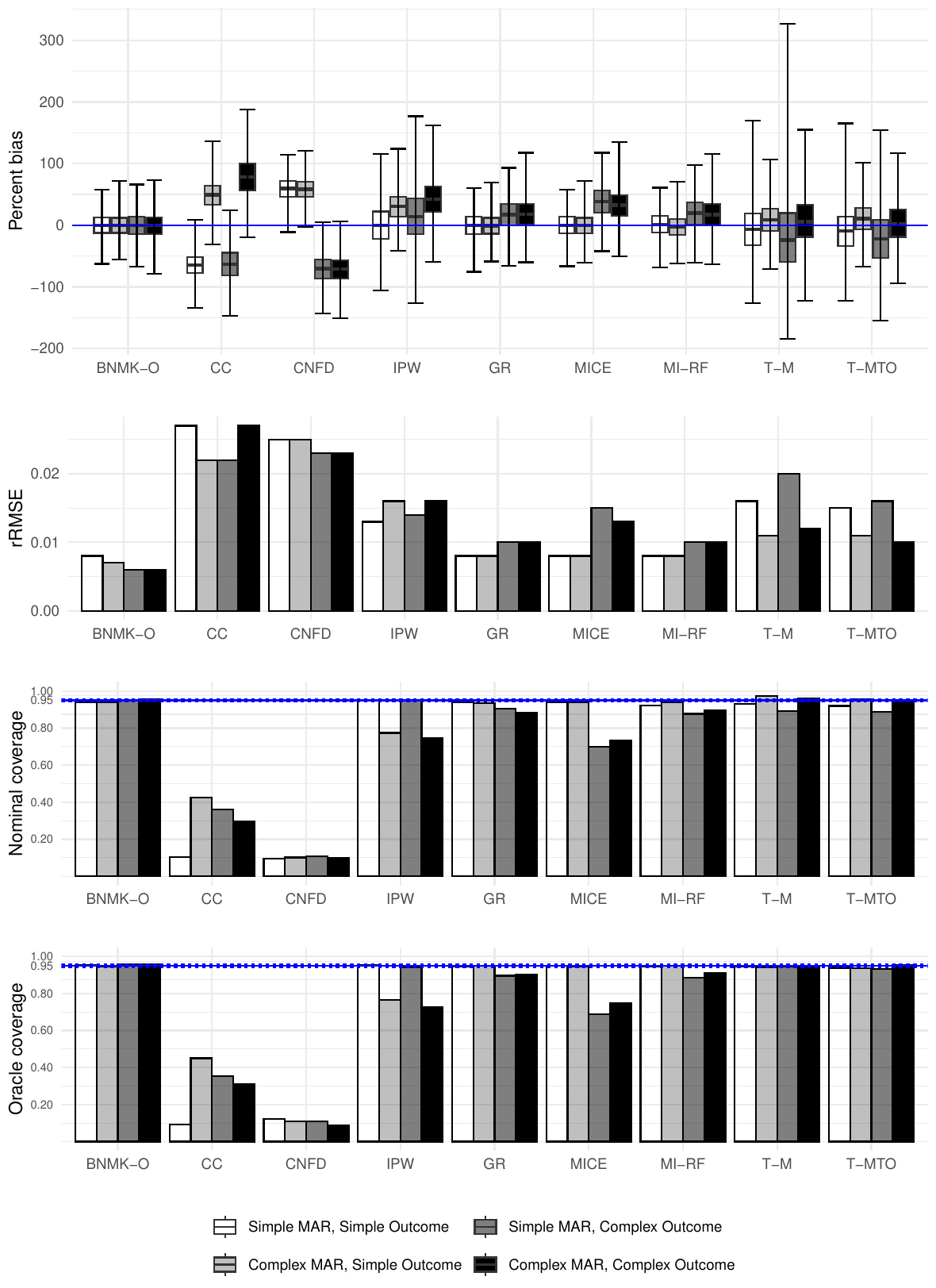}
\label{fig:mRD_mar_oracle_highout_lowmissing}
\end{figure}
\clearpage
\newpage
\subsubsection{Base Case, mRR}
\newpage

\begin{figure}[!htb]
\caption[Missing at random, census truth, high outcome, low missing]{\textbf{Synthetic Data MAR Simulation: Census mlogRR}. Comparing estimators of the census estimand with \textbf{40\% confounder missingness} and \textbf{12\% outcome proportion}. \textbf{Top graph}: \%Bias (median, IQR, min and max of converged simulations); \textbf{Middle graph}: Robust RMSE (rRMSE), using median bias and MAD; \textbf{Bottom graphs}: Nominal and oracle coverage, respectively, with blue confidence bands at $ .95 \pm 1.96 \sqrt{\frac{.05\cdot .95}{2500}}$. True values of mlogRR are 0.334 and 0.315 for simple and complex outcome models, respectively.}

\includegraphics[scale=0.65]{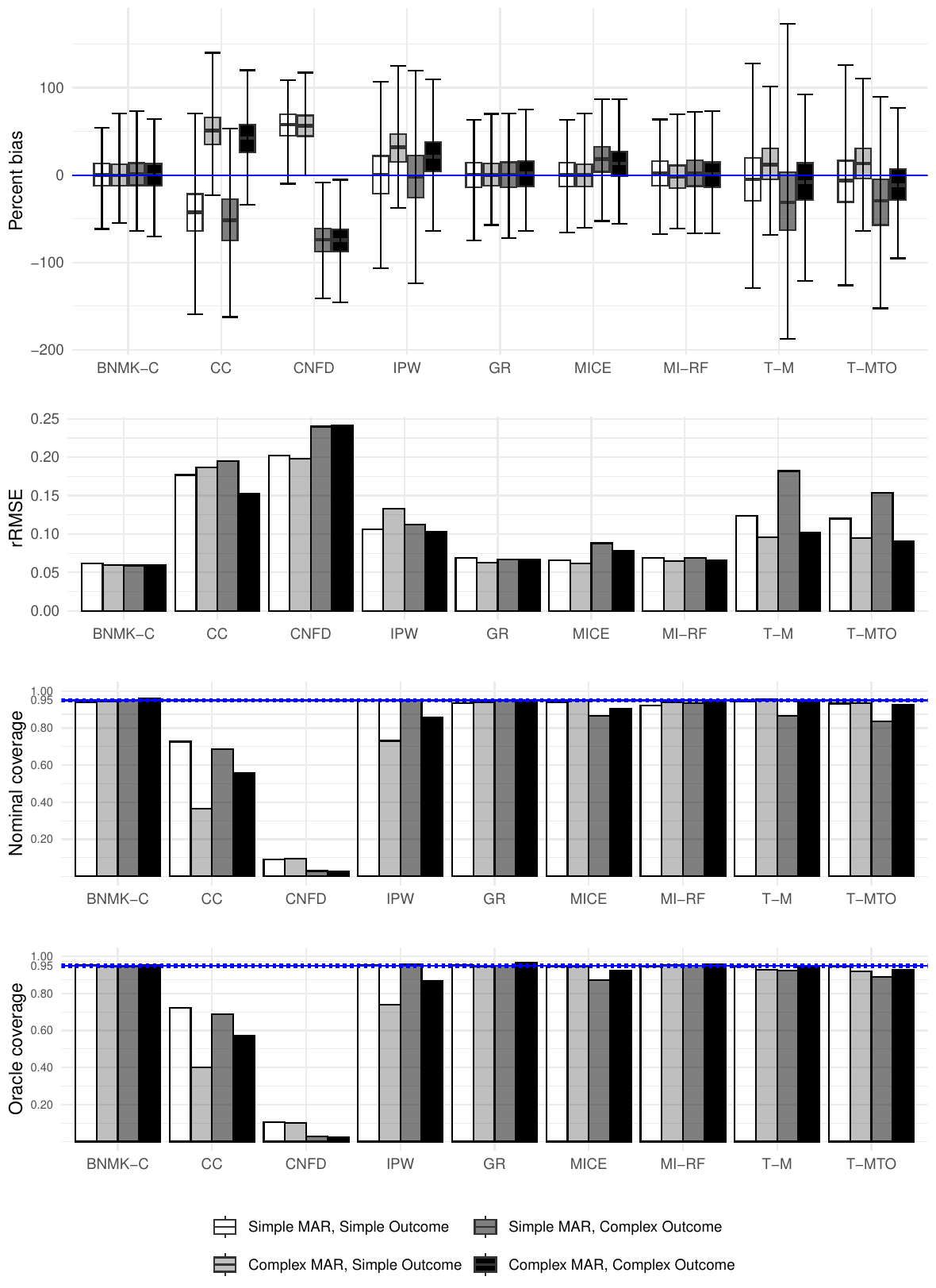}
\label{fig:mlogRR_mar_census_highout_lowmissing}
\end{figure}
\clearpage

\begin{figure}[!htb]
\caption[Missing at random, oracle truth, high outcome, low missing]{\textbf{Synthetic Data MAR Simulation: Oracle mlogRR}. Comparing estimators of the oracle estimand with \textbf{40\% confounder missingness} and \textbf{12\% outcome proportion}. \textbf{Top graph}: \%Bias (median, IQR, min and max of converged simulations); \textbf{Middle graph}: Robust RMSE (rRMSE), using median bias and MAD; \textbf{Bottom graphs}: Nominal and oracle coverage, respectively, with blue confidence bands at $ .95 \pm 1.96 \sqrt{\frac{.05\cdot .95}{2500}}$. True values of mlogRR are 0.334 and 0.271 for simple and complex outcome models, respectively.}

\includegraphics[scale=0.65]{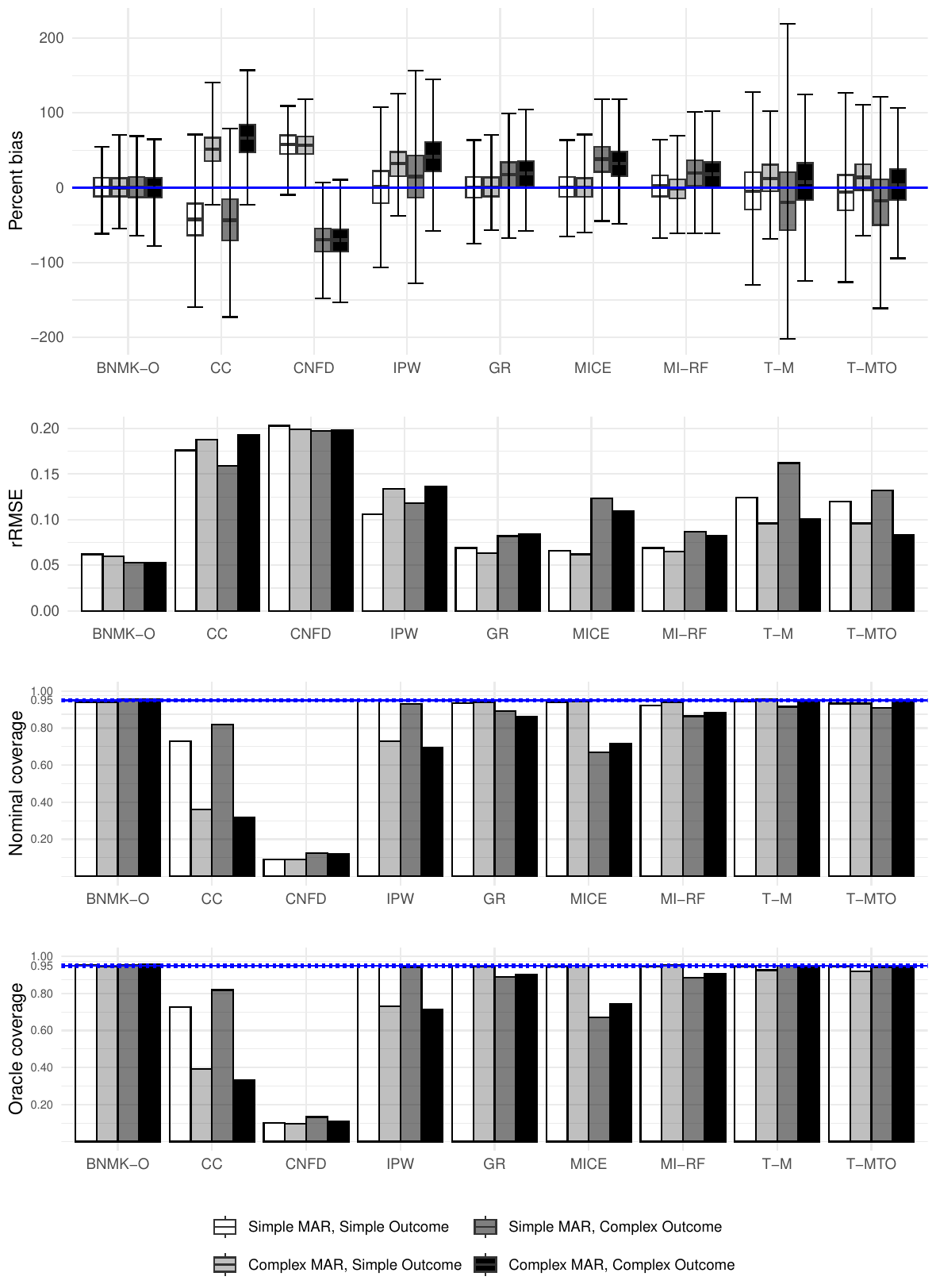}
\label{fig:mlogRR_mar_oracle_highout_lowmissing}
\end{figure}

\clearpage
\newpage

\subsubsection{12\% Outcome, 80\% MAR, cOR}
\newpage

\begin{figure}[!htb]
\caption[Missing at random, census truth, high outcome, high missing]{\textbf{Synthetic Data MAR Simulation: Census clogOR}. Comparing estimators of the census estimand with \textbf{80\% missingness} and \textbf{12\% outcome proportion}. \textbf{Top graph}: \%Bias (median, IQR, min and max of converged simulations); \textbf{Middle graph}: Robust RMSE (rRMSE), using median bias and MAD; \textbf{Bottom graphs}: Nominal and oracle coverage, respectively, with blue confidence bands at $ .95 \pm 1.96 \sqrt{\frac{.05\cdot .95}{2500}}$. True clogOR values are 0.405 and 0.371 for simple and complex outcome models, respectively.}

\includegraphics[scale=0.65]{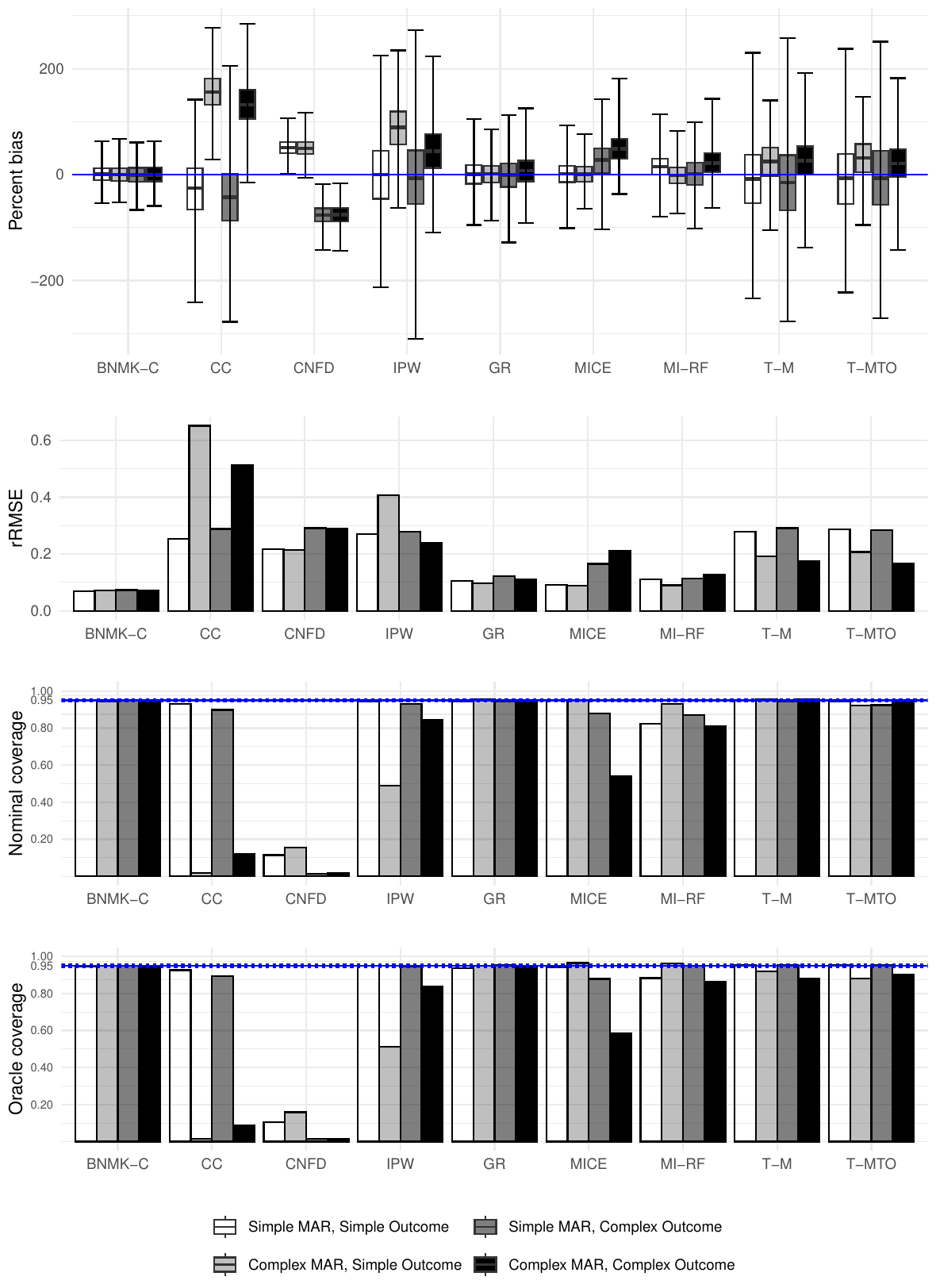}
\label{fig:clogOR_mar_census_highout_highmissing}
\end{figure}
\clearpage


\begin{figure}[!htb]
\caption[Missing at random, oracle truth, high outcome, high missing]{\textbf{Synthetic Data MAR Simulation: Oracle clogOR}. Comparing estimators of the oracle estimand with \textbf{80\% confounder missingness MAR} and \textbf{12\% outcome proportion}. \textbf{Top graph}: \%Bias (median, IQR, min and max of converged simulations); \textbf{Middle graph}: Robust RMSE (rRMSE), using median bias and MAD; \textbf{Bottom graphs}: Nominal and oracle coverage, respectively, with blue confidence bands at $ .95 \pm 1.96 \sqrt{\frac{.05\cdot .95}{2500}}$. The true clogOR value is 0.405 for both simple and complex outcome models.}

\includegraphics[scale=0.65]{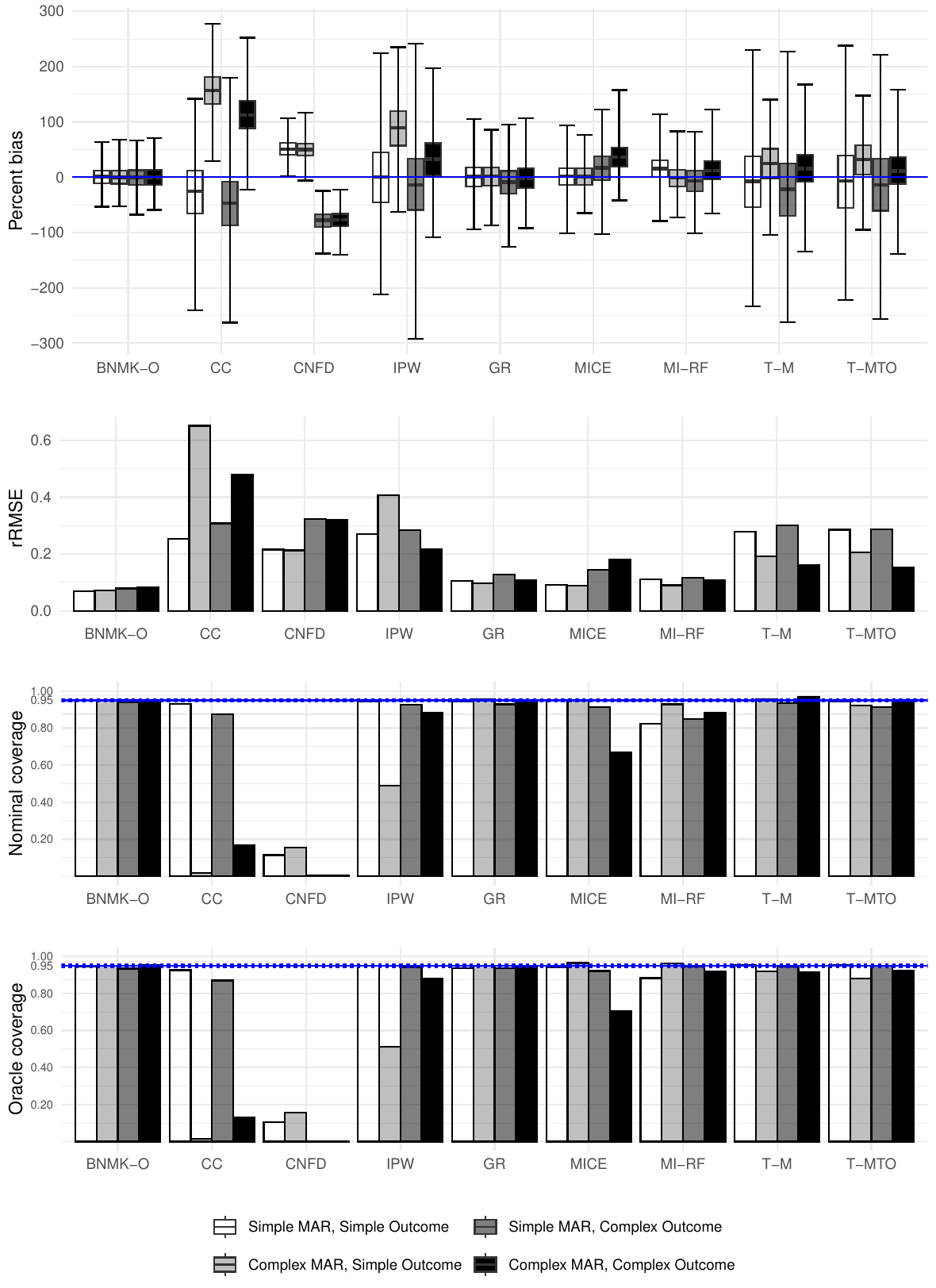}
\label{fig:clogOR_mar_oracle_highout_highmissing}
\end{figure}
\clearpage
\newpage

\subsubsection{12\% Outcome, 80\% MAR, mRD}
\newpage

\begin{figure}[!htb]
\caption[Missing at random, census truth, high outcome, high missing]{\textbf{Synthetic Data MAR Simulation: Census mRD}. Comparing estimators of the census truth with \textbf{80\% confounder missingness} and \textbf{12\% outcome proportion}. \textbf{Top graph}: \%Bias (median, IQR, min and max of converged simulations); \textbf{Middle graph}: Robust RMSE (rRMSE), using median bias and MAD; \textbf{Bottom graphs}: Nominal and oracle coverage, respectively, with blue confidence bands at $ .95 \pm 1.96 \sqrt{\frac{.05\cdot .95}{2500}}$. True mRD values are 0.040 and 0.037 for simple and complex outcome models, respectively.}

\includegraphics[scale=0.65]{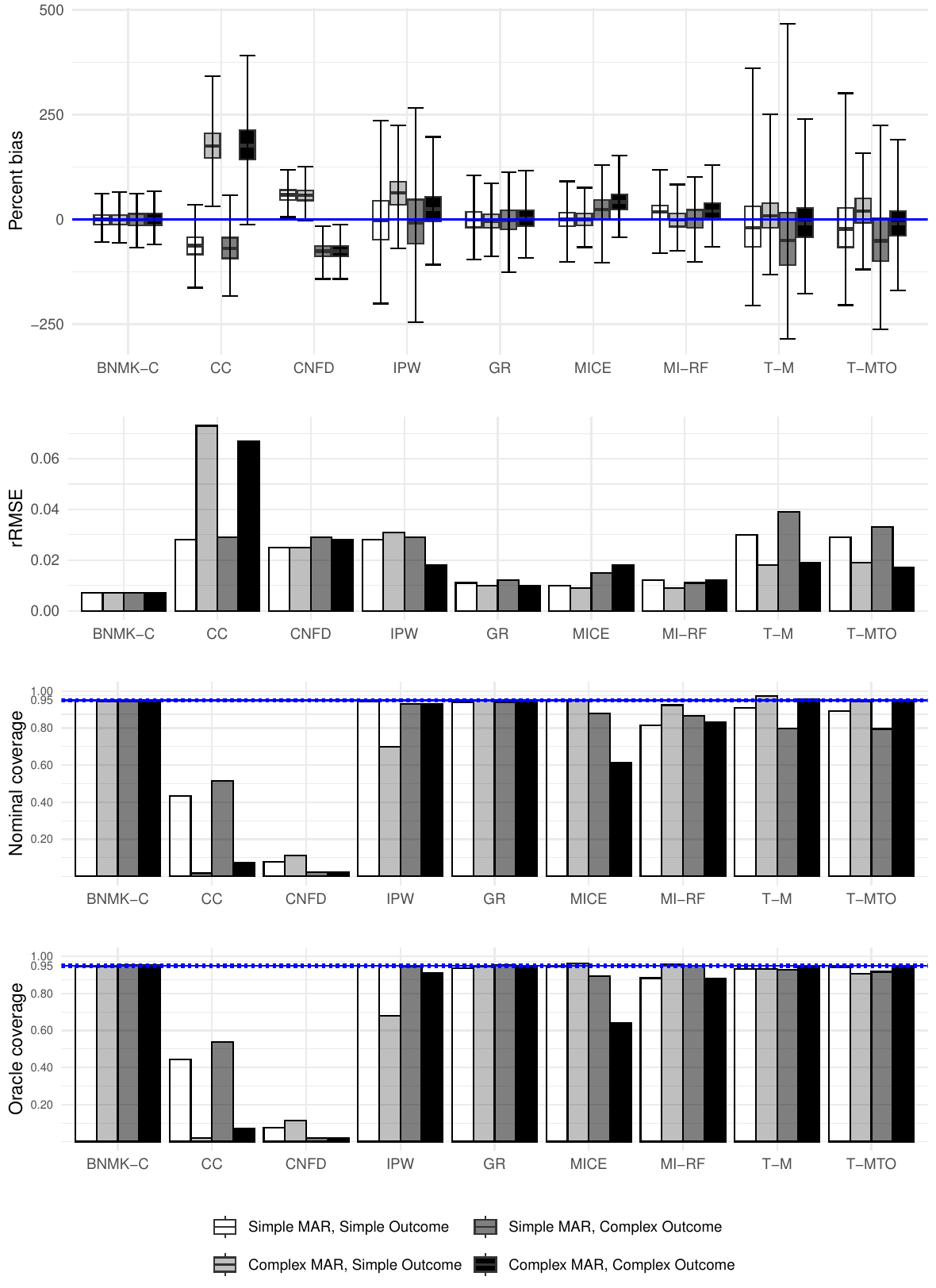}
\label{fig:mRD_mar_census_highout_highmissing}
\end{figure}
\clearpage

\begin{figure}[!htb]
\caption[Missing at random, oracle truth, high outcome, high missing]{\textbf{Synthetic Data MAR Simulation: Oracle mRD}. Comparing estimators of the oracle estimand with \textbf{80\% confounder missingness} and \textbf{12\% outcome proportion}. \textbf{Top graph}: \%Bias (median, IQR, min and max of converged simulations); \textbf{Middle graph}: Robust RMSE (rRMSE), using median bias and MAD; \textbf{Bottom graphs}: Nominal and oracle coverage, respectively, with blue confidence bands at $ .95 \pm 1.96 \sqrt{\frac{.05\cdot .95}{2500}}$. True mRD values are 0.040 and 0.031 for simple and complex outcome models, respectively.}

\includegraphics[scale=0.65]{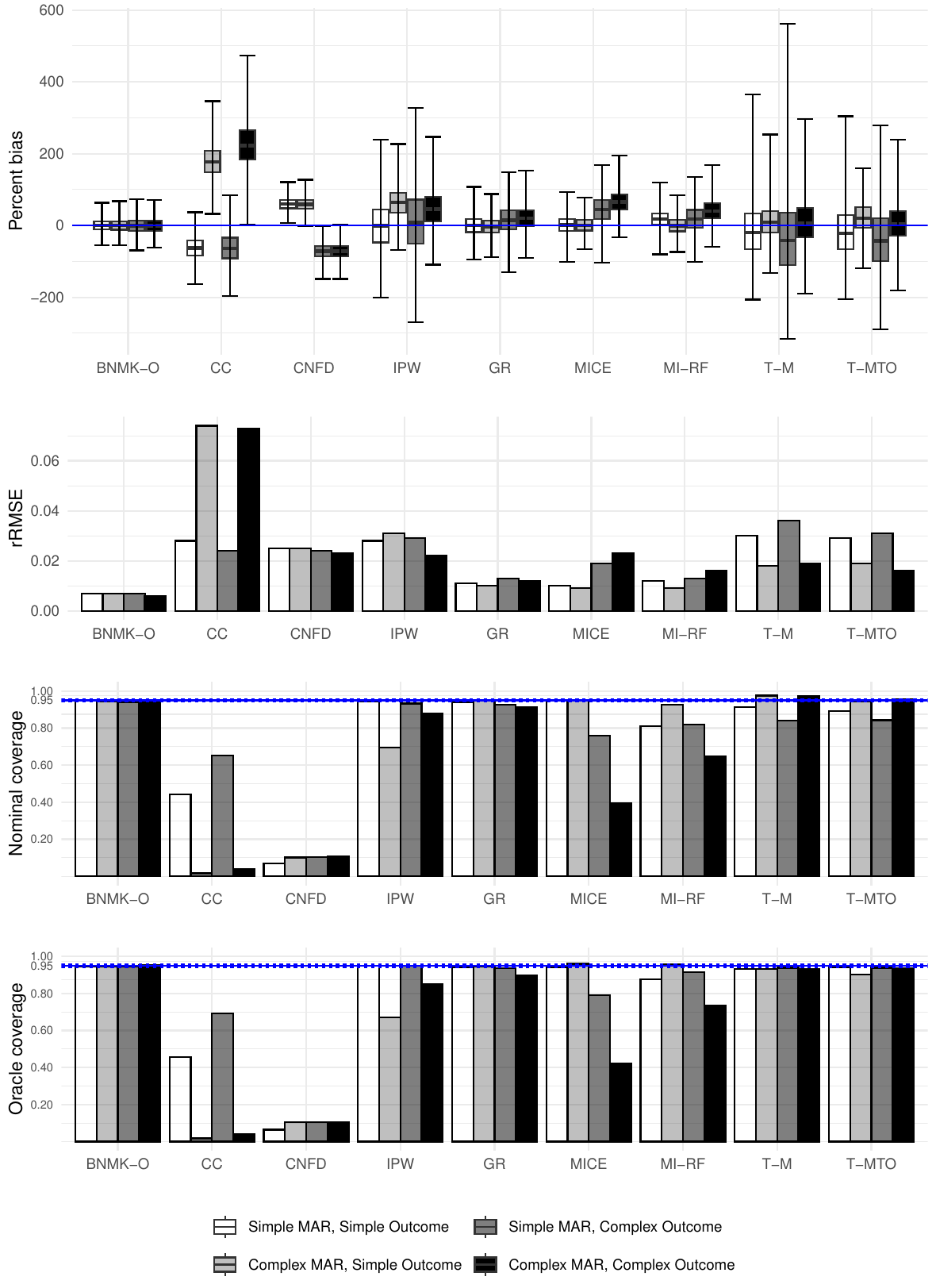}
\label{fig:mRD_mar_oracle_highout_highmissing}
\end{figure}
\clearpage
\newpage
\subsubsection{12\% Outcome, 80\% MAR, mRR}
\newpage

\begin{figure}[!htb]
\caption[Missing at random, census truth, high outcome, high missing]{\textbf{Synthetic Data MAR Simulation: Census mlogRR}. Comparing estimators of the census truth with \textbf{80\% confounder missingness} and \textbf{12\% outcome proportion}. \textbf{Top graph}: \%Bias (median, IQR, min and max of converged simulations); \textbf{Middle graph}: Robust RMSE (rRMSE), using median bias and MAD; \textbf{Bottom graphs}: Nominal and oracle coverage, respectively, with blue confidence bands at $ .95 \pm 1.96 \sqrt{\frac{.05\cdot .95}{2500}}$. True values of mlogRR are 0.334 and 0.315 for simple and complex outcome models, respectively.}

\includegraphics[scale=0.65]{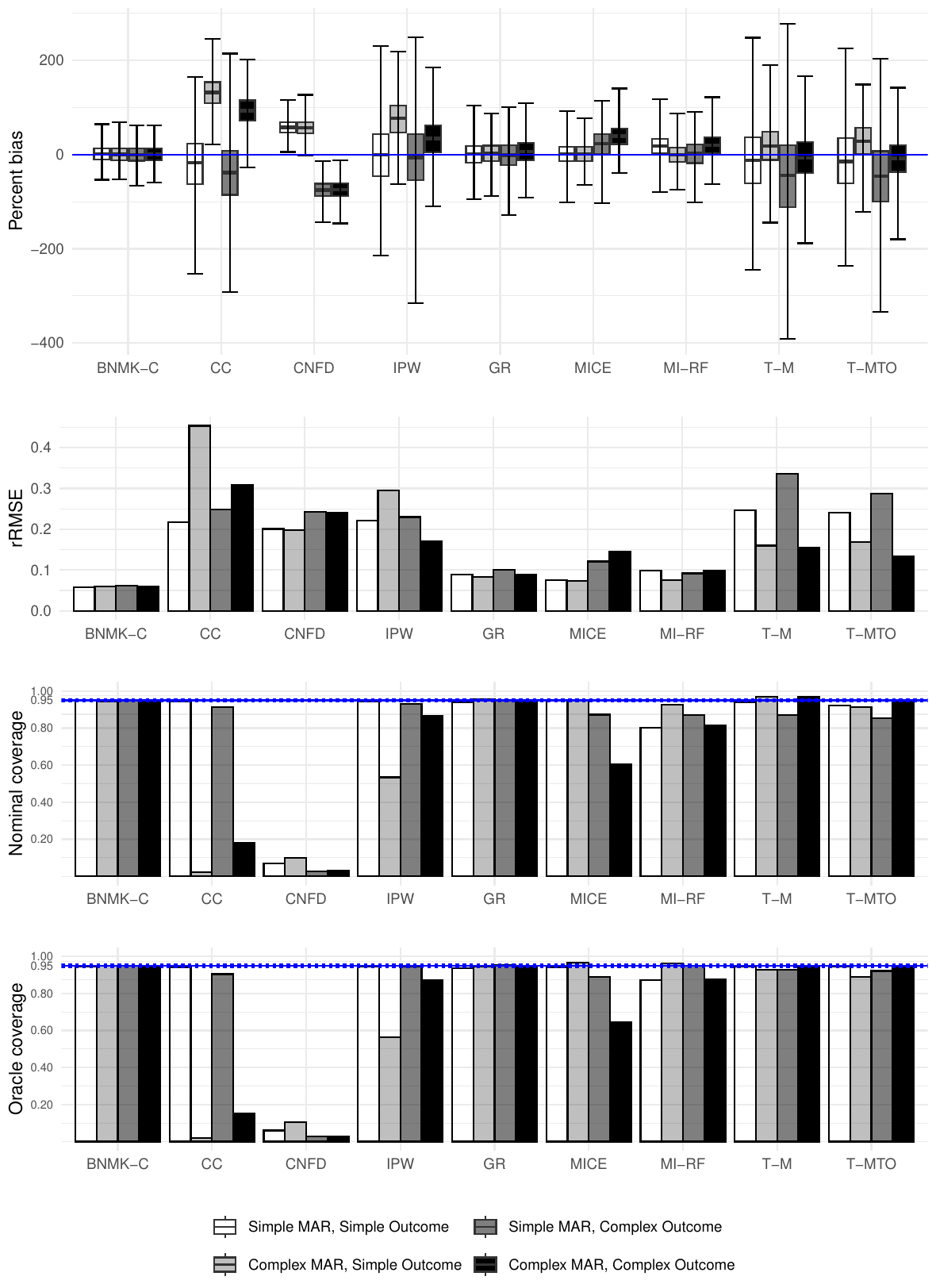}
\label{fig:mlogRR_mar_census_highout_highmissing}
\end{figure}
\clearpage

\begin{figure}[!htb]
\caption[Missing at random, oracle truth, high outcome, high missing]{\textbf{Synthetic Data MAR Simulation: Oracle mlogRR}. Comparing estimators of the oracle estimand with \textbf{80\% confounder missingness} and \textbf{12\% outcome proportion}. \textbf{Top graph}: \%Bias (median, IQR, min and max of converged simulations); \textbf{Middle graph}: Robust RMSE (rRMSE), using median bias and MAD; \textbf{Bottom graphs}: Nominal and oracle coverage, respectively, with blue confidence bands at $ .95 \pm 1.96 \sqrt{\frac{.05\cdot .95}{2500}}$. True values of mlogRR are 0.334 and 0.271 for simple and complex outcome models, respectively.}

\includegraphics[scale=0.65]{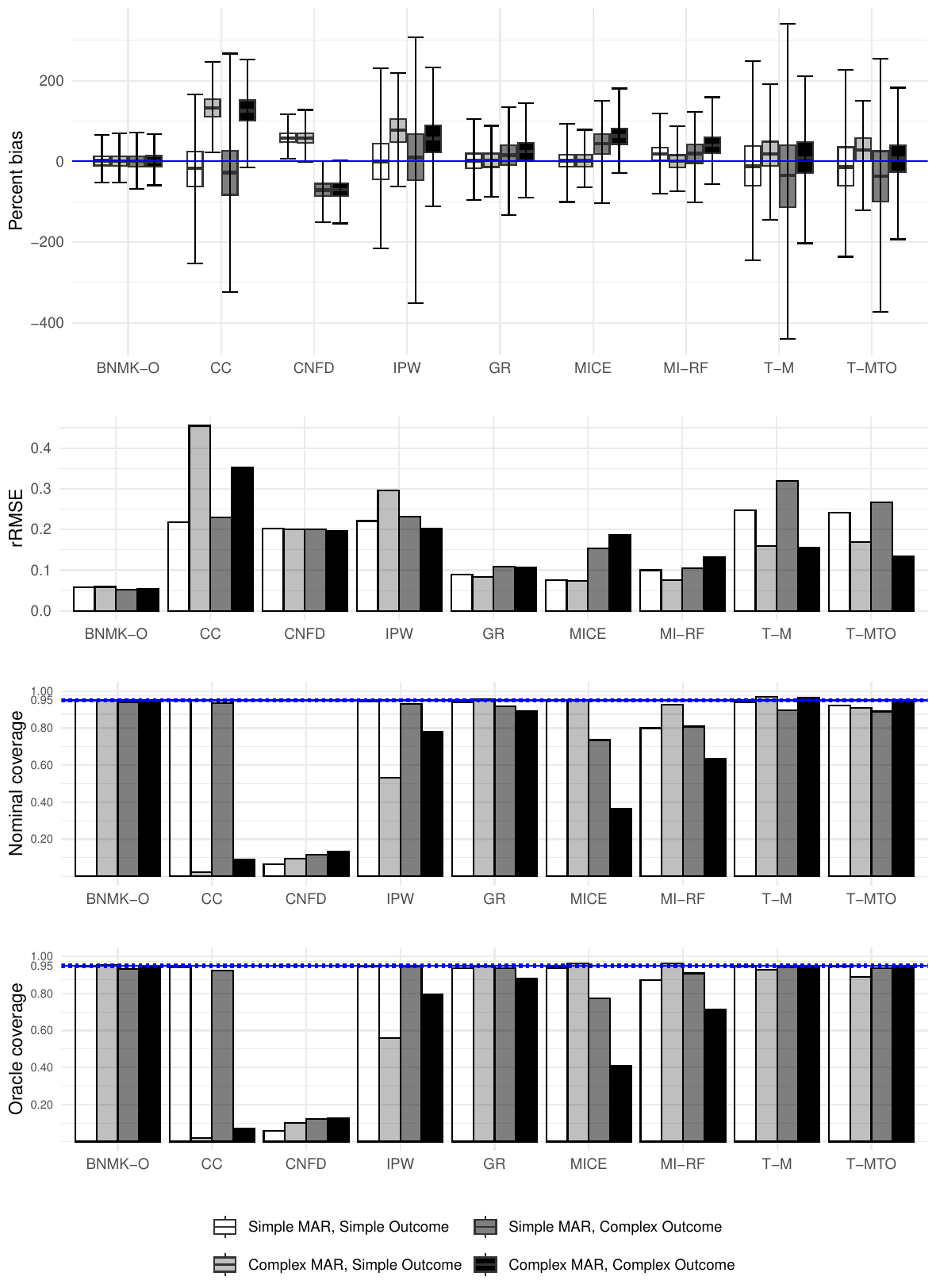}
\label{fig:mlogRR_mar_oracle_highout_highmissing}
\end{figure}
\clearpage
\newpage

\subsubsection{5\% Outcome, 80\% MAR, cOR}
\newpage

\begin{figure}[!htb]
\caption[Missing at random, census truth, low outcome, high missing]{\textbf{Synthetic Data MAR Simulation: Census clogOR}. Comparing estimators of the census estimand with \textbf{80\% confounder missingness MAR} and \textbf{5\% outcome proportion}. \textbf{Top graph}: \%Bias (median, IQR, min and max of converged simulations); \textbf{Middle graph}: Robust RMSE (rRMSE), using median bias and MAD; \textbf{Bottom graphs}: Nominal and oracle coverage, respectively, with blue confidence bands at $ .95 \pm 1.96 \sqrt{\frac{.05\cdot .95}{2500}}$. True clogOR values are 0.405 and 0.380 for simple and complex outcome models, respectively.}

\includegraphics[scale=0.65]{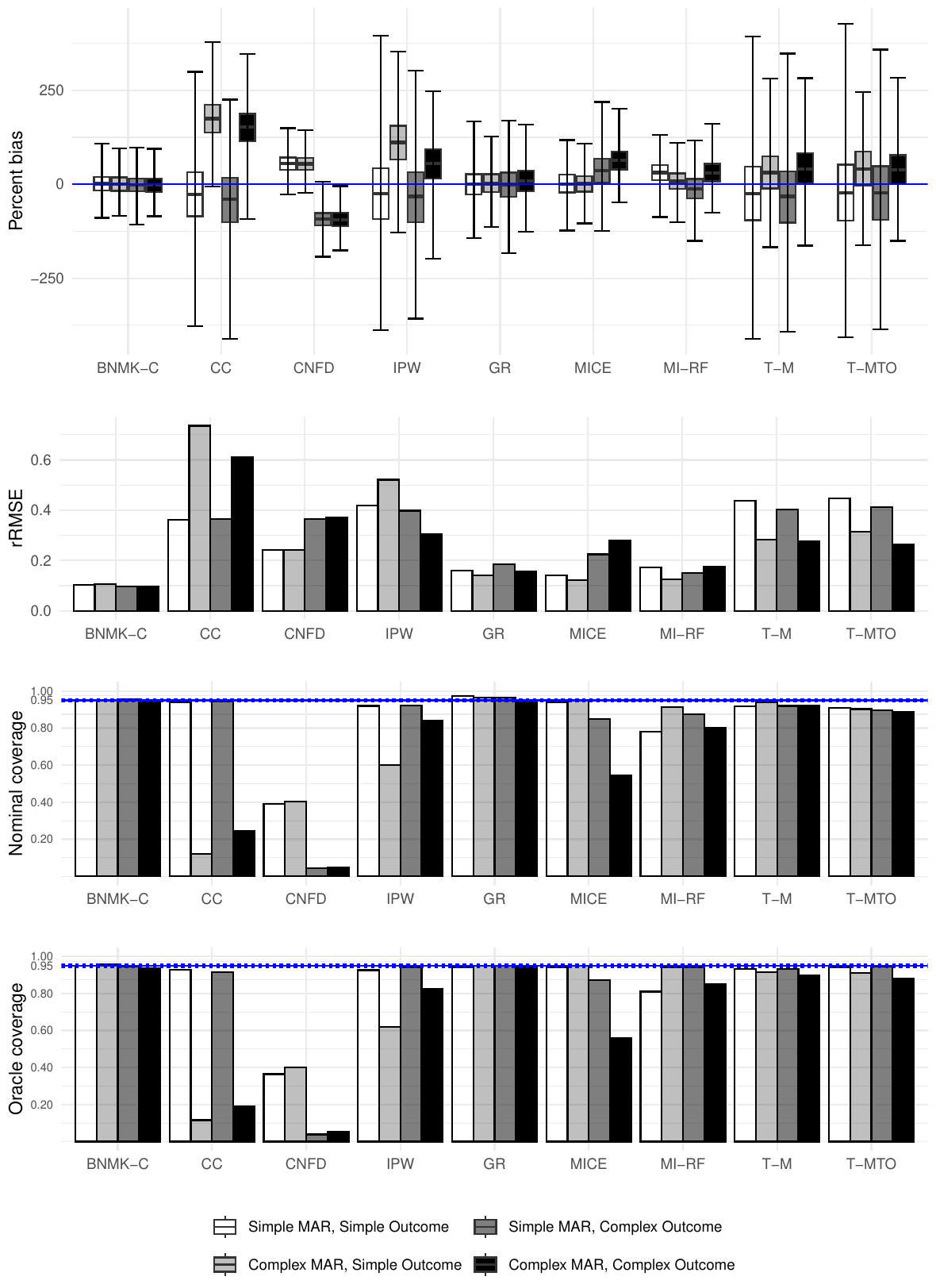}
\label{fig:clogOR_mar_census_lowout_highmissing}
\end{figure}
\clearpage

\begin{figure}[!htb]
\caption[Missing at random, oracle truth, low outcome, high missing]{\textbf{Synthetic Data MAR Simulation: Oracle clogOR}. Comparing estimators of the oracle estimand with \textbf{80\% confounder missingness MAR} and \textbf{5\% outcome proportion}. \textbf{Top graph}: \%Bias (median, IQR, min and max of converged simulations); \textbf{Middle graph}: Robust RMSE (rRMSE), using median bias and MAD; \textbf{Bottom graphs}: Nominal and oracle coverage, respectively, with blue confidence bands at $ .95 \pm 1.96 \sqrt{\frac{.05\cdot .95}{2500}}$. The true clogOR value is 0.405 for both simple and complex outcome models.}

\includegraphics[scale=0.65]{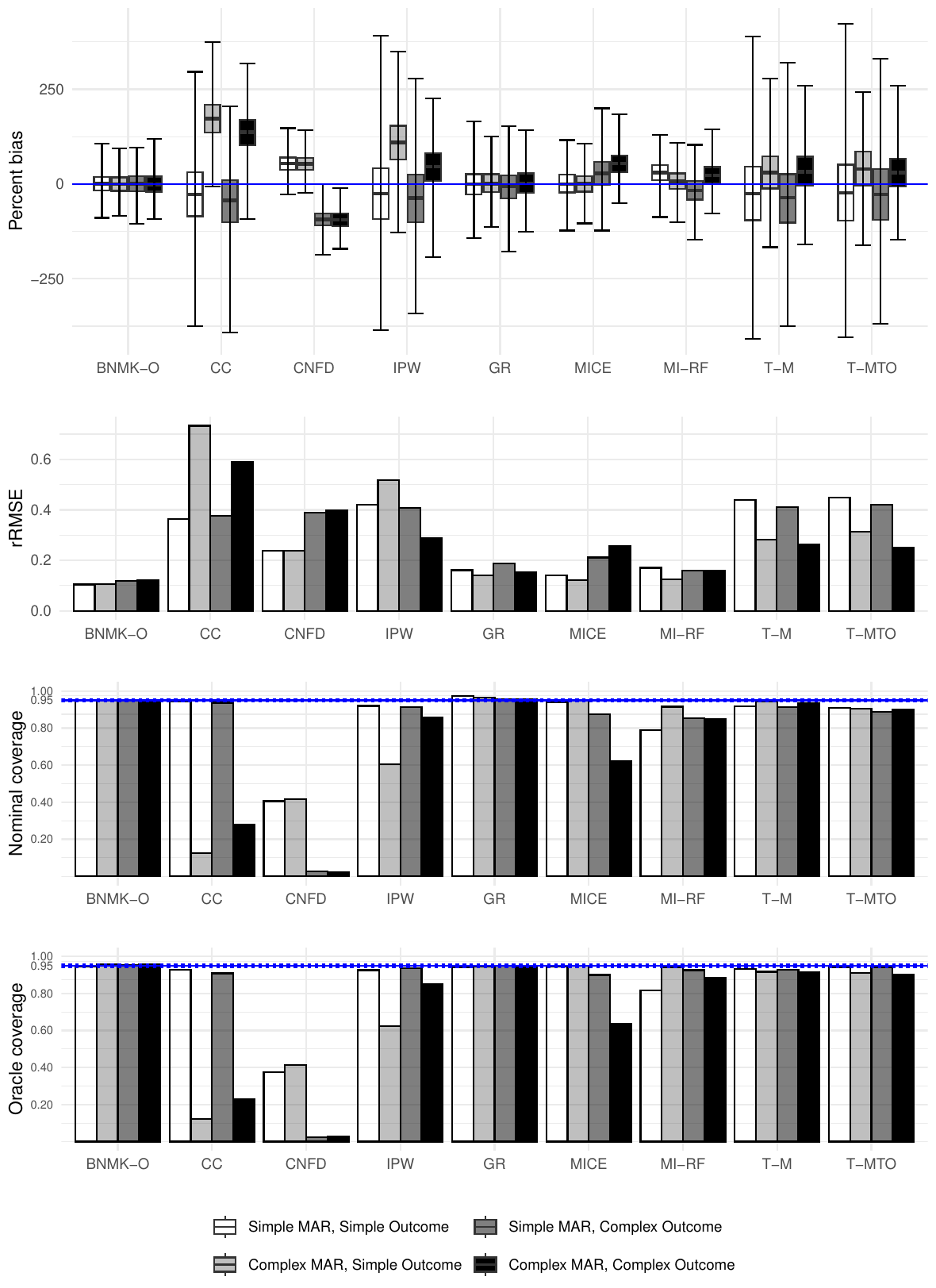}
\label{fig:clogOR_mar_oracle_lowout_highmissing}
\end{figure}
\clearpage
\newpage

\subsubsection{5\% Outcome, 80\% MAR, mRD}
\newpage

\begin{figure}[!htb]
\caption[Missing at random, census truth, low outcome, high missing]{\textbf{Synthetic Data MAR Simulation: Census mRD}. Comparing estimators of the census estimand with \textbf{80\% confounder missingness} and \textbf{5\% outcome proportion}. \textbf{Top graph}: \%Bias (median, IQR, min and max of converged simulations); \textbf{Middle graph}: Robust RMSE (rRMSE), using median bias and MAD; \textbf{Bottom graphs}: Nominal and oracle coverage, respectively, with blue confidence bands at $ .95 \pm 1.96 \sqrt{\frac{.05\cdot .95}{2500}}$. The true mRD value is 0.019 for both simple and complex outcome models.}

\includegraphics[scale=0.65]{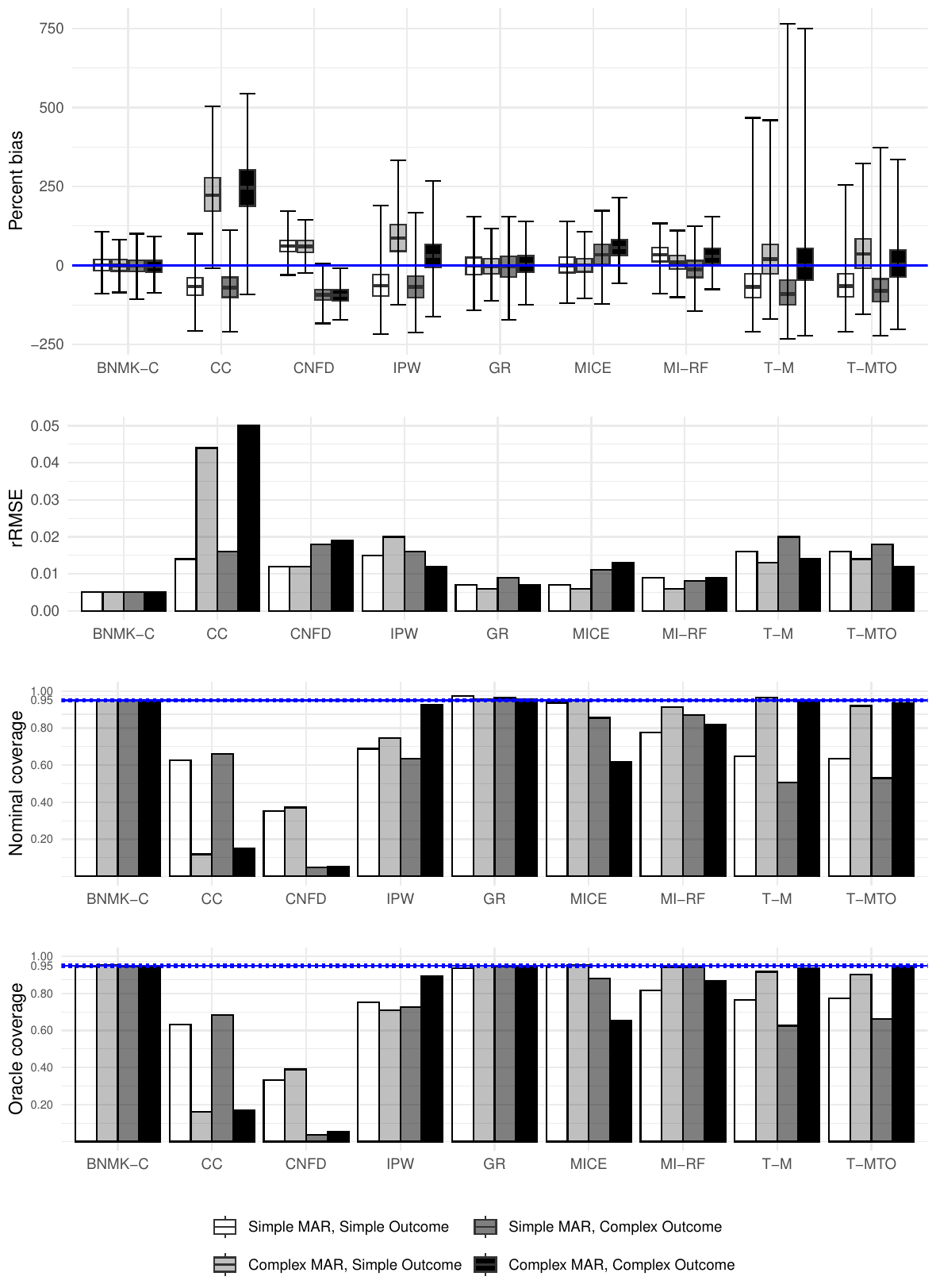}
\label{fig:mRD_mar_census_lowout_highmissing}
\end{figure}
\clearpage

\begin{figure}[!htb]
\caption[Missing at random, oracle truth, low outcome, high missing]{\textbf{Synthetic Data MAR Simulation: Oracle mRD}. Comparing estimators of the oracle estimand with \textbf{80\% confounder missingness} and \textbf{5\% outcome proportion}. \textbf{Top graph}: \%Bias (median, IQR, min and max of converged simulations); \textbf{Middle graph}: Robust RMSE (rRMSE), using median bias and MAD; \textbf{Bottom graphs}: Nominal and oracle coverage, respectively, with blue confidence bands at $ .95 \pm 1.96 \sqrt{\frac{.05\cdot .95}{2500}}$. True mRD values are 0.019 and 0.015 for simple and complex outcome models, respectively.}

\includegraphics[scale=0.65]{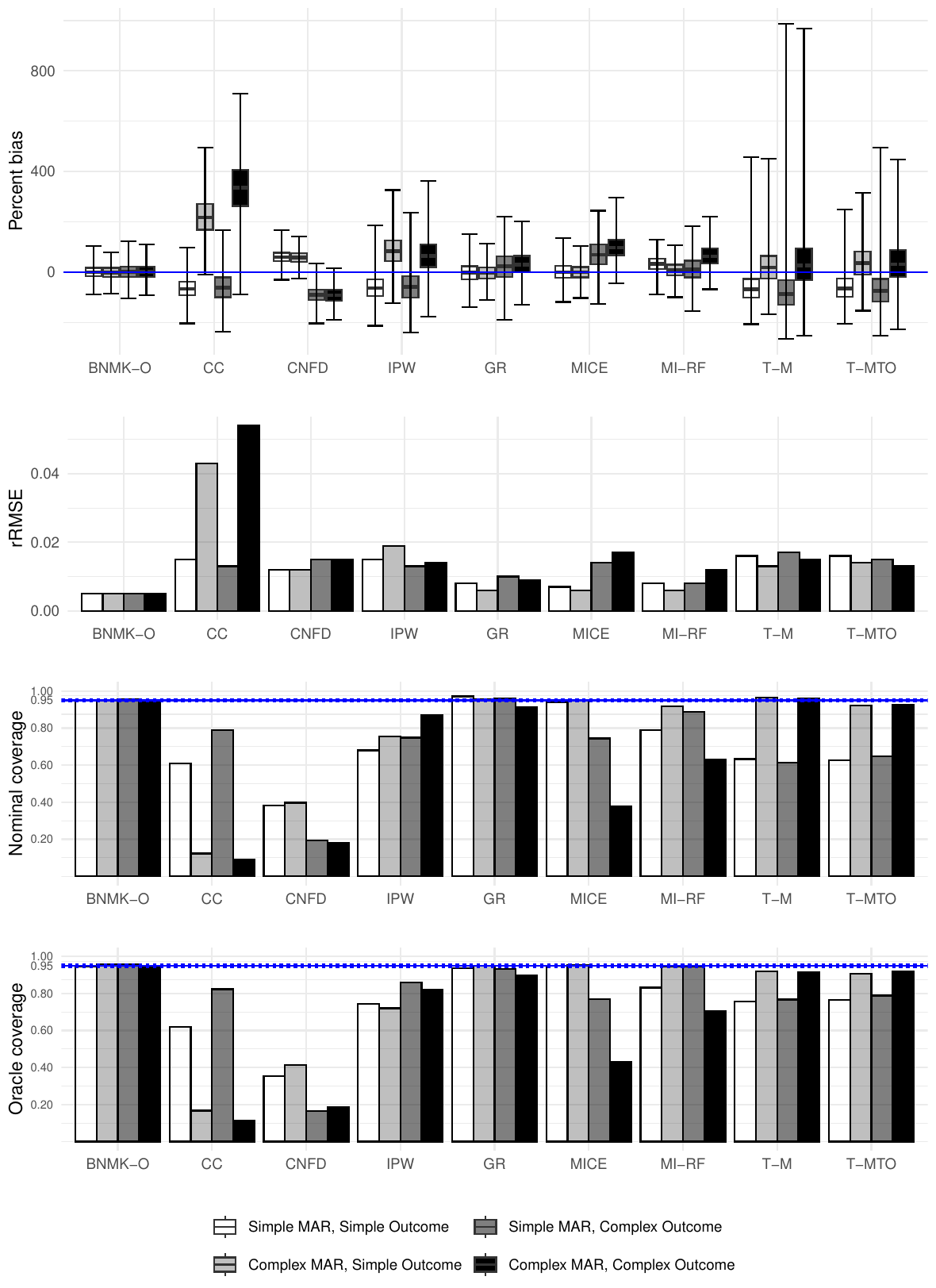}
\label{fig:mRD_mar_oracle_lowout_higmissing}
\end{figure}
\clearpage

\newpage

\subsubsection{5\% Outcome, 80\% MAR, mRR}
\newpage


\begin{figure}[!htb]
\caption[Missing at random, census truth, low outcome, high missing]{\textbf{Synthetic Data MAR Simulation: Census mlogRR}. Comparing estimators of the census estimand with \textbf{80\% confounder missingness} and \textbf{5\% outcome proportion}. \textbf{Top graph}: \%Bias (median, IQR, min and max of converged simulations); \textbf{Middle graph}: Robust RMSE (rRMSE), using median bias and MAD; \textbf{Bottom graphs}: Nominal and oracle coverage, respectively, with blue confidence bands at $ .95 \pm 1.96 \sqrt{\frac{.05\cdot .95}{2500}}$. True values of mlogRR are 0.370 and 0.349 for simple and complex outcome models, respectively.}

\includegraphics[scale=0.65]{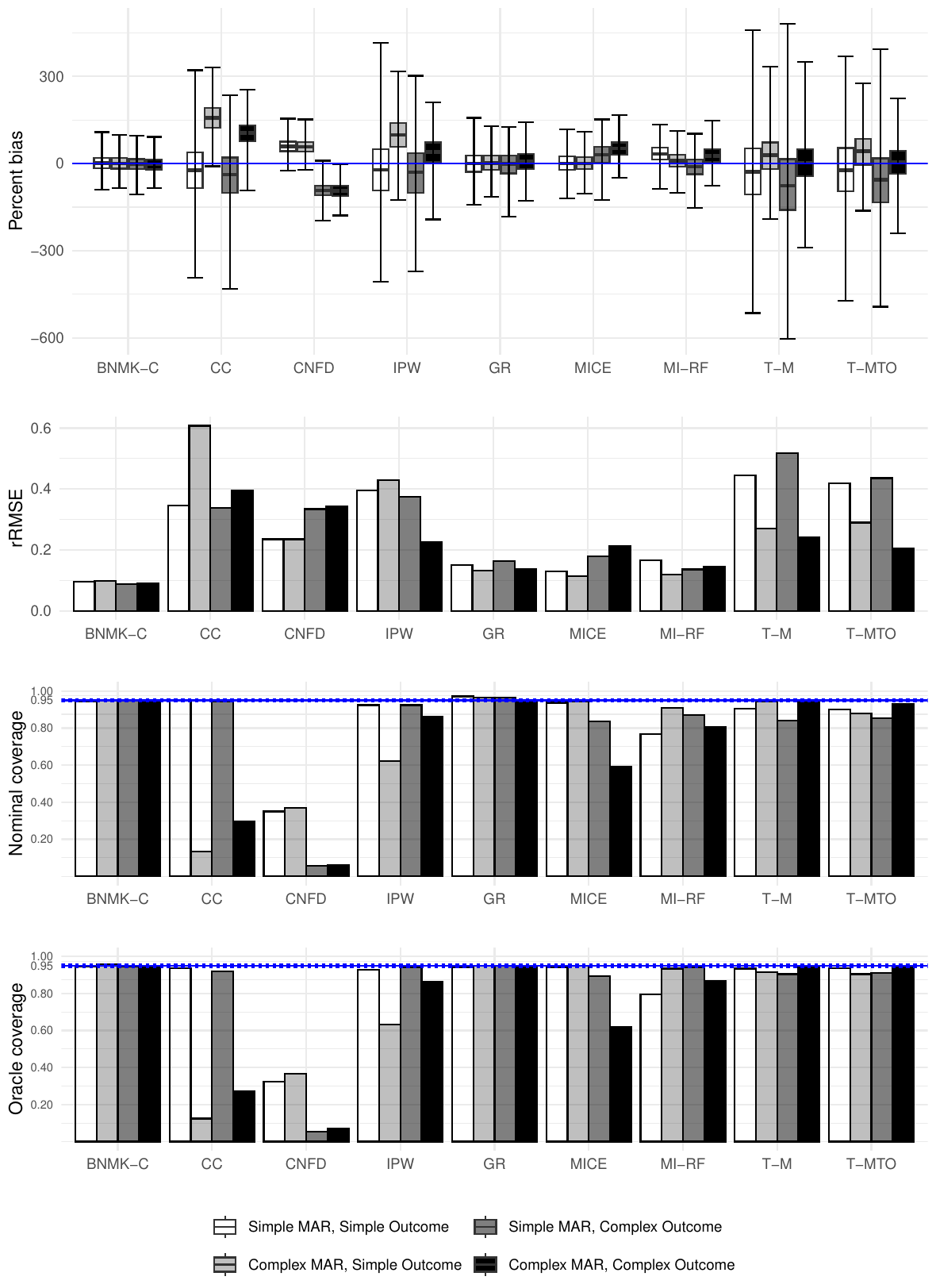}
\label{fig:mlogRR_mar_census_lowout_highmissing}
\end{figure}


\begin{figure}[!htb]
\caption[Missing at random, oracle truth, low outcome, high missing]{\textbf{Synthetic Data MAR Simulation: Oracle mlogRR}. Comparing estimators of the oracle estimand with \textbf{80\% confounder missingness} and \textbf{5\% outcome proportion}. \textbf{Top graph}: \%Bias (median, IQR, min and max of converged simulations); \textbf{Middle graph}: Robust RMSE (rRMSE), using median bias and MAD; \textbf{Bottom graphs}: Nominal and oracle coverage, respectively, with blue confidence bands at $ .95 \pm 1.96 \sqrt{\frac{.05\cdot .95}{2500}}$. True values of mlogRR are 0.370 and 0.281 for simple and complex outcome models, respectively.}

\includegraphics[scale=0.65]{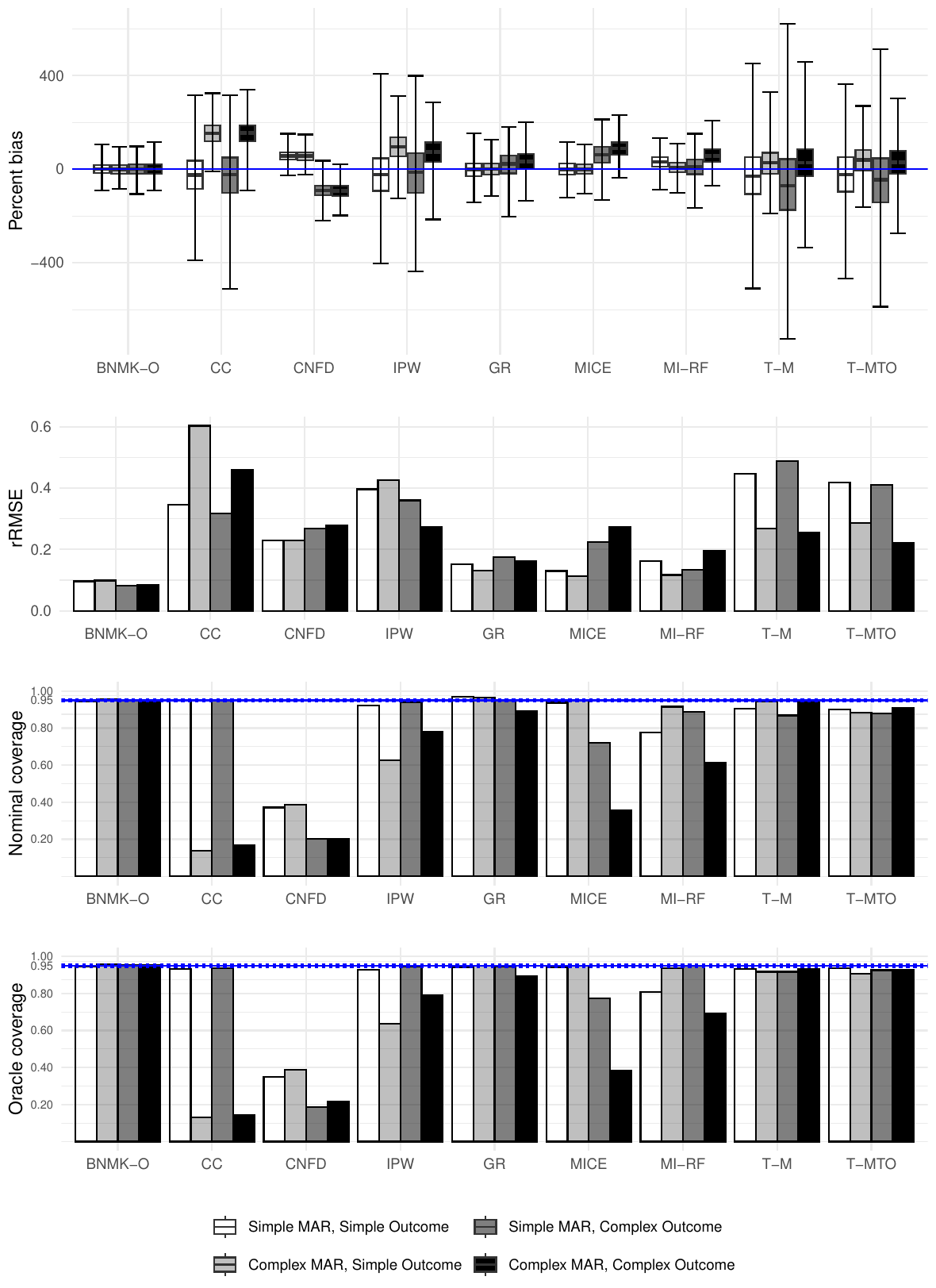}
\label{fig:mlogRR_mar_oracle_lowout_highmissing}
\end{figure}
\clearpage
\newpage
\subsubsection{12\% Outcome, 40\% MNAR, mRD}

\newpage

\begin{figure}[!htb]
\caption[Missing not at random, census truth, high outcome, low missing]{\textbf{Synthetic Data MNAR Simulation: Census mRD}. Comparing estimators of the census estimand with \textbf{40\% confounder missingness} and \textbf{12\% outcome proportion}. \textbf{Top graph}: \%Bias (median, IQR, min and max of converged simulations); \textbf{Middle graph}: Robust RMSE (rRMSE), using median bias and MAD; \textbf{Bottom graphs}: Nominal and oracle coverage, respectively, with blue confidence bands at $ .95 \pm 1.96 \sqrt{\frac{.05\cdot .95}{2500}}$. The true mRD values are 0.040 for simple outcome models, and 0.038 and 0.037 for complex outcome models for MNAR unobserved and MNAR value, respectively.}

\includegraphics[scale=0.65]{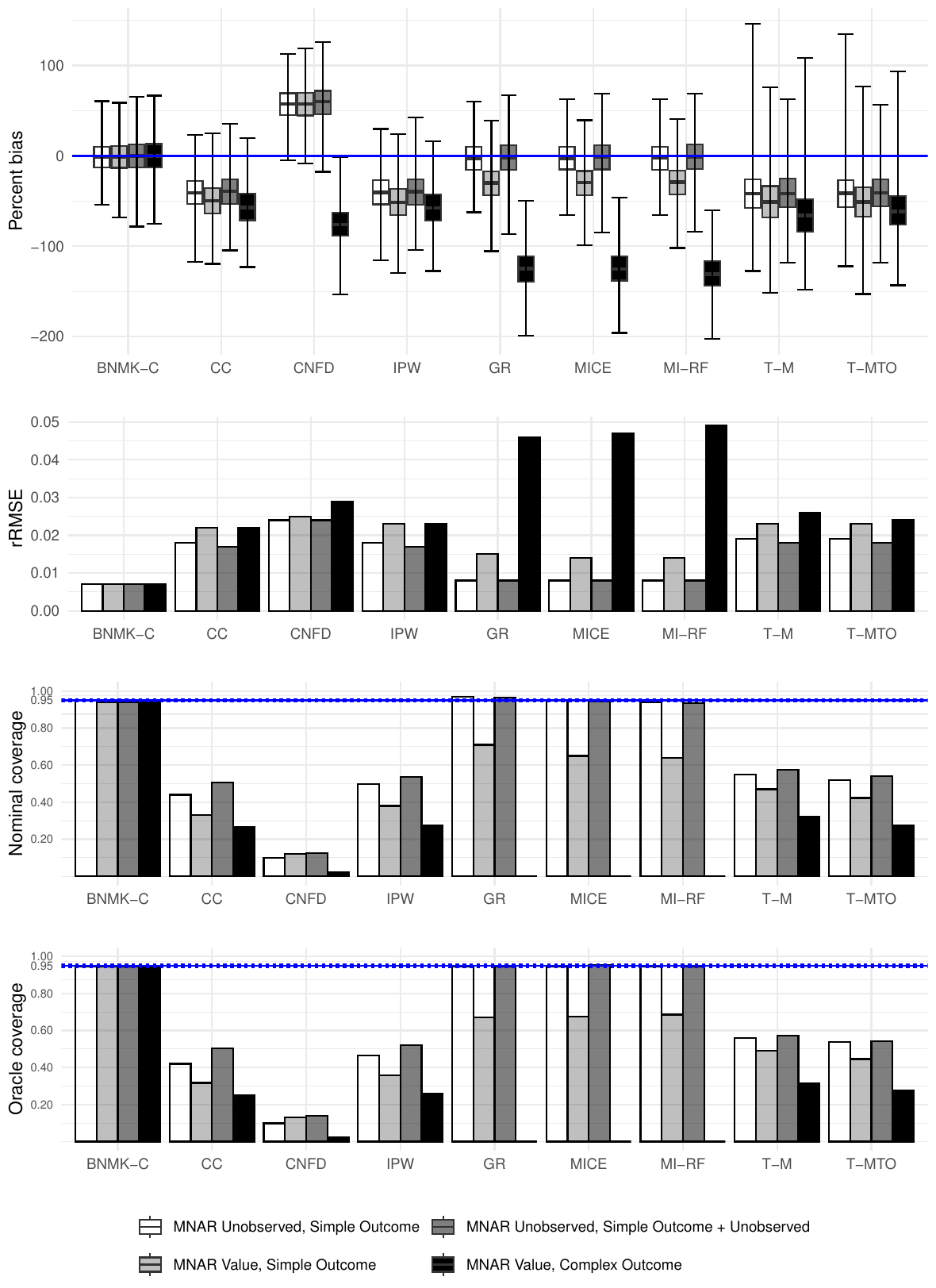}
\label{fig:mRD_mnar_census_highout_lowmissing}
\end{figure}


\begin{figure}[!htb]
\caption[Missing not at random, oracle truth, high outcome, low missing]{\textbf{Synthetic Data MNAR Simulation: Oracle mRD}. Comparing estimators of the oracle truth with \textbf{40\% confounder missingness} and \textbf{12\% outcome proportion}. \textbf{Top graph}: \%Bias (median, IQR, min and max of converged simulations); \textbf{Middle graph}: Robust RMSE (rRMSE), using median bias and MAD; \textbf{Bottom graphs}: Nominal and oracle coverage, respectively, with blue confidence bands at $ .95 \pm 1.96 \sqrt{\frac{.05\cdot .95}{2500}}$. The true mRD values are 0.040 for simple outcome models, and 0.038 and 0.031 for complex outcome models for MNAR unobserved and MNAR value, respectively.}

\includegraphics[scale=0.65]{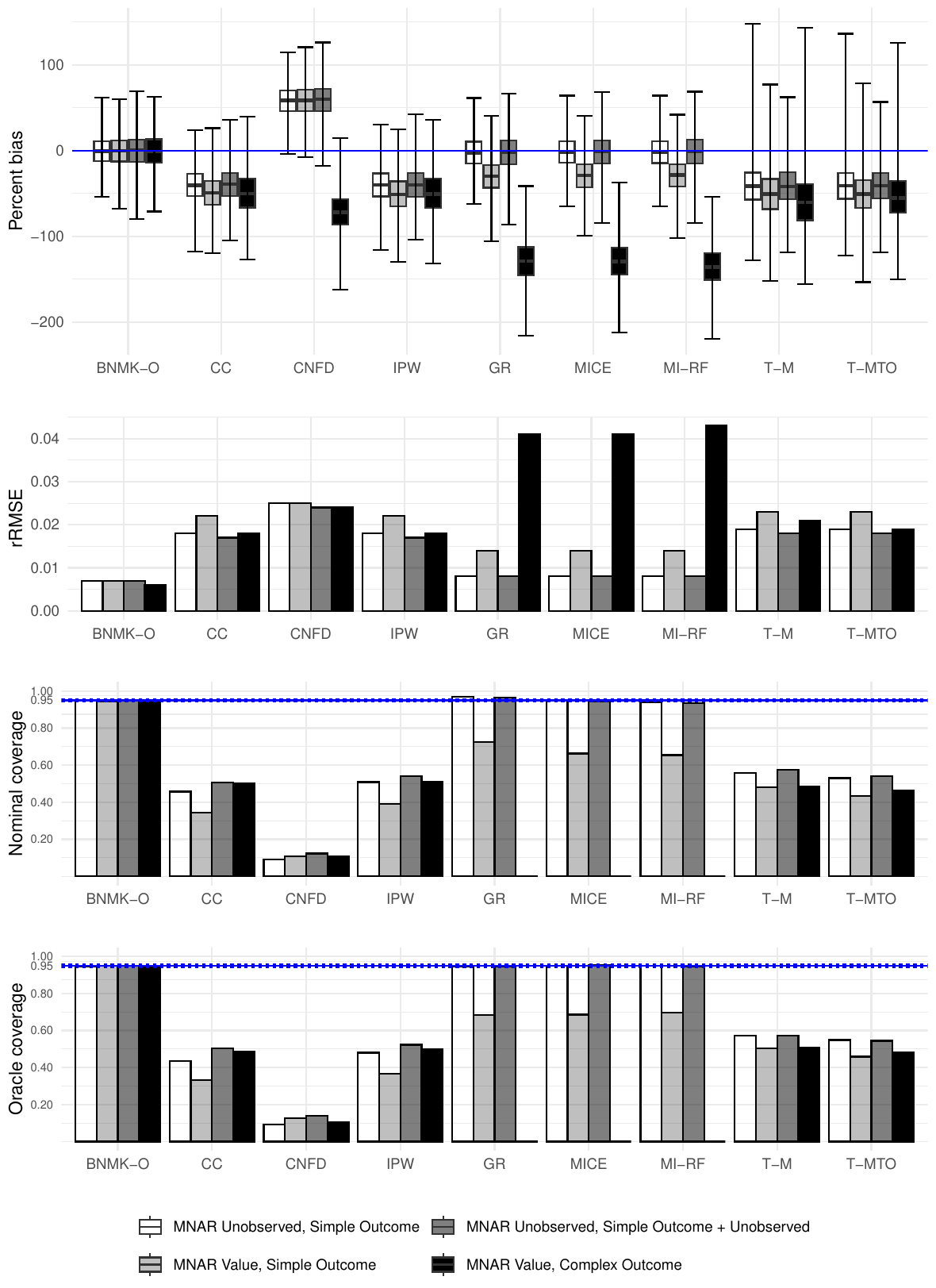}
\label{fig:mRD_mnar_oracle_highout_lowmissing}
\end{figure}
\clearpage
\newpage

\subsubsection{ 12\% Outcome, 40\% MNAR, mRR}
\newpage


\begin{figure}[!htb]
\caption[Missing not at random, census truth, high outcome, low missing]{\textbf{Synthetic Data MNAR Simulation: Census mlogRR}. Comparing estimators of the census estimand with \textbf{40\% confounder missingness} and \textbf{12\% outcome proportion}. \textbf{Top graph}: \%Bias (median, IQR, min and max of converged simulations); \textbf{Middle graph}: Robust RMSE (rRMSE), using median bias and MAD; \textbf{Bottom graphs}: Nominal and oracle coverage, respectively, with blue confidence bands at $ .95 \pm 1.96 \sqrt{\frac{.05\cdot .95}{2500}}$. True values of mlogRR are 0.334 for simple outcome models and 0.339 and 0.315 for complex outcome models for MNAR unobserved and MNAR value, respectively.}

\includegraphics[scale=0.65]{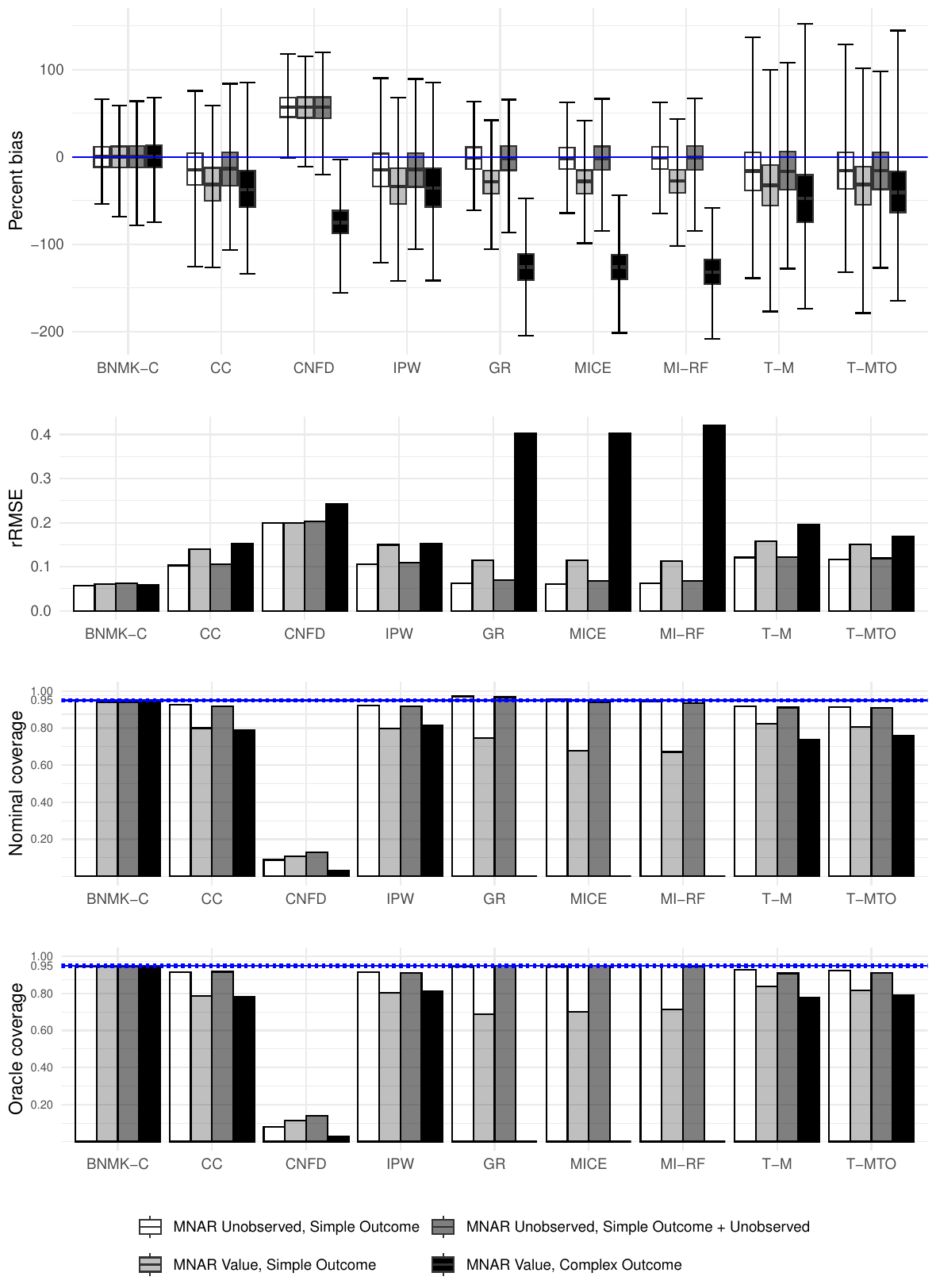}
\label{fig:mlogRR_mnar_census_highout_lowmissing}
\end{figure}


\begin{figure}[!htb]
\caption[Missing not at random, oracle truth, high outcome, low missing]{\textbf{Synthetic Data MNAR Simulation: Oracle mlogRR}. Comparing estimators of the oracle estimand with \textbf{40\% confounder missingness} and \textbf{12\% outcome proportion}. \textbf{Top graph}: \%Bias (median, IQR, min and max of converged simulations); \textbf{Middle graph}: Robust RMSE (rRMSE), using median bias and MAD; \textbf{Bottom graphs}: Nominal and oracle coverage, respectively, with blue confidence bands at $ .95 \pm 1.96 \sqrt{\frac{.05\cdot .95}{2500}}$. True values of mlogRR are 0.334 for simple outcome models and 0.339 and 0.271 for complex outcome models for MNAR unobserved and MNAR value, respectively.}

\includegraphics[scale=0.65]{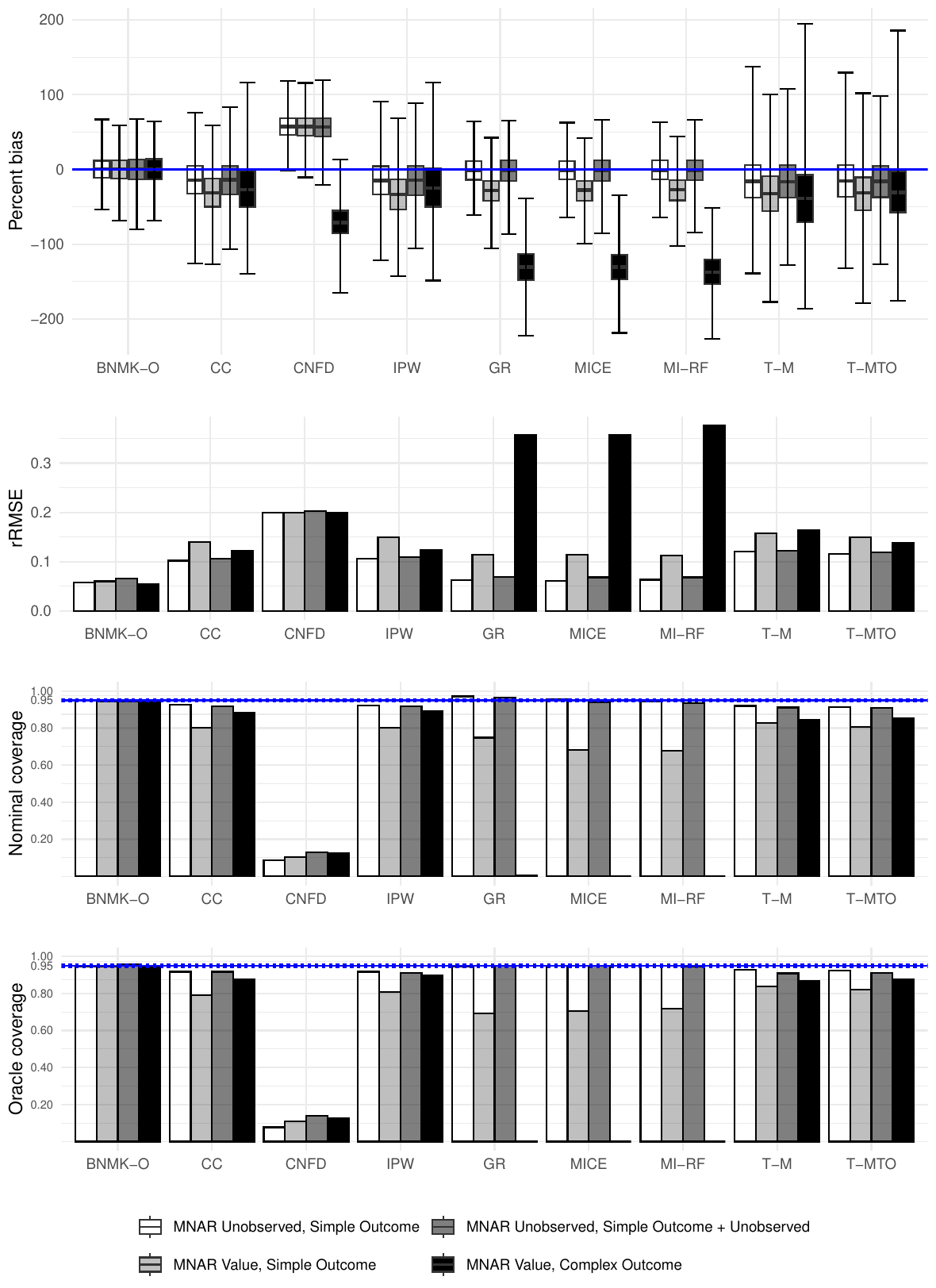}
\label{fig:mlogRR_mnar_oracle_highout_lowmissing}
\end{figure}

\clearpage
\newpage
\subsubsection{ 12\% Outcome, 80\% MNAR, cOR}

\newpage

\begin{figure}[!htb]
\caption[Missing not at random, census truth, high outcome, high missing]{\textbf{Synthetic Data MNAR Simulation: Census cOR}. Comparing estimators of the census estimand with \textbf{80\% confounder missingness} and \textbf{12\% outcome proportion}. \textbf{Top graph}: \%Bias (median, IQR, min and max of converged simulations); \textbf{Middle graph}: Robust RMSE (rRMSE), using median bias and MAD; \textbf{Bottom graphs}: Nominal and oracle coverage, respectively, with blue confidence bands at $ .95 \pm 1.96 \sqrt{\frac{.05\cdot .95}{2500}}$. True clogOR values are 0.405 for the simple outcome models and 0.404 and 0.371 for complex outcome models for MNAR unobserved and MNAR value, respectively.}

\includegraphics[scale=0.65]{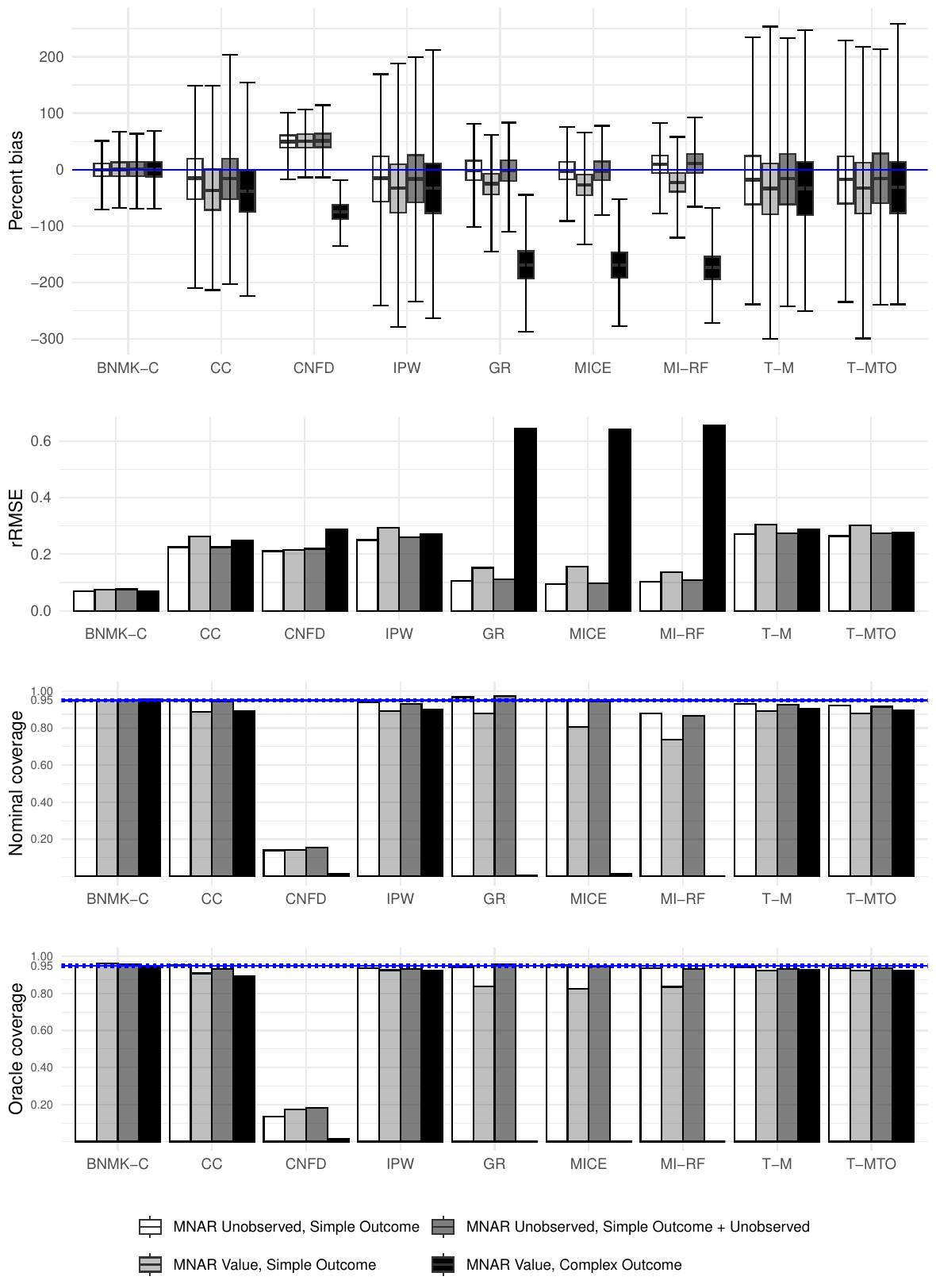}
\label{fig:cOR_mnar_census_highout_highmissing}
\end{figure}

\clearpage

\begin{figure}[!htb]
\caption[Missing not at random, oracle truth, high outcome, high missing]{\textbf{Synthetic Data MNAR Simulation: Oracle cOR}. Comparing estimators of the oracle truth with \textbf{80\% confounder missingness} and \textbf{12\% outcome proportion}. \textbf{Top graph}: \%Bias (median, IQR, min and max of converged simulations); \textbf{Middle graph}: Robust RMSE (rRMSE), using median bias and MAD; \textbf{Bottom graphs}: Nominal and oracle coverage, respectively, with blue confidence bands at $ .95 \pm 1.96 \sqrt{\frac{.05\cdot .95}{2500}}$. The true cOR value is 0.405 in all scenarios.}

\includegraphics[scale=0.65]{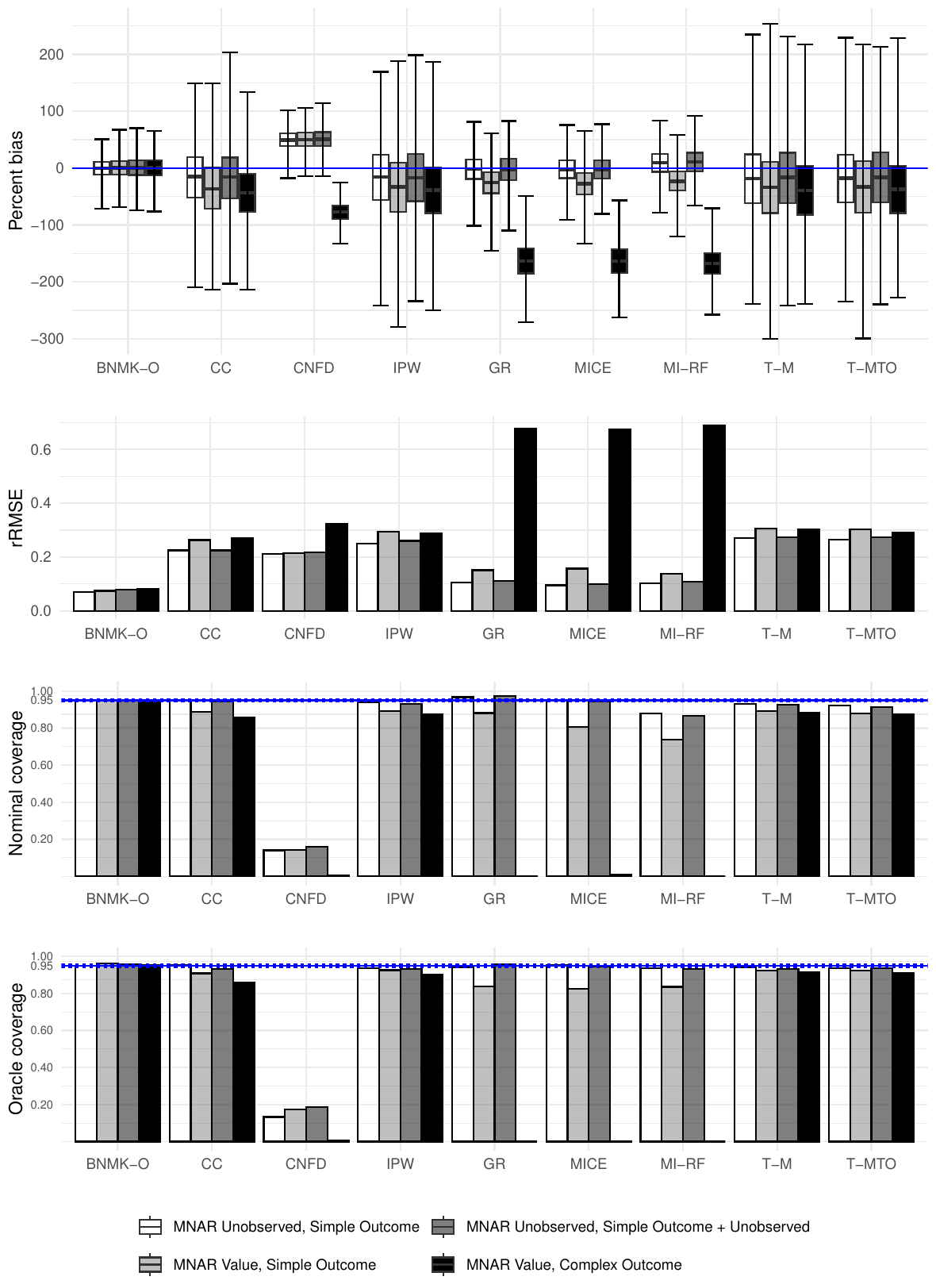}
\label{fig:cOR_mnar_oracle_highout_highmissing}
\end{figure}
\clearpage
\newpage

\subsubsection{ 12\% Outcome, 80\% MNAR, mRD}

\newpage

\begin{figure}[!htb]
\caption[Missing not at random, census truth, high outcome, high missing]{\textbf{Synthetic Data MNAR Simulation: Census mRD}. Comparing estimators of the census estimand with \textbf{80\% confounder missingness} and \textbf{12\% outcome proportion}. \textbf{Top graph}: \%Bias (median, IQR, min and max of converged simulations); \textbf{Middle graph}: Robust RMSE (rRMSE), using median bias and MAD; \textbf{Bottom graphs}: Nominal and oracle coverage, respectively, with blue confidence bands at $ .95 \pm 1.96 \sqrt{\frac{.05\cdot .95}{2500}}$. The true mRD values are 0.040 for simple outcome models, and 0.038 and 0.037 for complex outcome models for MNAR unobserved and MNAR value, respectively.}

\includegraphics[scale=0.65]{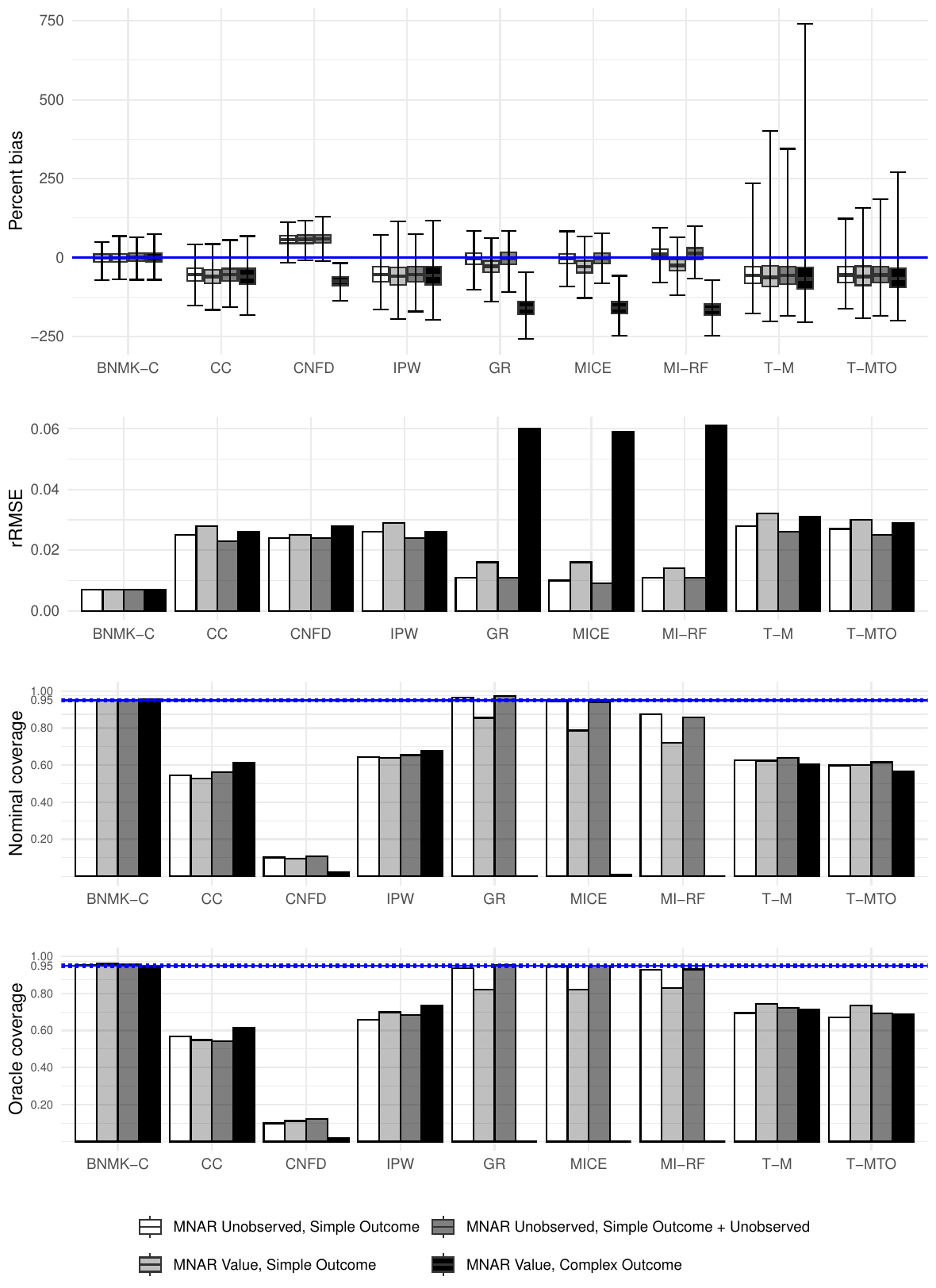}
\label{fig:mRD_mnar_census_highout_highmissing}
\end{figure}


\begin{figure}[!htb]
\caption[Missing not at random, oracle truth, high outcome, high missing]{\textbf{Synthetic Data MNAR Simulation: Oracle mRD}. Comparing estimators of the oracle truth with \textbf{80\% confounder missingness} and \textbf{12\% outcome proportion}. \textbf{Top graph}: \%Bias (median, IQR, min and max of converged simulations); \textbf{Middle graph}: Robust RMSE (rRMSE), using median bias and MAD; \textbf{Bottom graphs}: Nominal and oracle coverage, respectively, with blue confidence bands at $ .95 \pm 1.96 \sqrt{\frac{.05\cdot .95}{2500}}$. The true mRD values are 0.040 for simple outcome models, and 0.038 and 0.031 for complex outcome models for MNAR unobserved and MNAR value, respectively.}

\includegraphics[scale=0.65]{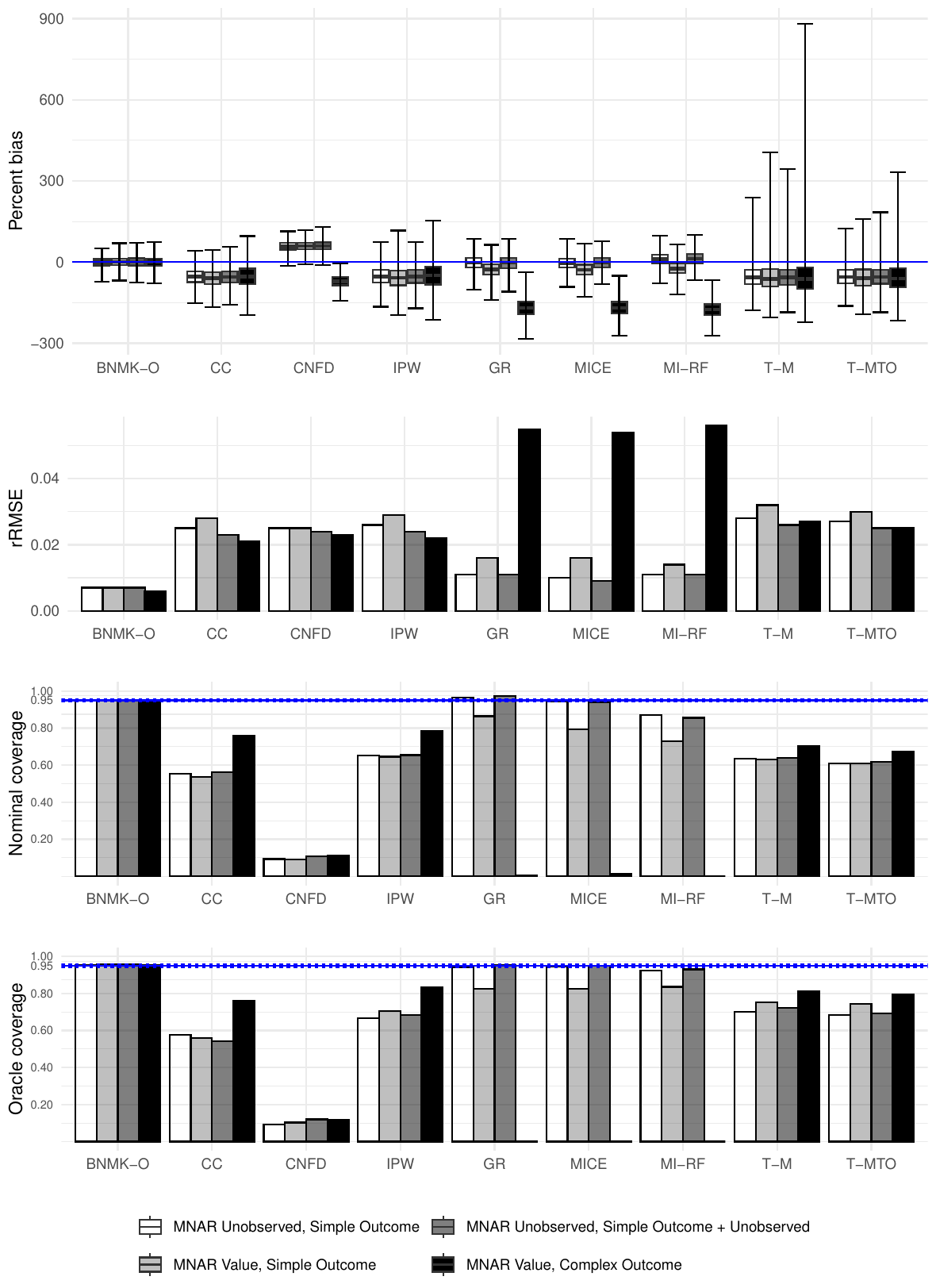}
\label{fig:mRD_mnar_oracle_highout_highmissing}
\end{figure}
\clearpage
\newpage

\subsubsection{ 12\% Outcome, 80\% MNAR, mRR}
\newpage


\begin{figure}[!htb]
\caption[Missing not at random, census truth, high outcome, high missing]{\textbf{Synthetic Data MNAR Simulation: Census mlogRR}. Comparing estimators of the census estimand with \textbf{80\% confounder missingness} and \textbf{12\% outcome proportion}. \textbf{Top graph}: \%Bias (median, IQR, min and max of converged simulations); \textbf{Middle graph}: Robust RMSE (rRMSE), using median bias and MAD; \textbf{Bottom graphs}: Nominal and oracle coverage, respectively, with blue confidence bands at $ .95 \pm 1.96 \sqrt{\frac{.05\cdot .95}{2500}}$. True values of mlogRR are 0.334 for simple outcome models and 0.339 and 0.315 for complex outcome models for MNAR unobserved and MNAR value, respectively.}

\includegraphics[scale=0.65]{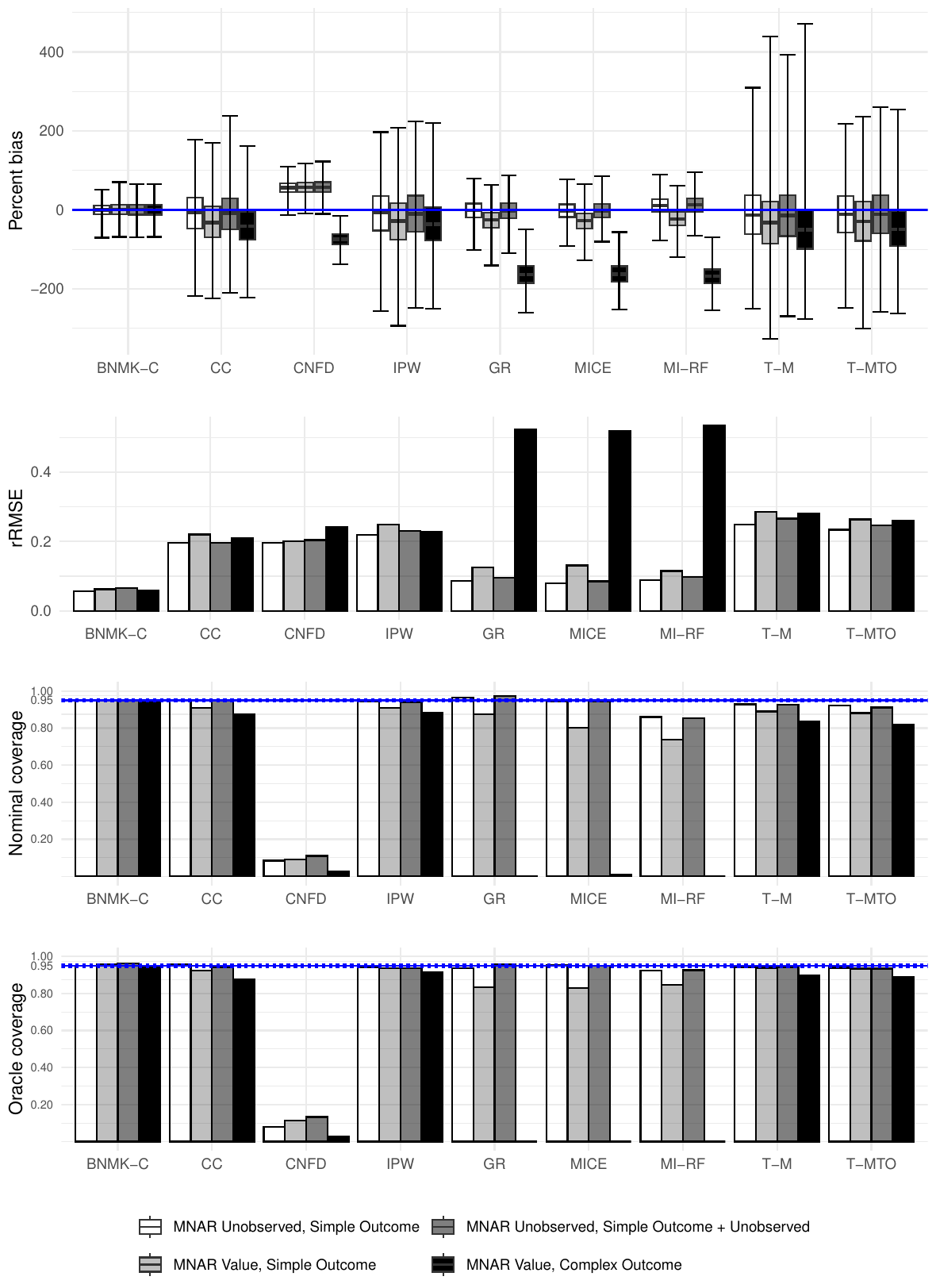}
\label{fig:mlogRR_mnar_census_highout_highmissing}
\end{figure}
\clearpage

\begin{figure}[!htb]
\caption[Missing not at random, oracle truth, high outcome, high missing]{\textbf{Synthetic Data MNAR Simulation: Oracle mlogRR}. Comparing estimators of the oracle estimand with \textbf{80\% confounder missingness} and \textbf{12\% outcome proportion}. \textbf{Top graph}: \%Bias (median, IQR, min and max of converged simulations); \textbf{Middle graph}: Robust RMSE (rRMSE), using median bias and MAD; \textbf{Bottom graphs}: Nominal and oracle coverage, respectively, with blue confidence bands at $ .95 \pm 1.96 \sqrt{\frac{.05\cdot .95}{2500}}$. True values of mlogRR are 0.334 for simple outcome models and 0.339 and 0.271 for complex outcome models for MNAR unobserved and MNAR value, respectively.}

\includegraphics[scale=0.65]{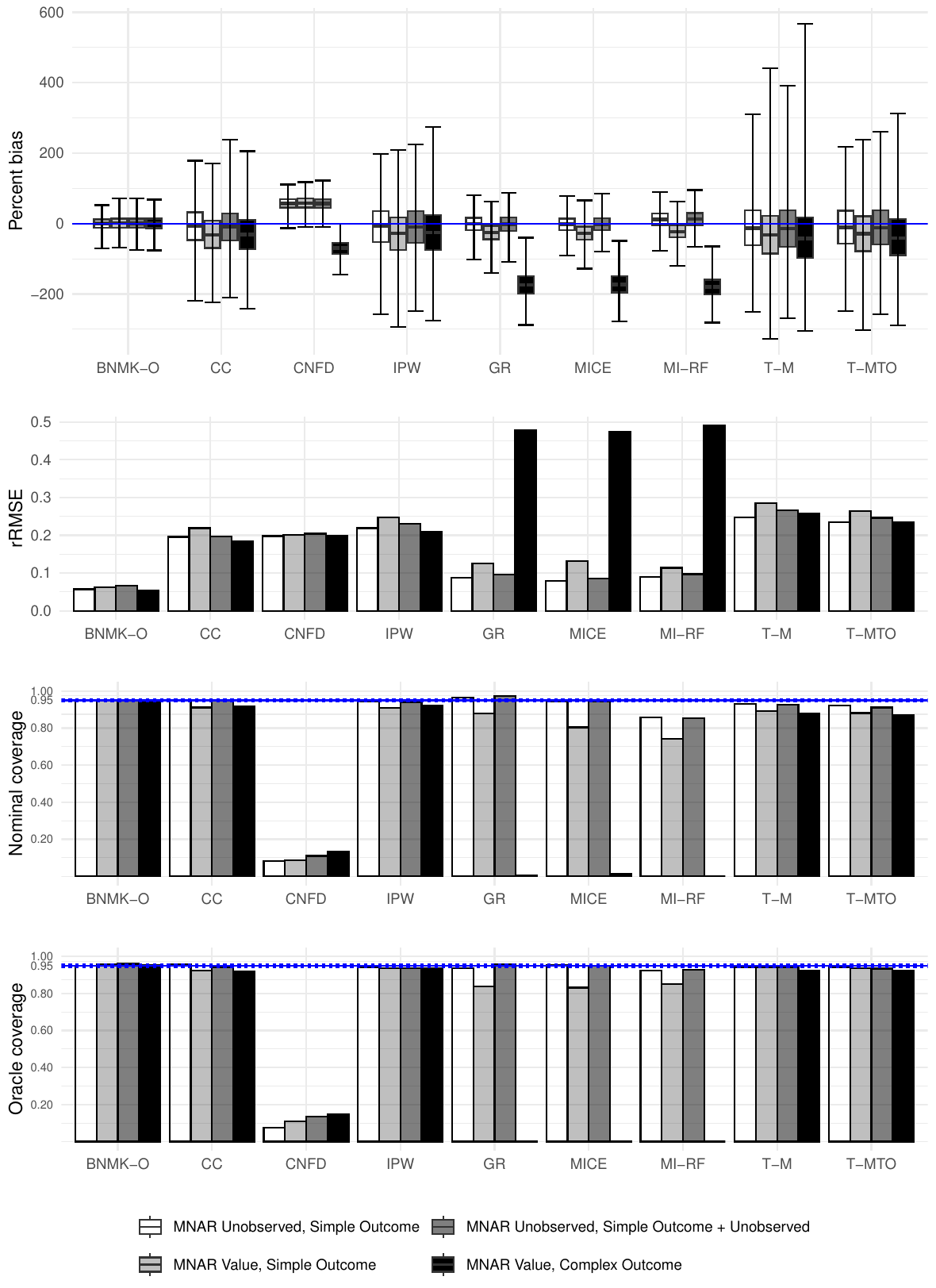}
\label{fig:mlogRR_mnar_oracle_highout_highmissing}
\end{figure}

\clearpage

\newpage
\subsubsection{ 5\% Outcome, 80\% MNAR, cOR}

\newpage

\begin{figure}[!htb]
\caption[Missing not at random, census truth, low outcome, high missing]{\textbf{Synthetic Data MNAR Simulation: Census cOR}. Comparing estimators of the census estimand with \textbf{80\% confounder missingness} and \textbf{5\% outcome proportion}. \textbf{Top graph}: \%Bias (median, IQR, min and max of converged simulations); \textbf{Middle graph}: Robust RMSE (rRMSE), using median bias and MAD; \textbf{Bottom graphs}: Nominal and oracle coverage, respectively, with blue confidence bands at $ .95 \pm 1.96 \sqrt{\frac{.05\cdot .95}{2500}}$. True clogOR values are 0.405 for the simple outcome models and 0.402 and 0.38 for complex outcome models for MNAR unobserved and MNAR value, respectively.}

\includegraphics[scale=0.65]{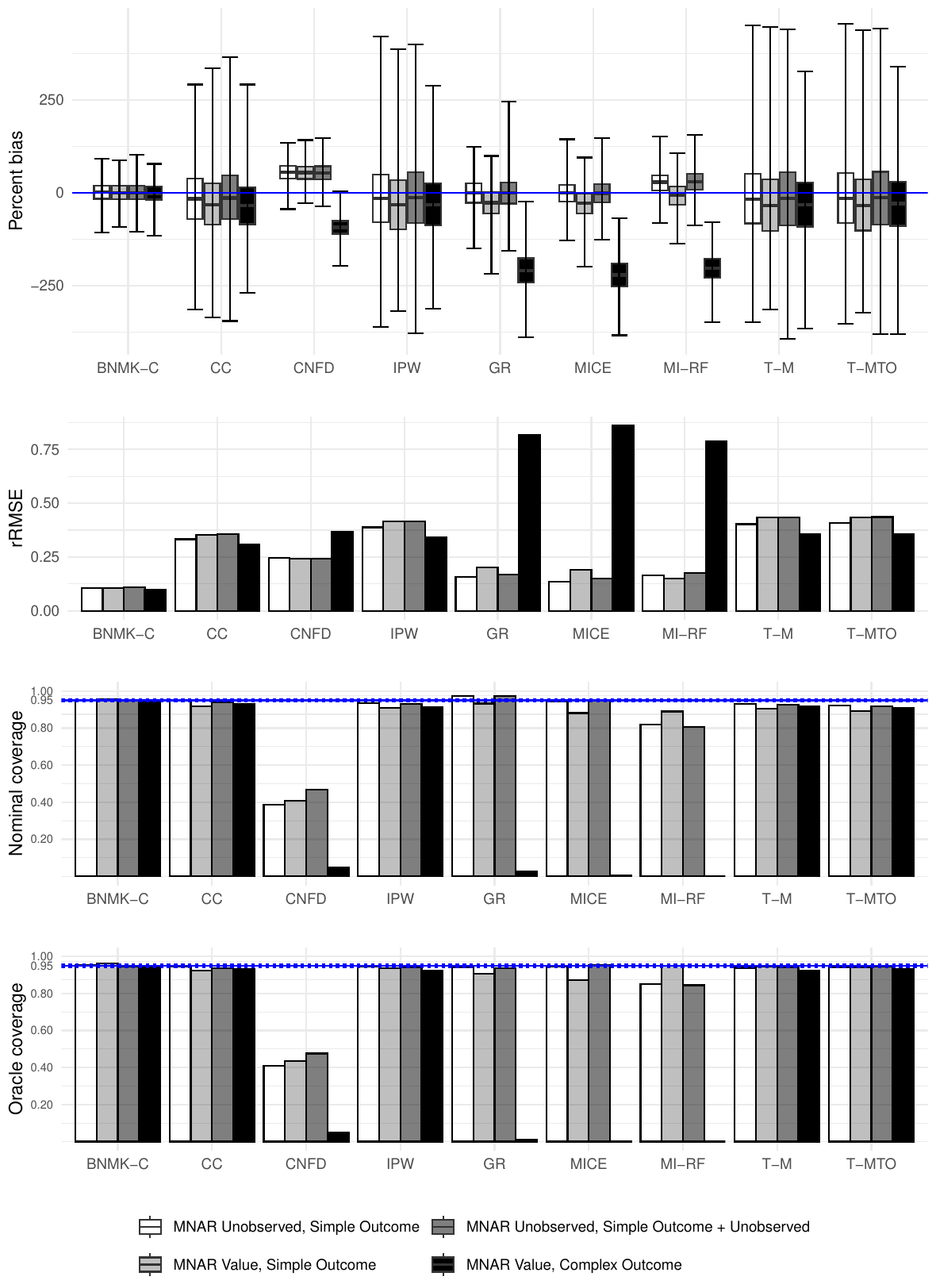}
\label{fig:cOR_mnar_census_lowout_highmissing}
\end{figure}


\begin{figure}[!htb]
\caption[Missing not at random, oracle truth, low outcome, high missing]{\textbf{Synthetic Data MNAR Simulation: Oracle cOR}. Comparing estimators of the oracle truth with \textbf{80\% confounder missingness} and \textbf{5\% outcome proportion}. \textbf{Top graph}: \%Bias (median, IQR, min and max of converged simulations); \textbf{Middle graph}: Robust RMSE (rRMSE), using median bias and MAD; \textbf{Bottom graphs}: Nominal and oracle coverage, respectively, with blue confidence bands at $ .95 \pm 1.96 \sqrt{\frac{.05\cdot .95}{2500}}$. The true cOR value is 0.405 in all scenarios.}

\includegraphics[scale=0.65]{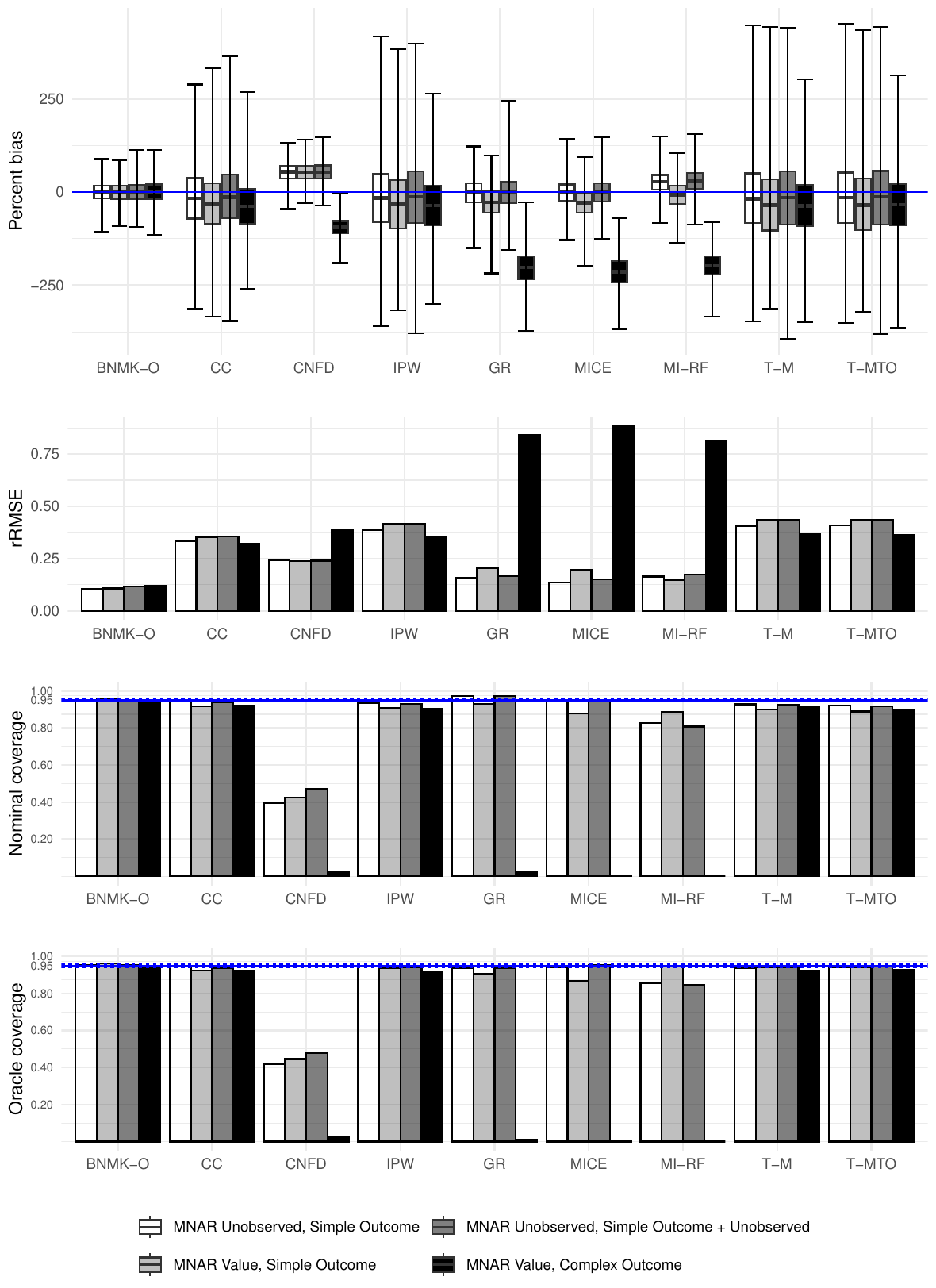}
\label{fig:cOR_mnar_oracle_lowout_highmissing}
\end{figure}
\clearpage
\newpage

\subsubsection{ 5\% Outcome, 80\% MNAR, mRD}

\newpage

\begin{figure}[!htb]
\caption[Missing not at random, census truth, low outcome, high missing]{\textbf{Synthetic Data MNAR Simulation: Census mRD}. Comparing estimators of the census estimand with \textbf{80\% confounder missingness} and \textbf{5\% outcome proportion}. \textbf{Top graph}: \%Bias (median, IQR, min and max of converged simulations); \textbf{Middle graph}: Robust RMSE (rRMSE), using median bias and MAD; \textbf{Bottom graphs}: Nominal and oracle coverage, respectively, with blue confidence bands at $ .95 \pm 1.96 \sqrt{\frac{.05\cdot .95}{2500}}$. The true mRD values are 0.019 for simple outcome models, and 0.016 and 0.019 for complex outcome models for MNAR unobserved and MNAR value, respectively.}

\includegraphics[scale=0.65]{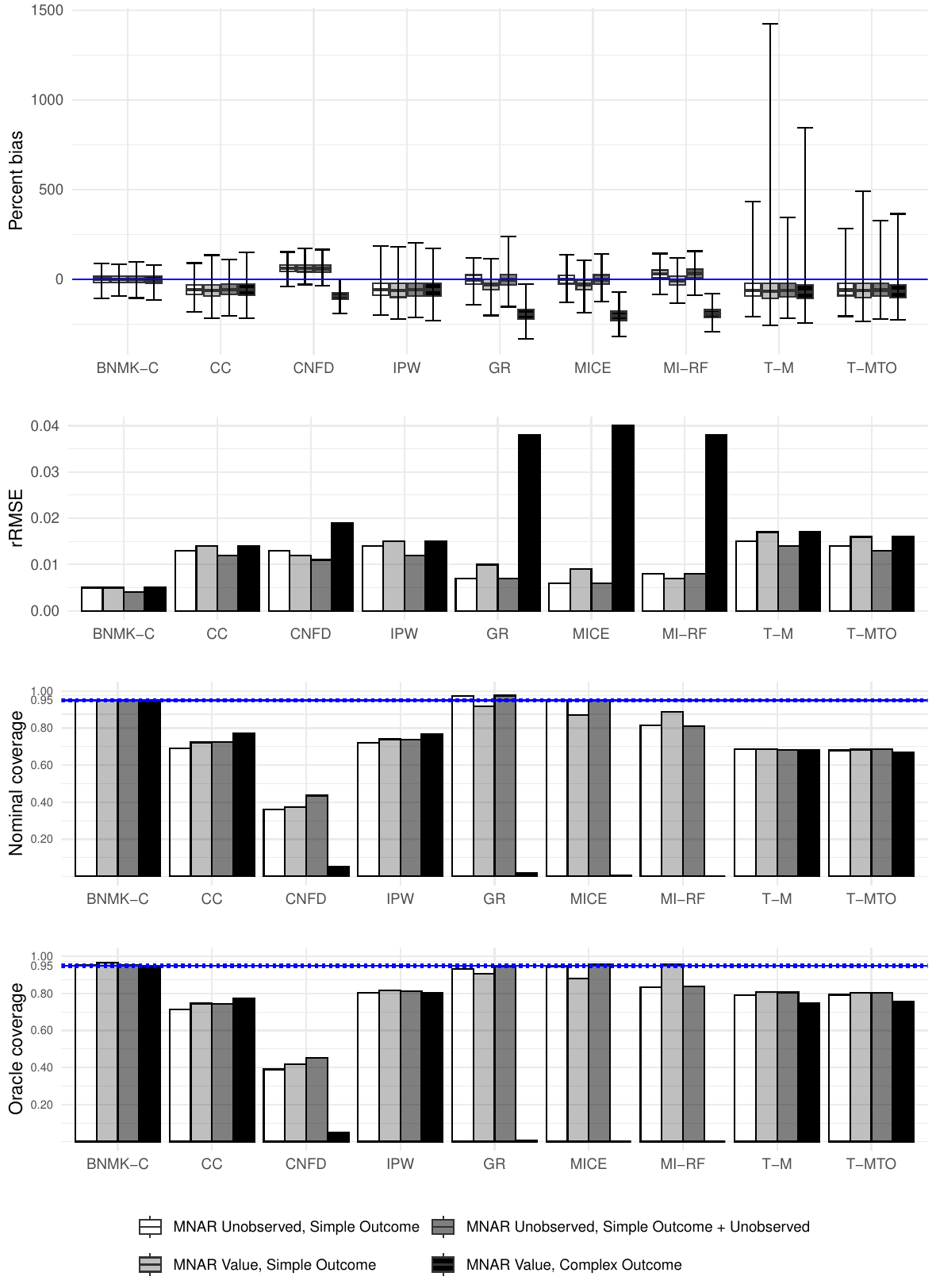}
\label{fig:mRD_mnar_census_lowout_highmissing}
\end{figure}


\begin{figure}[!htb]
\caption[Missing not at random, oracle truth, low outcome, high missing]{\textbf{Synthetic Data MNAR Simulation: Oracle mRD}. Comparing estimators of the oracle truth with \textbf{80\% confounder missingness} and \textbf{5\% outcome proportion}. \textbf{Top graph}: \%Bias (median, IQR, min and max of converged simulations); \textbf{Middle graph}: Robust RMSE (rRMSE), using median bias and MAD; \textbf{Bottom graphs}: Nominal and oracle coverage, respectively, with blue confidence bands at $ .95 \pm 1.96 \sqrt{\frac{.05\cdot .95}{2500}}$. The true mRD values are 0.019 for simple outcome models, and 0.016 and 0.015 for complex outcome models for MNAR unobserved and MNAR value, respectively.}

\includegraphics[scale=0.65]{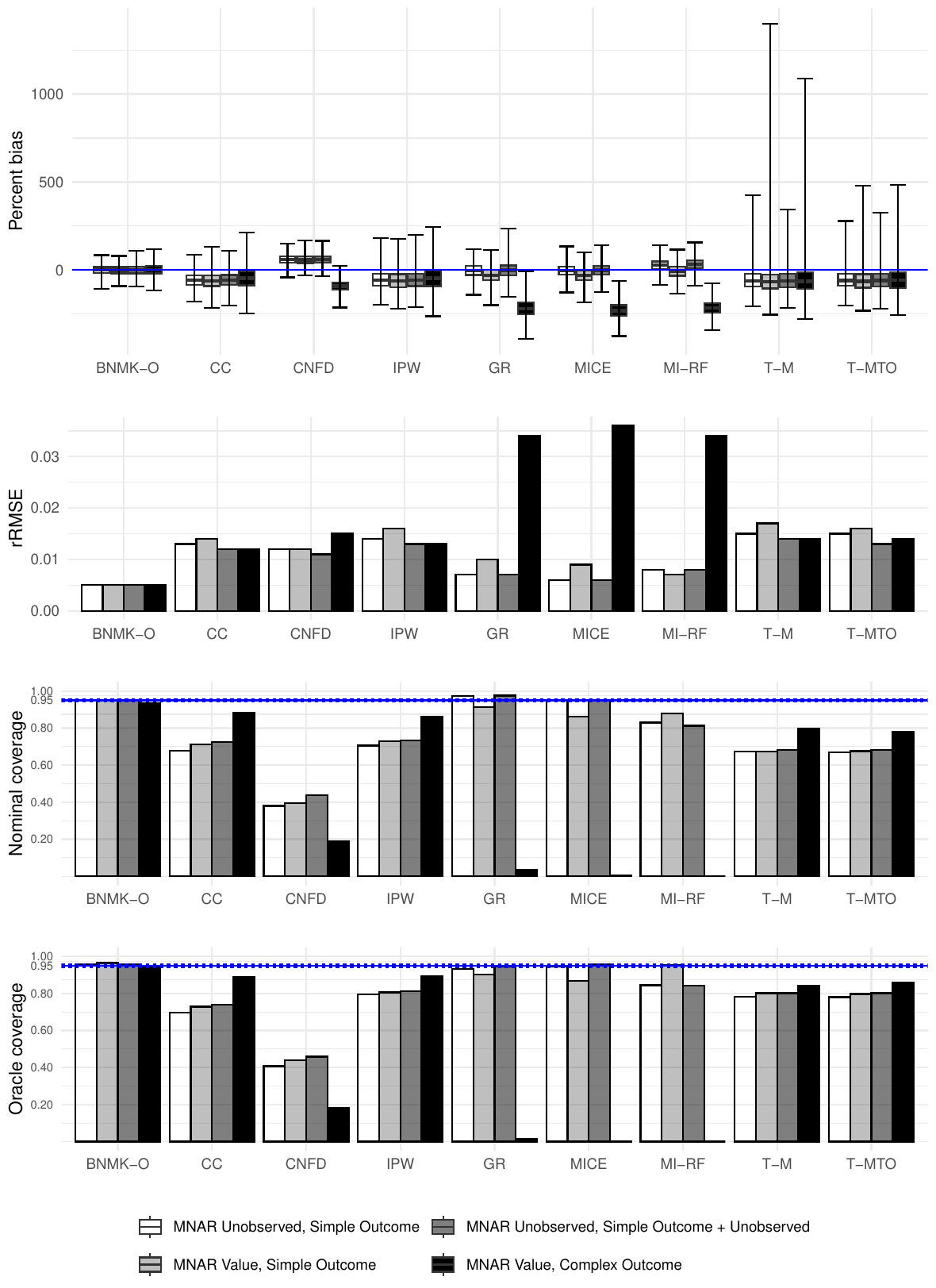}
\label{fig:mRD_mnar_oracle_lowout_highmissing}
\end{figure}
\clearpage
\newpage

\subsubsection{ 5\% Outcome, 80\% MNAR, mRR}
\newpage


\begin{figure}[!htb]
\caption[Missing not at random, census truth, low outcome, high missing]{\textbf{Synthetic Data MNAR Simulation: Census mlogRR}. Comparing estimators of the census estimand with \textbf{80\% confounder missingness} and \textbf{5\% outcome proportion}. \textbf{Top graph}: \%Bias (median, IQR, min and max of converged simulations); \textbf{Middle graph}: Robust RMSE (rRMSE), using median bias and MAD; \textbf{Bottom graphs}: Nominal and oracle coverage, respectively, with blue confidence bands at $ .95 \pm 1.96 \sqrt{\frac{.05\cdot .95}{2500}}$. True values of mlogRR are 0.369 for simple outcome models and 0.374 and 0.281 for complex outcome models for MNAR unobserved and MNAR value, respectively.}

\includegraphics[scale=0.65]{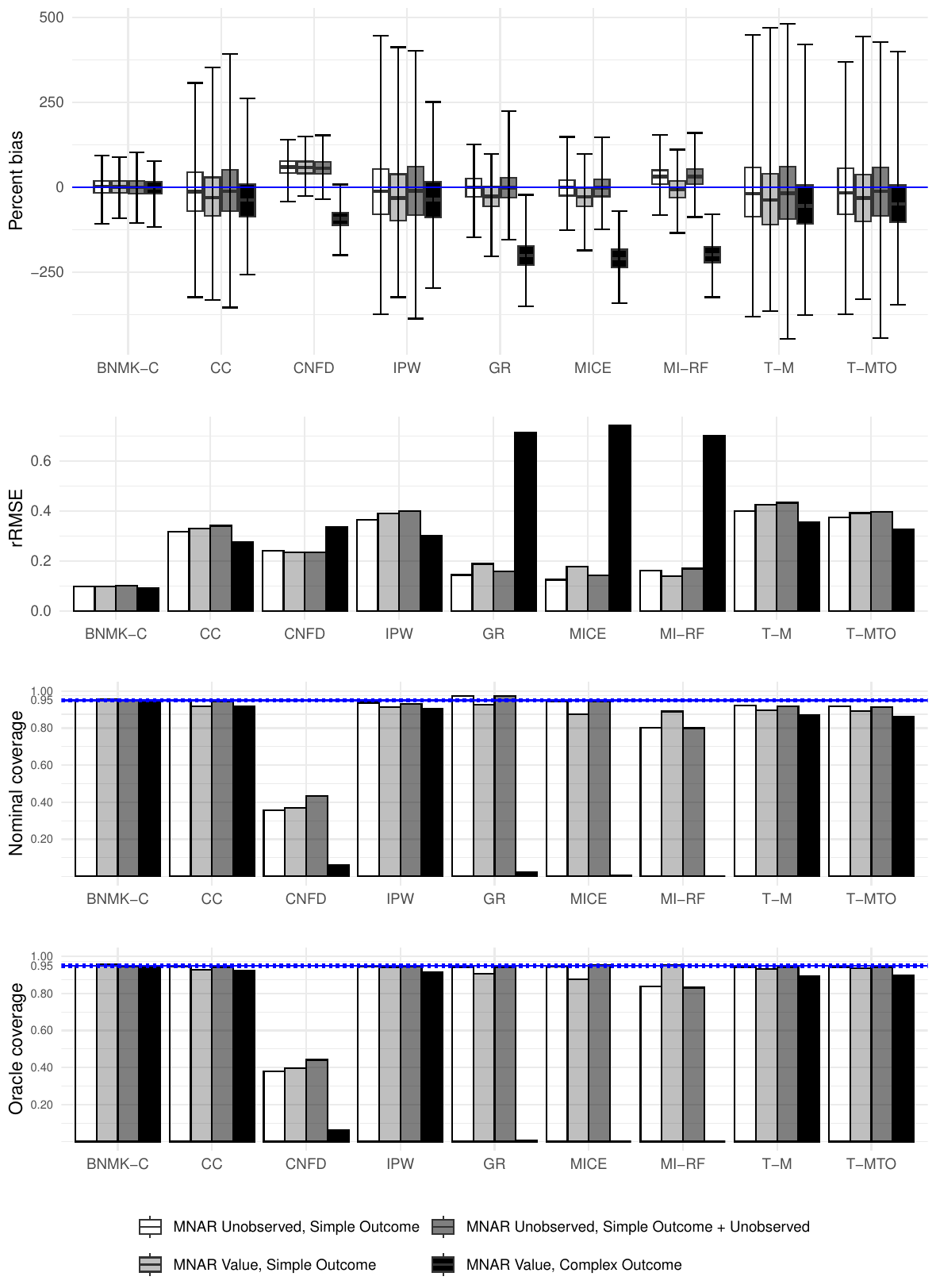}
\label{fig:mlogRR_mnar_census_lowout_highmissing}
\end{figure}


\begin{figure}[!htb]
\caption[Missing not at random, oracle truth, low outcome, high missing]{\textbf{Synthetic Data MNAR Simulation: Oracle mlogRR}. Comparing estimators of the oracle estimand with \textbf{80\% confounder missingness} and \textbf{5\% outcome proportion}. \textbf{Top graph}: \%Bias (median, IQR, min and max of converged simulations); \textbf{Middle graph}: Robust RMSE (rRMSE), using median bias and MAD; \textbf{Bottom graphs}: Nominal and oracle coverage, respectively, with blue confidence bands at $ .95 \pm 1.96 \sqrt{\frac{.05\cdot .95}{2500}}$. True values of mlogRR are 0.374 for simple outcome models and 0.369 and 0.349 for complex outcome models for MNAR unobserved and MNAR value, respectively.}

\includegraphics[scale=0.65]{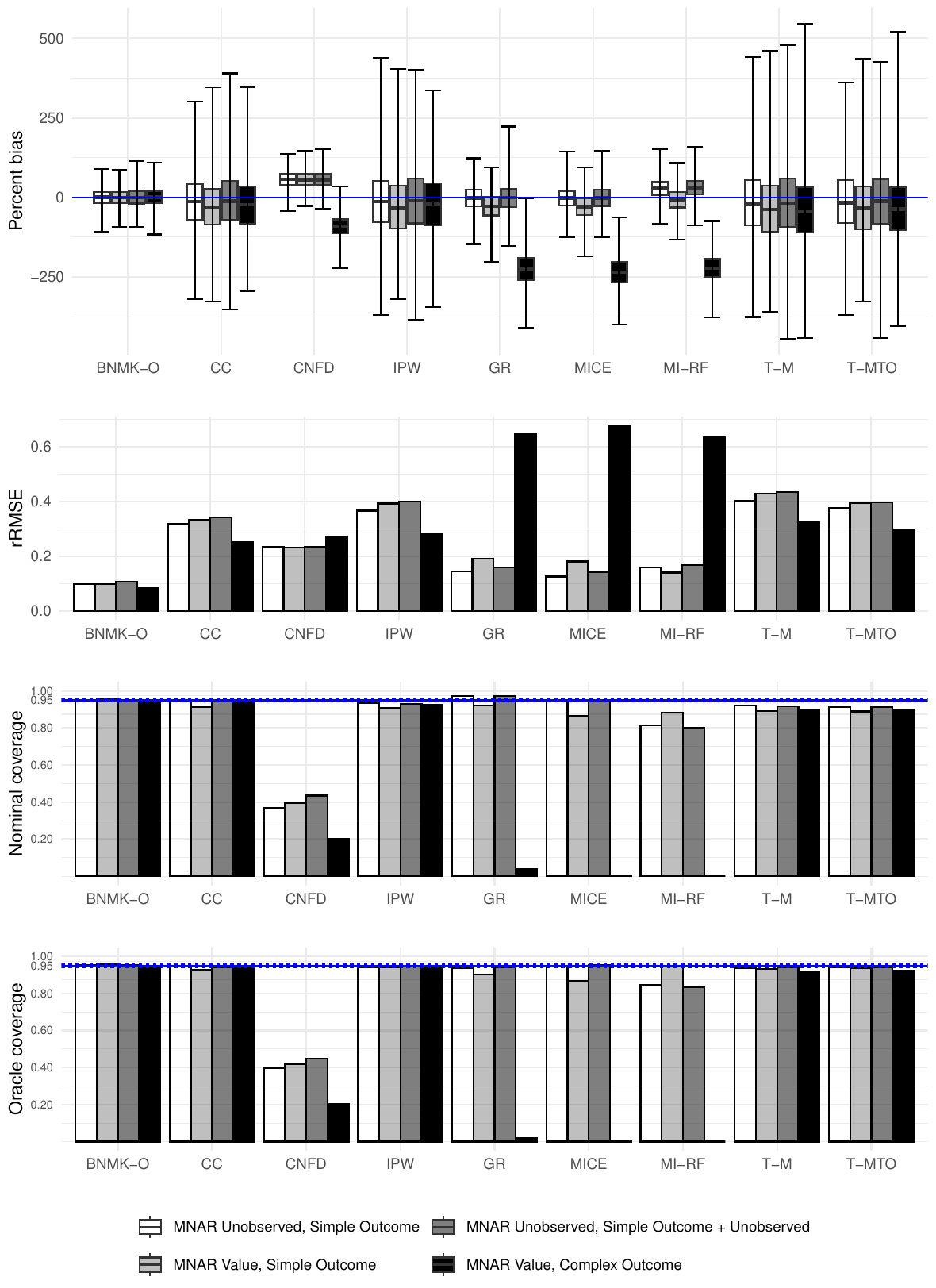}
\label{fig:mlogRR_mnar_oracle_lowout_highmissing}
\end{figure}

\clearpage

\newpage


\section{Plasmode-data Simulation}\label{sec:plasmode-gen}
\subsection{Source Data}
We extracted electronic health records data for individuals aged 13 years and older initiating depression treatment at Kaiser Permanente Washington (KPWA) from January 1, 2008 to December 31, 2018. This includes 112,770 individuals initiating either antidepressant medication or psychotherapy, defined as individuals enrolled in KPWA for a full 365 days prior to treatment start with no antidepressant fills or psychotherapy visits in the prior 365 days and with at least 1 diagnosis of depression in the prior 365 or up to 14 days post treatment initiations. We exclude individuals with diagnoses of bipolar disorder or psychotic disorder in the prior 365 days. 

The variables we will use as covariates in this study are: sex (mostly represents sex assigned at birth), age in years at time of treatment initiation, Charlson comorbidity index total score, presence of an anxiety diagnosis in the past year, presence of alcohol use disorder in the past year, recoded self-harm in the prior 6 months, and the 9 item Patient Health Questionnaire (PHQ-9). We use the sum of first 8 items to summarize depressive symptoms and the 9th item as a measure of suicidal ideation.  This patient reported outcome is routinely collected in the KPWA health system and is available through electronic health records, but would not be available for Sentinel data contributing sites with access only to insurance claims information. In our sample the PHQ is available on 50,337 people (45\% of the cohort). Table \ref{tab:plasmodetab1} describes the study cohort overall and by treatment initiation status.
In this complete-data cohort, there are 5,193 individuals who experience the composite outcome of self-harm (fatal or non-fatal) or hospitalization with a mental health diagnosis in the 5 years following treatment initiation (a rate of 10.3\%) and 358 people who have a medically attended self-harm injury or poisoning (a rate of 0.7\%)  in the 365 days following treatment initiation.
\clearpage
\newpage

\subsection{Description of Plasmode Cohort Data}\label{sec:supp-plasmode-cohort}

\begin{table}[!ht]
\small
    \centering
    \caption{Description of data for plasmode simulations; individuals starting antidepressant or psychotherapy treatment for depression.  Means and (standard deviations) provided for continuous variable; proportions and (numbers) for categorical variable. $^*$at time of treatment initiation; $^\dagger$based on diagnoses in the past year $^\ddagger$based on diagnoses in the last 6 months. MH=mental health}
    \label{tab:plasmodetab1}
    \begin{tabular}{l|c|c|c}
         & \textbf{Antidepressant } &  &     \\ 
        \textbf{Characteristic}&\textbf{medication} & \textbf{Psychotherapy} & \textbf{Overall} \\
        \hline
Female & 67.1 (18369) & 62.0 (14245) &64.8 (32618)\\
Age in years$^*$ & 44.4 (19.0)& 38.6 (18.4) & 41.8 (18.9)\\
\: \: 13 to 17& 5.0 (1369) & 13.5 (3179) & 9.0 (4548)\\
\: \: 18 to 29& 21.7 (5936)& 24.6 (5659) & 23.0 (11595)\\
\: \: 30 to 44& 25.2 (6903)& 25.1 (5752)& 25.1 (12655)\\
\: \: 45 to 64& 32.5 (8906)& 26.0 (5963)& 29.5 (14869)\\
\: \: 65 or older & 15.6 (4263)& 10.5 (2407)& 13.3 (6670)\\
Charlson$^\dagger$ &&&\\
\: \: 0  & 75.4 (20654) & 79.1 (18168) & 77.1 (38822)\\
\: \: 1  & 14.1 (3852) & 13.1 (3018)& 13.6 (6810)\\
\: \: 2  & 5.0 (1364) & 4.0 (919) &  4.5(2283)\\
\: \: 3 or more  & 5.5 (1507) & 3.7 (855)& 4.7 (2364) \\
Anxiety disorder$^\dagger$ & 13.2 (3611) & 18.7 (4284) & 15.7 (7895)\\
Alcohol use disorder$^\dagger$ &2.0 (544) & 2.4(558) &2.2 (1102)\\
Prior Self-harm$^\ddagger$ & 0.3 (90) & 0.7 (172) & 0.5 (262)\\
Prior hospitalization with MH diagnosis$^\ddagger$ & 6.8 (1858)& 4.9 (1116) & 5.9 (2974)\\
PHQ8 total score$^*$ &14.7 (5.0) & 11.7 (5.8) & 13.3 (5.9)\\
\: \: 0-5&&& 9.7 (4868)\\
\: \: 5-10 (Mild symptoms)&4.4 (1200)&16.0 (3668)& 21.7 (10933)\\
\: \: 11-15 (Moderate symptoms)&16.6 (4540)&27.8 (6393)& 31.1 (15634)\\
\: \: 16-20 (Moderate/Severe symptoms)&33.4 (8934)&19.9(4575)& 26.8 (13509)\\
\: \: 21-24 (Severe symptoms)&13.0 (3569)&7.9 (1824)& 10.7 (5393)\\
PHQ item 9$^*$ &&&\\
\: \: 0 (none of the days)& 66.5 (18198)&67.2(15422)&5.1 (2566)\\
\: \: 1 (several days) &20.1(5507)&20.4(4687)& 7.9 (3957)\\
\: \: 2 (more than half the days) &8.4(2296)&7.2(1661)&20.3 (10194)\\
\: \: 3 (nearly every day) &5.0(1376)&5.2(1190)&66.8 (33620)\\
    \end{tabular}
\end{table}
\clearpage
\newpage

\newpage

\subsection{Plasmode Data Generating Models}\label{sec:supp-plasmode-models}

\begin{table}[!ht]
    \centering
    \caption{Description of data generating models for missing PHQ data; exposure (antidepressants or psychotherapy [PT]); self-harm or hospitalization with a mental health diagnosis in the following 5 years (5-year Hosp/SH, first outcome); and self-harm in the following year (1-year SH, second outcome). Coefficient values from logistic regression data generating models (i.e. coefficients are on the logodds scale).\\
    $^*$unit used for age is 10 years; $^\dagger$based on diagnoses in the past year; $^\ddagger$based on diagnoses in the last 6 months. MH=mental health; Hosp=hospitalization with a MH diagnosis; SH=self-harm}
    \label{tab:plasmodetab2}
    \begin{tabular}{l|c|c|c|c}
    &\multicolumn{3}{c}{\textbf{Coefficient in logistic regression}}\\
&\textbf{Missing } & \textbf{Receipt} & \textbf{5-year}&\textbf{1-year}\\
 \textbf{Characteristic}&\textbf{PHQ} & \textbf{ of PT} & \textbf{Hosp/SH} & \textbf{SH}\\
\hline
Intercept & 0.15 & 2.60 & -1.28 & -2.27\\
Psychotherapy & 0 & 0 & -0.21 &  0.10\\
Female sex & 0.02 & -0.24 & 0.36 & 0.47\\
Age at initiation$^*$ & -0.90 & -0.33 & -0.65 & -0.18\\ 
Age at initiation squared$^*$ & 0.002 & 0.01 & 0.08 & 0.00\\
Charlson  1  & 0.16 & 0.64 & 1.51& 0.34\\
Charlson  2  & 0.11 & 0.43& 0.43& 0.37\\
Charlson  3+ & 0.28 & 0.12 & -0.08&0.55\\
Anxiety disorder$^\dagger$ & 0.11 & 0.51 & 0.84&0.41\\ 
Alcohol use disorder$^\dagger$ & 0.08 & 0.24 & 0.15&0.77\\
Prior self-harm$^\ddagger$ & 0 & 0.14 & 1.96&1.38\\
Prior hospitalization with MH diagnosis$^\ddagger$ & 0 & -0.32 & 0.91&0.93\\ 
PHQ8 total score$^*$: 5-10 & 0 & -0.88 & -0.03&-0.39\\
PHQ8 total score$^*$: 11-15 & 0 & -1.67& 0.21&-0.07\\
PHQ8 total score$^*$: 16-20 & 0 & -2.07& 0.34&-0.01\\
PHQ8 total score$^*$: 21-24 & 0 & -2.13& 0.35& -0.14\\
PHQ item 9*: 1 & 0 & 0.14 & 0.22&1.14\\
PHQ item 9*: 2 & 0 & 0.12 & 0.30&1.44\\
PHQ item 9*: 3 & 0 & 0.45& 0.55&2.00\\
Charlson score 1 \& anxiety disorder  & -0.13 & 0.11&  0.25&0.52\\
Charlson score 2 \& anxiety disorder  & -0.08 & -0.09&  0.15&0.08\\
Charlson score 3+ \& anxiety disorder & -0.02 & -0.25&  -0.16&0.05\\
Age at initiation$^*$ \& female & -0.16 & 0.003 & -0.07 &-0.01\\
Female \& prior self-harm &  -0.38 & 0.15 & -0.01 &0.41\\
Age at initiation$^*$ \& prior self-harm & 0.30 & -0.03 & -0.20&0.04\\
Charlson score 1 \& age at initiation$^*$  & -0.004 & -0.09 & -0.15&-0.01\\
Charlson score 2 \& age at initiation$^*$  & -0.002 & -0.07 & -0.07&-0.01\\
Charlson score 3+ \& age at initiation$^*$ & -0.005 & -0.01& 0.02&-0.02\\
PHQ item 9 score 1 \& female & 0 & 0.09 & -0.04&-0.16\\
PHQ item 9 score 2 \& female & 0 & 0.05& -0.06&-0.34\\
PHQ item 9 score 3 \& female & 0 & 0.03& 0.06&0.03\\
PHQ item 9 score 1 \& prior self-harm & 0 & 0.50& -0.22& -0.39\\
PHQ item 9 score 2 \& prior self-harm & 0 & 0.89& -0.49&-0.91\\
PHQ item 9 score 3 \& prior self-harm & 0 & 0.33& -0.53&-1.56\\
    \end{tabular}
\end{table}

\clearpage

\newpage


\begin{figure}[htb]
\caption{Plasmode treatment assignment tree model}
\includegraphics[height=3in,width=5in]{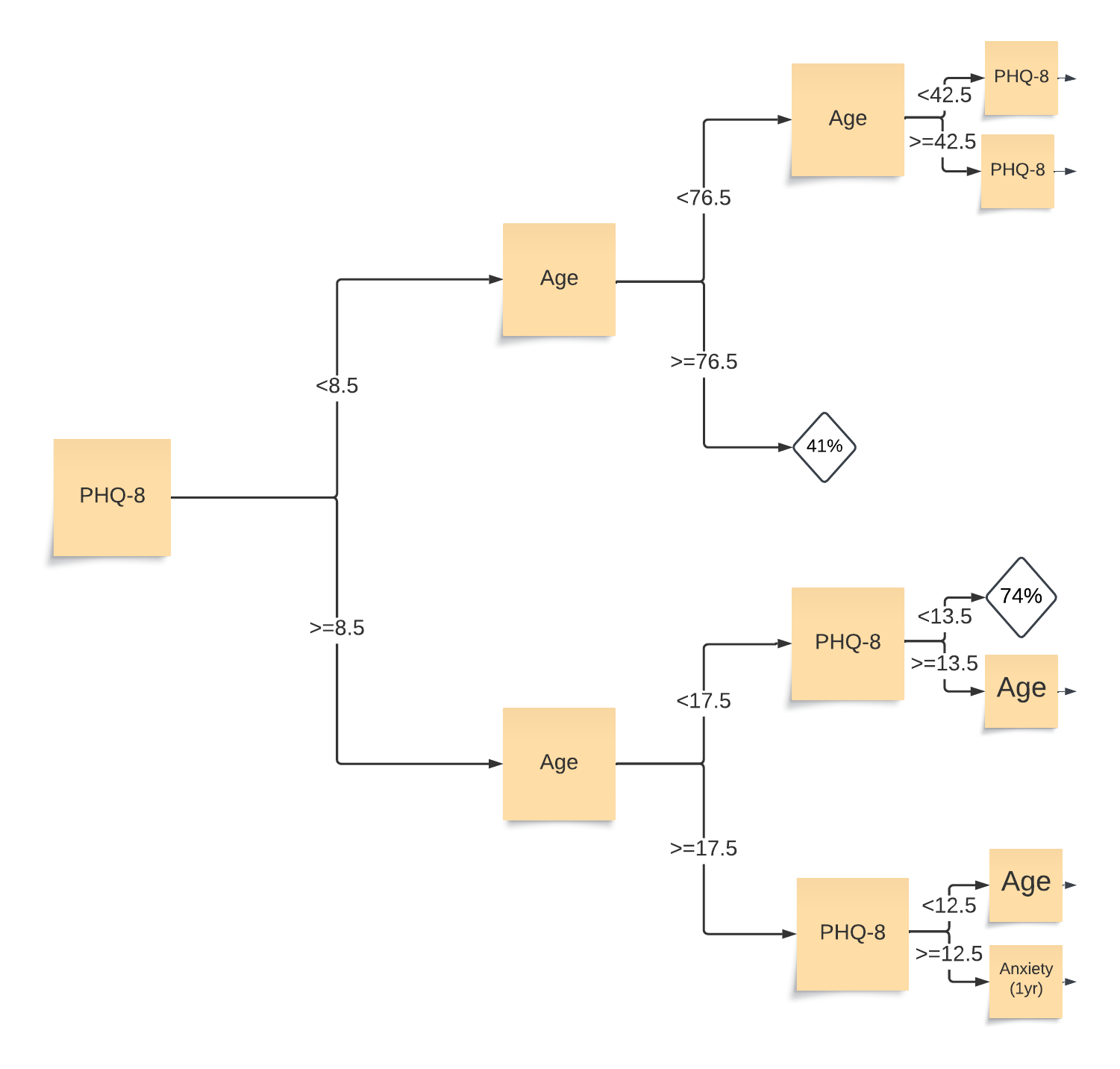}
\floatfoot{ The full tree has a maximum depth of 15 nodes.  Splitting variables not seen in the first four layers include the Patient Health Questionnaire (PHQ) item 9 score, anxiety, Charlson score, mental health inpatient use (MH IP), anxiety, alcohol use, sex, and prior self-harm.  Assignment likelihoods range from 9.7\% to 91\%.}
\label{fig:treatTree}
\end{figure}


\begin{figure}[htb]
\caption{Plasmode  5-year self-harm or hospitalization and 1-year self-harm outcome tree model}
\begin{subfigure}{\textwidth }
\caption{5-year outcome}
\includegraphics[height=3in,width=5in]{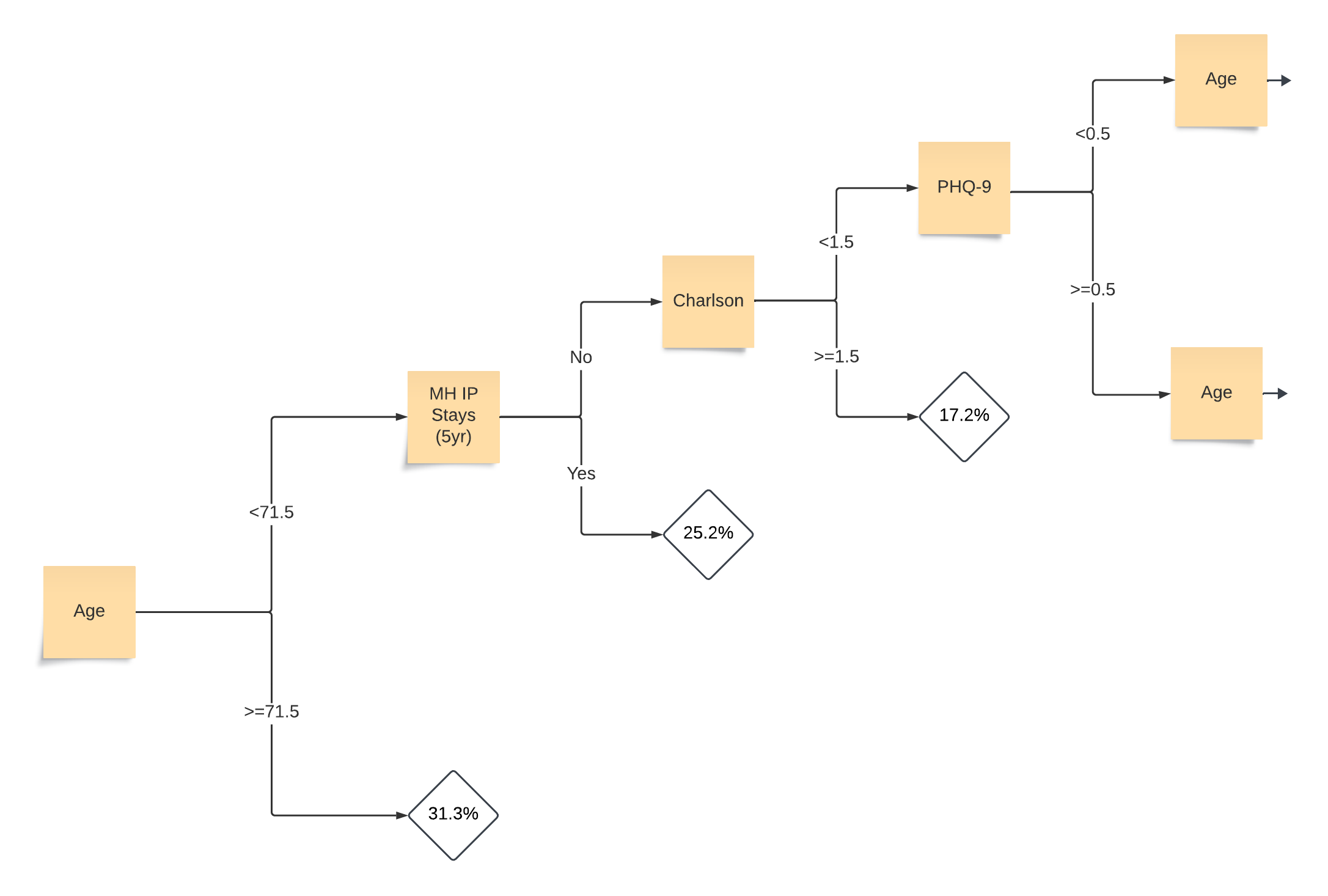}
\label{fig:outTree5yr}
\end{subfigure}

\begin{subfigure}{\textwidth}
\caption{1-year outcome}
\includegraphics[height=3in,width=5in]{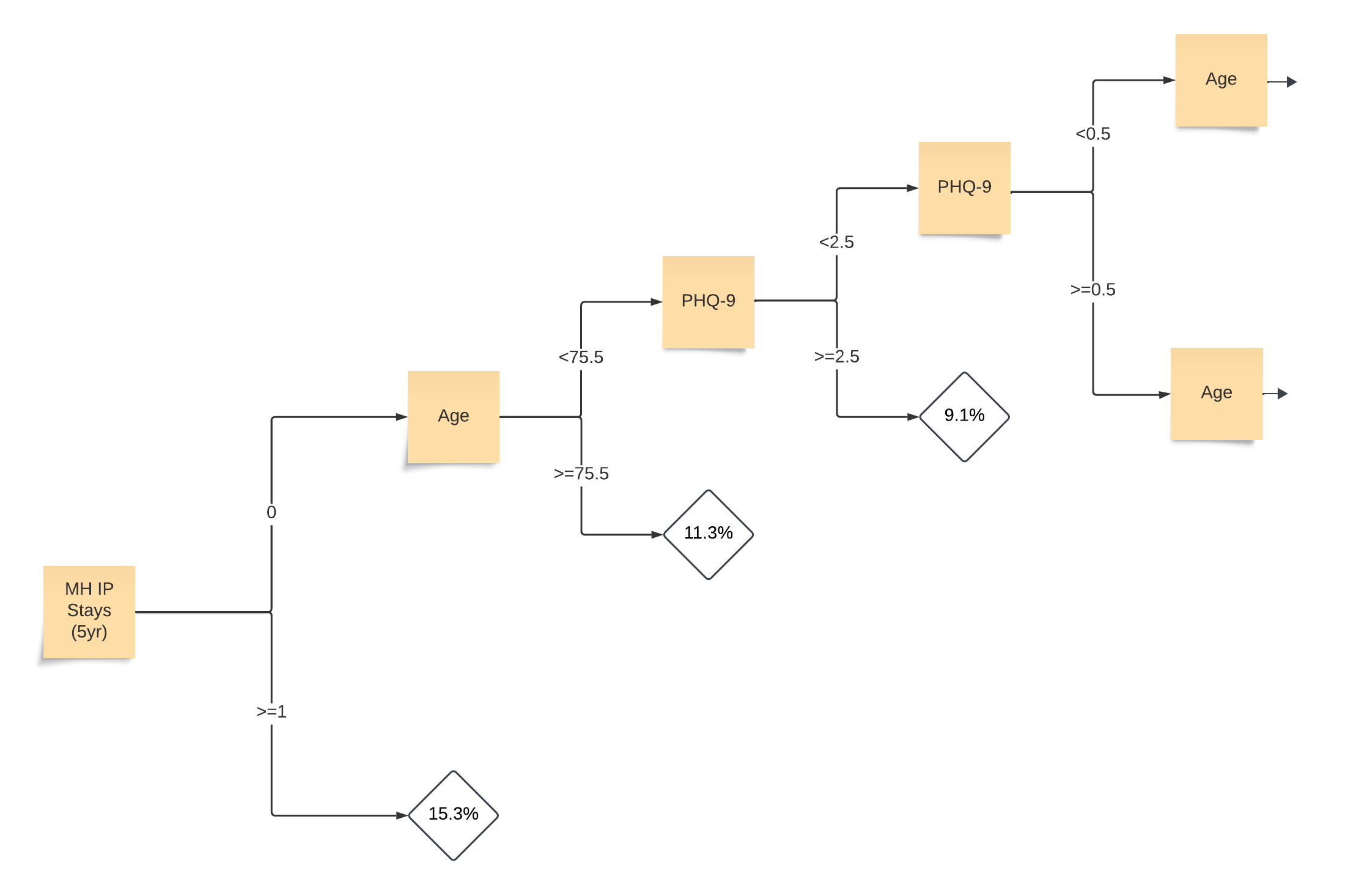}
\floatfoot{ }
\label{fig:outTree1yr}
\end{subfigure}
\floatfoot{ (a) The full tree for the 5-year outcome has a maximum depth of 13 nodes.  Splitting variables not seen in the first four layers include prior self-harm and alcohol use.  Outcome likelihoods range from 2.6\% to 40.1\%. (b) The full tree for the 1-year outcome has a maximum depth of 14 nodes. Splitting variables not seen in the first four layers include Charlson comorbidity score, alcohol use disorder, PHQ-8 total score, and treatment type.  Outcome likelihoods range from 15.3\% to 0.9\%.}
\end{figure}

\begin{figure}[htb]
\caption{Plasmode missingness assignment tree model}
\includegraphics[height=3in,width=5in]{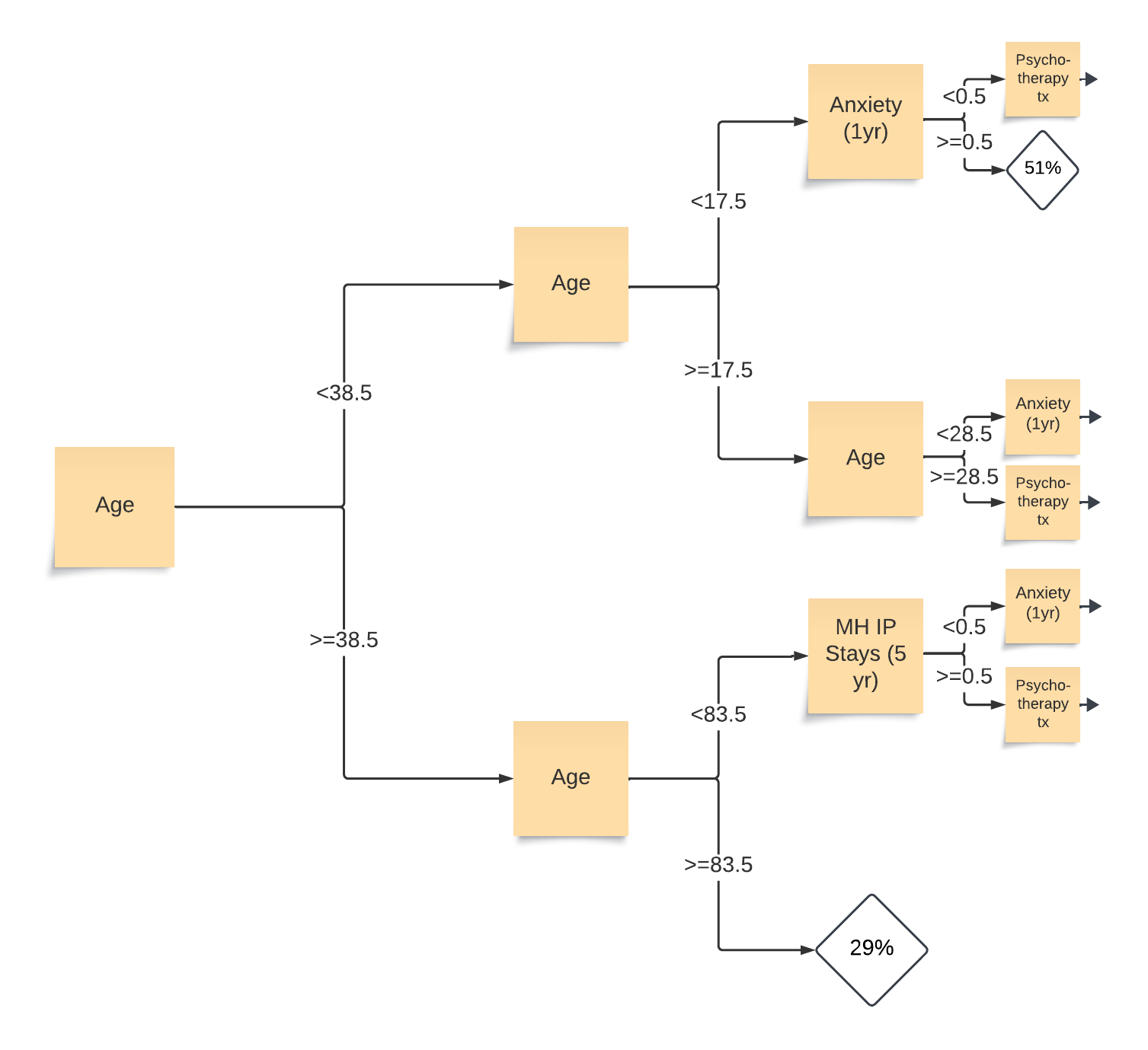}
\floatfoot{ The full tree has a maximum depth of 16 nodes.  Splitting variables not seen in the first four layers include mental health inpatient use (MH IP), Charlson score, alcohol use, and prior self-harm.  Missingness likelihoods range from 21\% to 75\%.}
\label{fig:missTree}
\end{figure}

\clearpage
\newpage

\normalsize

\subsection{Plasmode Simulation Results Tables}\label{sec:supp-tables-plasmode}
\clearpage

\subsubsection{mRD}
\clearpage

\begin{table}
\centering
\caption{\label{tab:xNA_mNA_yglm_SH_365day_mRD_census}\textbf{Plasmode data simulation: 1-year self-harm or hospitalization, regression functions are glms, census marginal risk difference (mRD)}. Relative performance of estimators with sample size n = 50,337 and 1000 simulation replications. Note: Bias, ESE, ASE, MAD, and RMSE scaled by a factor of 10 to facilitate comparisons across estimands. The value of the estimand is 0.001. ESE = empirical standard error, ASE = asymptotic standard error, MAD = mean absolute deviation, RMSE = root mean squared error, rRMSE = robust RMSE (using median bias and MAD), Oracle coverage = coverage of a confidence interval based on the ESE, Nominal coverage = coverage of a confidence interval based on the ASE. Estimators that are mismatched with the estimand (i.e., are estimating a different parameter) are emphasized using a star.}
\centering
\fontsize{9}{11}\selectfont
\begin{tabular}[t]{>{\raggedright\arraybackslash}p{2cm}|>{\raggedright\arraybackslash}p{1cm}|>{\raggedleft\arraybackslash}p{1cm}|>{\raggedleft\arraybackslash}p{1cm}|>{\raggedleft\arraybackslash}p{1cm}|>{\raggedleft\arraybackslash}p{1cm}|>{\raggedright\arraybackslash}p{1cm}|>{\raggedright\arraybackslash}p{1cm}|>{\raggedleft\arraybackslash}p{1cm}|>{\raggedleft\arraybackslash}p{1cm}|>{\raggedleft\arraybackslash}p{1cm}|>{\raggedleft\arraybackslash}p{1cm}|>{}p{1cm}}
\hline
Estimator & Mean bias & Median bias & ESE & ASE & MAD & RMSE & rRMSE & Oracle coverage & Nominal coverage & Power & Prop. completed\\
\hline
Benchmark model & 0 & 0.000 & 0.008 & 0.008 & 0.008 & 0.008 & 0.008 & 0.944 & 0.937 & 0.178 & 98.6\\
\hline
Complete-case & 0 & 0.000 & 0.012 & 0.012 & 0.013 & 0.012 & 0.013 & 0.956 & 0.948 & 0.113 & 98.6\\
\hline
Confounded model & -0.004 & -0.004 & 0.008 & 0.008 & 0.008 & 0.008 & 0.009 & 0.922 & 0.918 & 0.087 & 98.6\\
\hline
IPW & 0 & -0.001 & 0.012 & 0.011 & 0.013 & 0.011 & 0.013 & 0.956 & 0.950 & 0.101 & 98.6\\
\hline
Raking (vanilla) & 0 & 0.000 & 0.008 & 0.008 & 0.009 & 0.008 & 0.009 & 0.950 & 0.950 & 0.148 & 99.2\\
\hline
MICE & -0.001 & -0.001 & 0.008 & 0.008 & 0.008 & 0.008 & 0.008 & 0.941 & 0.937 & 0.147 & 98.6\\
\hline
MI-RF & 0 & 0.000 & 0.008 & 0.008 & 0.008 & 0.008 & 0.008 & 0.941 & 0.946 & 0.150 & 98.6\\
\hline
IPCW-TMLE-M${}^*$ & 0 & 0.000 & 0.012 & 0.011 & 0.012 & 0.011 & 0.012 & 0.954 & 0.953 & 0.106 & 100.0\\
\hline
IPCW-TMLE-MTO${}^*$ & 0 & 0.000 & 0.011 & 0.011 & 0.012 & 0.011 & 0.012 & 0.953 & 0.949 & 0.102 & 100.0\\
\hline
r-IPCW-TMLE-MTO${}^*$ & 0 & -0.001 & 0.011 & 0.011 & 0.012 & 0.011 & 0.012 & 0.951 & 0.947 & 0.105 & 100.0\\
\hline
\end{tabular}
\end{table}

\begin{table}
\centering
\caption{\label{tab:xNA_mNA_ytree_SH_365day_mRD_census}\textbf{Plasmode data simulation: 1-year self-harm or hospitalization, regression functions are trees, census marginal risk difference (mRD)}. Relative performance of estimators with sample size n = 50,337 and 1000 simulation replications. Note: Bias, ESE, ASE, MAD, and RMSE scaled by a factor of 10 to facilitate comparisons across estimands. The value of the estimand is 0. ESE = empirical standard error, ASE = asymptotic standard error, MAD = mean absolute deviation, RMSE = root mean squared error, rRMSE = robust RMSE (using median bias and MAD), Oracle coverage = coverage of a confidence interval based on the ESE, Nominal coverage = coverage of a confidence interval based on the ASE. Estimators that are mismatched with the estimand (i.e., are estimating a different parameter) are emphasized using a star.}
\centering
\fontsize{9}{11}\selectfont
\begin{tabular}[t]{>{\raggedright\arraybackslash}p{2cm}|>{\raggedright\arraybackslash}p{1cm}|>{\raggedleft\arraybackslash}p{1cm}|>{\raggedleft\arraybackslash}p{1cm}|>{\raggedleft\arraybackslash}p{1cm}|>{\raggedleft\arraybackslash}p{1cm}|>{\raggedright\arraybackslash}p{1cm}|>{\raggedright\arraybackslash}p{1cm}|>{\raggedleft\arraybackslash}p{1cm}|>{\raggedleft\arraybackslash}p{1cm}|>{\raggedleft\arraybackslash}p{1cm}|>{\raggedleft\arraybackslash}p{1cm}|>{}p{1cm}}
\hline
Estimator & Mean bias & Median bias & ESE & ASE & MAD & RMSE & rRMSE & Oracle coverage & Nominal coverage & Power & Prop. completed\\
\hline
Benchmark model & 0 & 0.000 & 0.008 & 0.008 & 0.008 & 0.008 & 0.008 & 0.954 & 0.954 & 0.052 & 98.1\\
\hline
Complete-case & 0.005 & 0.005 & 0.012 & 0.012 & 0.012 & 0.013 & 0.013 & 0.928 & 0.925 & 0.085 & 98.1\\
\hline
Confounded model & -0.004 & -0.004 & 0.008 & 0.008 & 0.007 & 0.009 & 0.008 & 0.922 & 0.924 & 0.062 & 98.1\\
\hline
IPW & 0.004 & 0.005 & 0.011 & 0.011 & 0.011 & 0.012 & 0.012 & 0.933 & 0.929 & 0.084 & 98.1\\
\hline
Raking (vanilla) & 0.004 & 0.004 & 0.008 & 0.009 & 0.008 & 0.009 & 0.009 & 0.920 & 0.938 & 0.084 & 98.1\\
\hline
MICE & 0.003 & 0.003 & 0.008 & 0.008 & 0.008 & 0.009 & 0.008 & 0.931 & 0.923 & 0.094 & 98.1\\
\hline
MI-RF & 0.003 & 0.003 & 0.008 & 0.008 & 0.007 & 0.008 & 0.008 & 0.939 & 0.943 & 0.073 & 98.1\\
\hline
IPCW-TMLE-M${}^*$ & 0.002 & 0.002 & 0.011 & 0.012 & 0.011 & 0.012 & 0.011 & 0.943 & 0.969 & 0.035 & 100.0\\
\hline
IPCW-TMLE-MTO${}^*$ & -0.004 & -0.003 & 0.010 & 0.011 & 0.010 & 0.012 & 0.011 & 0.932 & 0.951 & 0.047 & 100.0\\
\hline
r-IPCW-TMLE-MTO${}^*$ & -0.004 & -0.004 & 0.011 & 0.011 & 0.011 & 0.012 & 0.011 & 0.928 & 0.946 & 0.045 & 100.0\\
\hline
\end{tabular}
\end{table}

\begin{table}
\centering
\caption{\label{tab:xNA_mNA_yglm_SH_HOSP_1826day_mRD_census}\textbf{Plasmode data simulation: 5-year self-harm or hospitalization, regression functions are glms, census marginal risk difference (mRD)}. Relative performance of estimators with sample size n = 50,337 and 1000 simulation replications. Note: Bias, ESE, ASE, MAD, and RMSE scaled by a factor of 10 to facilitate comparisons across estimands. The value of the estimand is -0.016. ESE = empirical standard error, ASE = asymptotic standard error, MAD = mean absolute deviation, RMSE = root mean squared error, rRMSE = robust RMSE (using median bias and MAD), Oracle coverage = coverage of a confidence interval based on the ESE, Nominal coverage = coverage of a confidence interval based on the ASE.}
\centering
\fontsize{9}{11}\selectfont
\begin{tabular}[t]{>{\raggedright\arraybackslash}p{2cm}|>{\raggedright\arraybackslash}p{1cm}|>{\raggedleft\arraybackslash}p{1cm}|>{\raggedleft\arraybackslash}p{1cm}|>{\raggedleft\arraybackslash}p{1cm}|>{\raggedleft\arraybackslash}p{1cm}|>{\raggedright\arraybackslash}p{1cm}|>{\raggedright\arraybackslash}p{1cm}|>{\raggedleft\arraybackslash}p{1cm}|>{\raggedleft\arraybackslash}p{1cm}|>{\raggedleft\arraybackslash}p{1cm}|>{\raggedleft\arraybackslash}p{1cm}|>{}p{1cm}}
\hline
Estimator & Mean bias & Median bias & ESE & ASE & MAD & RMSE & rRMSE & Oracle coverage & Nominal coverage & Power & Prop. completed\\
\hline
Benchmark model & -0.006 & -0.005 & 0.028 & 0.028 & 0.026 & 0.028 & 0.027 & 0.940 & 0.934 & 1.000 & 100.0\\
\hline
Complete-case & 0.001 & 0.002 & 0.041 & 0.040 & 0.043 & 0.04 & 0.043 & 0.953 & 0.946 & 0.981 & 100.0\\
\hline
Confounded model & -0.074 & -0.073 & 0.027 & 0.027 & 0.025 & 0.079 & 0.077 & 0.192 & 0.179 & 1.000 & 100.0\\
\hline
IPW & -0.004 & -0.004 & 0.043 & 0.042 & 0.042 & 0.042 & 0.042 & 0.948 & 0.942 & 0.978 & 100.0\\
\hline
Raking (vanilla) & -0.005 & -0.005 & 0.030 & 0.029 & 0.027 & 0.03 & 0.028 & 0.939 & 0.935 & 0.999 & 99.2\\
\hline
MICE & -0.006 & -0.007 & 0.030 & 0.028 & 0.027 & 0.029 & 0.028 & 0.938 & 0.930 & 0.999 & 100.0\\
\hline
MI-XGB & 0.197 & 0.194 & 0.082 & 0.042 & 0.082 & 0.201 & 0.211 & 0.337 & 0.074 & 0.357 & 100.0\\
\hline
MI-RF & -0.037 & -0.036 & 0.028 & 0.028 & 0.025 & 0.047 & 0.044 & 0.757 & 0.760 & 1.000 & 100.0\\
\hline
IPCW-TMLE-M & -0.005 & -0.004 & 0.043 & 0.043 & 0.043 & 0.043 & 0.043 & 0.953 & 0.951 & 0.983 & 100.0\\
\hline
IPCW-TMLE-MTO & -0.008 & -0.008 & 0.043 & 0.043 & 0.043 & 0.044 & 0.044 & 0.943 & 0.942 & 0.986 & 100.0\\
\hline
\end{tabular}
\end{table}

\begin{table}
\centering
\caption{\label{tab:xNA_mNA_ytree_SH_HOSP_1826day_mRD_census}\textbf{Plasmode data simulation: 5-year self-harm or hospitalization, regression functions are trees, census marginal risk difference (mRD)}. Relative performance of estimators with sample size n = 50,337 and 1000 simulation replications. Note: Bias, ESE, ASE, MAD, and RMSE scaled by a factor of 10 to facilitate comparisons across estimands. The value of the estimand is -0.006. ESE = empirical standard error, ASE = asymptotic standard error, MAD = mean absolute deviation, RMSE = root mean squared error, rRMSE = robust RMSE (using median bias and MAD), Oracle coverage = coverage of a confidence interval based on the ESE, Nominal coverage = coverage of a confidence interval based on the ASE. Estimators that are mismatched with the estimand (i.e., are estimating a different parameter) are emphasized using a star.}
\centering
\fontsize{9}{11}\selectfont
\begin{tabular}[t]{>{\raggedright\arraybackslash}p{2cm}|>{\raggedright\arraybackslash}p{1cm}|>{\raggedleft\arraybackslash}p{1cm}|>{\raggedleft\arraybackslash}p{1cm}|>{\raggedleft\arraybackslash}p{1cm}|>{\raggedleft\arraybackslash}p{1cm}|>{\raggedright\arraybackslash}p{1cm}|>{\raggedright\arraybackslash}p{1cm}|>{\raggedleft\arraybackslash}p{1cm}|>{\raggedleft\arraybackslash}p{1cm}|>{\raggedleft\arraybackslash}p{1cm}|>{\raggedleft\arraybackslash}p{1cm}|>{}p{1cm}}
\hline
Estimator & Mean bias & Median bias & ESE & ASE & MAD & RMSE & rRMSE & Oracle coverage & Nominal coverage & Power & Prop. completed\\
\hline
Benchmark model & 0.001 & 0.000 & 0.028 & 0.029 & 0.027 & 0.029 & 0.027 & 0.949 & 0.952 & 0.569 & 100.0\\
\hline
Complete-case & 0.009 & 0.008 & 0.041 & 0.041 & 0.042 & 0.042 & 0.042 & 0.951 & 0.951 & 0.243 & 100.0\\
\hline
Confounded model & -0.035 & -0.035 & 0.027 & 0.027 & 0.026 & 0.044 & 0.043 & 0.746 & 0.755 & 0.947 & 100.0\\
\hline
IPW & 0.007 & 0.005 & 0.042 & 0.043 & 0.043 & 0.043 & 0.043 & 0.949 & 0.948 & 0.248 & 100.0\\
\hline
Raking (vanilla) & 0.001 & 0.001 & 0.030 & 0.030 & 0.029 & 0.03 & 0.029 & 0.947 & 0.950 & 0.531 & 99.3\\
\hline
MICE & 0.002 & 0.003 & 0.030 & 0.029 & 0.029 & 0.029 & 0.029 & 0.945 & 0.941 & 0.531 & 100.0\\
\hline
MI-XGB & 0.046 & 0.047 & 0.086 & 0.042 & 0.085 & 0.062 & 0.097 & 0.907 & 0.594 & 0.355 & 100.0\\
\hline
MI-RF & -0.014 & -0.014 & 0.028 & 0.029 & 0.027 & 0.032 & 0.03 & 0.922 & 0.938 & 0.758 & 100.0\\
\hline
IPCW-TMLE-M${}^*$ & 0 & -0.001 & 0.037 & 0.043 & 0.036 & 0.043 & 0.036 & 0.944 & 0.973 & 0.267 & 100.0\\
\hline
IPCW-TMLE-MTO${}^*$ & -0.02 & -0.021 & 0.037 & 0.042 & 0.037 & 0.046 & 0.042 & 0.923 & 0.955 & 0.516 & 100.0\\
\hline
\end{tabular}
\end{table}

\clearpage

\begin{table}
\centering
\caption{\label{tab:xNA_mNA_yglm_SH_365day_mRD_oracle}\textbf{Plasmode data simulation: 1-year self-harm or hospitalization, regression functions are glms, oracle marginal risk difference (mRD)}. Relative performance of estimators with sample size n = 50,337 and 1000 simulation replications. Note: Bias, ESE, ASE, MAD, and RMSE scaled by a factor of 10 to facilitate comparisons across estimands. The value of the estimand is 0.001. ESE = empirical standard error, ASE = asymptotic standard error, MAD = mean absolute deviation, RMSE = root mean squared error, rRMSE = robust RMSE (using median bias and MAD), Oracle coverage = coverage of a confidence interval based on the ESE, Nominal coverage = coverage of a confidence interval based on the ASE. Estimators that are mismatched with the estimand (i.e., are estimating a different parameter) are emphasized using a star.}
\centering
\fontsize{9}{11}\selectfont
\begin{tabular}[t]{>{\raggedright\arraybackslash}p{2cm}|>{\raggedright\arraybackslash}p{1cm}|>{\raggedright\arraybackslash}p{1cm}|>{\raggedleft\arraybackslash}p{1cm}|>{\raggedleft\arraybackslash}p{1cm}|>{\raggedleft\arraybackslash}p{1cm}|>{\raggedright\arraybackslash}p{1cm}|>{\raggedright\arraybackslash}p{1cm}|>{\raggedleft\arraybackslash}p{1cm}|>{\raggedleft\arraybackslash}p{1cm}|>{\raggedleft\arraybackslash}p{1cm}|>{\raggedleft\arraybackslash}p{1cm}|>{}p{1cm}}
\hline
Estimator & Mean bias & Median bias & ESE & ASE & MAD & RMSE & rRMSE & Nominal coverage & Oracle coverage & Power & Prop. completed\\
\hline
Benchmark model & 0 & 0 & 0.008 & 0.008 & 0.008 & 0.008 & 0.008 & 0.937 & 0.947 & 0.173 & 98.6\\
\hline
Complete-case${}^*$ & 0 & 0 & 0.012 & 0.012 & 0.013 & 0.012 & 0.013 & 0.947 & 0.957 & 0.113 & 98.6\\
\hline
Confounded model${}^*$ & -0.004 & -0.004 & 0.008 & 0.008 & 0.008 & 0.008 & 0.009 & 0.920 & 0.925 & 0.087 & 98.6\\
\hline
IPW${}^*$ & 0 & 0 & 0.012 & 0.011 & 0.013 & 0.011 & 0.013 & 0.951 & 0.956 & 0.101 & 98.6\\
\hline
Raking (vanilla)${}^*$ & 0 & 0 & 0.008 & 0.008 & 0.009 & 0.008 & 0.009 & 0.950 & 0.952 & 0.148 & 99.2\\
\hline
MICE${}^*$ & -0.001 & -0.001 & 0.008 & 0.008 & 0.008 & 0.008 & 0.008 & 0.937 & 0.941 & 0.147 & 98.6\\
\hline
MI-RF${}^*$ & 0 & 0 & 0.008 & 0.008 & 0.008 & 0.008 & 0.008 & 0.947 & 0.942 & 0.150 & 98.6\\
\hline
IPCW-TMLE-M & 0 & 0 & 0.012 & 0.011 & 0.012 & 0.011 & 0.012 & 0.952 & 0.954 & 0.106 & 100.0\\
\hline
IPCW-TMLE-MTO & 0 & 0 & 0.011 & 0.011 & 0.012 & 0.011 & 0.012 & 0.950 & 0.953 & 0.102 & 100.0\\
\hline
r-IPCW-TMLE-MTO & 0 & 0 & 0.011 & 0.011 & 0.012 & 0.011 & 0.012 & 0.947 & 0.950 & 0.105 & 100.0\\
\hline
\end{tabular}
\end{table}

\begin{table}
\centering
\caption{\label{tab:xNA_mNA_ytree_SH_365day_mRD_oracle}\textbf{Plasmode data simulation: 1-year self-harm or hospitalization, regression functions are trees, oracle marginal risk difference (mRD)}. Relative performance of estimators with sample size n = 50,337 and 1000 simulation replications. Note: Bias, ESE, ASE, MAD, and RMSE scaled by a factor of 10 to facilitate comparisons across estimands. The value of the estimand is 0. ESE = empirical standard error, ASE = asymptotic standard error, MAD = mean absolute deviation, RMSE = root mean squared error, rRMSE = robust RMSE (using median bias and MAD), Oracle coverage = coverage of a confidence interval based on the ESE, Nominal coverage = coverage of a confidence interval based on the ASE. Estimators that are mismatched with the estimand (i.e., are estimating a different parameter) are emphasized using a star.}
\centering
\fontsize{9}{11}\selectfont
\begin{tabular}[t]{>{\raggedright\arraybackslash}p{2cm}|>{\raggedright\arraybackslash}p{1cm}|>{\raggedright\arraybackslash}p{1cm}|>{\raggedleft\arraybackslash}p{1cm}|>{\raggedleft\arraybackslash}p{1cm}|>{\raggedleft\arraybackslash}p{1cm}|>{\raggedright\arraybackslash}p{1cm}|>{\raggedright\arraybackslash}p{1cm}|>{\raggedleft\arraybackslash}p{1cm}|>{\raggedleft\arraybackslash}p{1cm}|>{\raggedleft\arraybackslash}p{1cm}|>{\raggedleft\arraybackslash}p{1cm}|>{}p{1cm}}
\hline
Estimator & Mean bias & Median bias & ESE & ASE & MAD & RMSE & rRMSE & Nominal coverage & Oracle coverage & Power & Prop. completed\\
\hline
Complete-case${}^*$ & 0.008 & 0.009 & 0.012 & 0.012 & 0.012 & 0.014 & 0.015 & 0.895 & 0.901 & 0.085 & 98.1\\
\hline
Confounded model${}^*$ & 0 & 0 & 0.008 & 0.008 & 0.007 & 0.008 & 0.007 & 0.950 & 0.943 & 0.062 & 98.1\\
\hline
IPW${}^*$ & 0.008 & 0.008 & 0.011 & 0.011 & 0.011 & 0.014 & 0.014 & 0.894 & 0.896 & 0.084 & 98.1\\
\hline
Raking (vanilla)${}^*$ & 0.008 & 0.008 & 0.008 & 0.009 & 0.008 & 0.011 & 0.011 & 0.866 & 0.846 & 0.084 & 98.1\\
\hline
MICE${}^*$ & 0.007 & 0.007 & 0.008 & 0.008 & 0.008 & 0.01 & 0.01 & 0.860 & 0.869 & 0.094 & 98.1\\
\hline
MI-RF${}^*$ & 0.006 & 0.006 & 0.008 & 0.008 & 0.007 & 0.01 & 0.01 & 0.887 & 0.867 & 0.073 & 98.1\\
\hline
IPCW-TMLE-M & 0.006 & 0.006 & 0.011 & 0.012 & 0.011 & 0.013 & 0.012 & 0.940 & 0.916 & 0.035 & 100.0\\
\hline
IPCW-TMLE-MTO & 0 & 0 & 0.010 & 0.011 & 0.010 & 0.011 & 0.01 & 0.967 & 0.946 & 0.047 & 100.0\\
\hline
r-IPCW-TMLE-MTO & 0 & 0 & 0.011 & 0.011 & 0.011 & 0.011 & 0.011 & 0.958 & 0.944 & 0.045 & 100.0\\
\hline
\end{tabular}
\end{table}

\begin{table}
\centering
\caption{\label{tab:xNA_mNA_yglm_SH_HOSP_1826day_mRD_oracle}\textbf{Plasmode data simulation: 5-year self-harm or hospitalization, regression functions are glms, oracle marginal risk difference (mRD)}. Relative performance of estimators with sample size n = 50,337 and 1000 simulation replications. Note: Bias, ESE, ASE, MAD, and RMSE scaled by a factor of 10 to facilitate comparisons across estimands. The value of the estimand is -0.017. ESE = empirical standard error, ASE = asymptotic standard error, MAD = mean absolute deviation, RMSE = root mean squared error, rRMSE = robust RMSE (using median bias and MAD), Oracle coverage = coverage of a confidence interval based on the ESE, Nominal coverage = coverage of a confidence interval based on the ASE.}
\centering
\fontsize{9}{11}\selectfont
\begin{tabular}[t]{>{\raggedright\arraybackslash}p{2cm}|>{\raggedright\arraybackslash}p{1cm}|>{\raggedright\arraybackslash}p{1cm}|>{\raggedleft\arraybackslash}p{1cm}|>{\raggedleft\arraybackslash}p{1cm}|>{\raggedleft\arraybackslash}p{1cm}|>{\raggedright\arraybackslash}p{1cm}|>{\raggedright\arraybackslash}p{1cm}|>{\raggedleft\arraybackslash}p{1cm}|>{\raggedleft\arraybackslash}p{1cm}|>{\raggedleft\arraybackslash}p{1cm}|>{\raggedleft\arraybackslash}p{1cm}|>{}p{1cm}}
\hline
Estimator & Mean bias & Median bias & ESE & ASE & MAD & RMSE & rRMSE & Nominal coverage & Oracle coverage & Power & Prop. completed\\
\hline
Benchmark model & -0.001 & -0.001 & 0.028 & 0.028 & 0.027 & 0.028 & 0.027 & 0.941 & 0.946 & 1.000 & 100.0\\
\hline
Complete-case & 0.012 & 0.012 & 0.041 & 0.040 & 0.043 & 0.042 & 0.045 & 0.943 & 0.950 & 0.981 & 100.0\\
\hline
Confounded model & -0.063 & -0.062 & 0.027 & 0.027 & 0.025 & 0.069 & 0.067 & 0.342 & 0.360 & 1.000 & 100.0\\
\hline
IPW & 0.006 & 0.007 & 0.043 & 0.042 & 0.042 & 0.043 & 0.043 & 0.944 & 0.954 & 0.978 & 100.0\\
\hline
Raking (vanilla) & 0.006 & 0.006 & 0.030 & 0.029 & 0.027 & 0.03 & 0.028 & 0.938 & 0.941 & 0.999 & 99.2\\
\hline
MICE & 0.004 & 0.004 & 0.030 & 0.028 & 0.027 & 0.029 & 0.027 & 0.926 & 0.946 & 0.999 & 100.0\\
\hline
MI-XGB & 0.207 & 0.205 & 0.082 & 0.042 & 0.082 & 0.212 & 0.221 & 0.056 & 0.290 & 0.357 & 100.0\\
\hline
MI-RF & -0.026 & -0.026 & 0.028 & 0.028 & 0.025 & 0.039 & 0.036 & 0.851 & 0.845 & 1.000 & 100.0\\
\hline
IPCW-TMLE-M & 0.006 & 0.007 & 0.043 & 0.043 & 0.043 & 0.043 & 0.043 & 0.955 & 0.958 & 0.983 & 100.0\\
\hline
IPCW-TMLE-MTO & 0.003 & 0.003 & 0.043 & 0.043 & 0.043 & 0.043 & 0.043 & 0.958 & 0.957 & 0.986 & 100.0\\
\hline
\end{tabular}
\end{table}

\begin{table}
\centering
\caption{\label{tab:xNA_mNA_ytree_SH_HOSP_1826day_mRD_oracle}\textbf{Plasmode data simulation: 5-year self-harm or hospitalization, regression functions are trees, oracle marginal risk difference (mRD)}. Relative performance of estimators with sample size n = 50,337 and 1000 simulation replications. Note: Bias, ESE, ASE, MAD, and RMSE scaled by a factor of 10 to facilitate comparisons across estimands. The value of the estimand is -0.008. ESE = empirical standard error, ASE = asymptotic standard error, MAD = mean absolute deviation, RMSE = root mean squared error, rRMSE = robust RMSE (using median bias and MAD), Oracle coverage = coverage of a confidence interval based on the ESE, Nominal coverage = coverage of a confidence interval based on the ASE. Estimators that are mismatched with the estimand (i.e., are estimating a different parameter) are emphasized using a star.}
\centering
\fontsize{9}{11}\selectfont
\begin{tabular}[t]{>{\raggedright\arraybackslash}p{2cm}|>{\raggedright\arraybackslash}p{1cm}|>{\raggedright\arraybackslash}p{1cm}|>{\raggedleft\arraybackslash}p{1cm}|>{\raggedleft\arraybackslash}p{1cm}|>{\raggedleft\arraybackslash}p{1cm}|>{\raggedright\arraybackslash}p{1cm}|>{\raggedright\arraybackslash}p{1cm}|>{\raggedleft\arraybackslash}p{1cm}|>{\raggedleft\arraybackslash}p{1cm}|>{\raggedleft\arraybackslash}p{1cm}|>{\raggedleft\arraybackslash}p{1cm}|>{}p{1cm}}
\hline
Estimator & Mean bias & Median bias & ESE & ASE & MAD & RMSE & rRMSE & Nominal coverage & Oracle coverage & Power & Prop. completed\\
\hline
Complete-case${}^*$ & 0.024 & 0.022 & 0.041 & 0.041 & 0.042 & 0.048 & 0.047 & 0.905 & 0.904 & 0.243 & 100.0\\
\hline
Confounded model${}^*$ & -0.02 & -0.02 & 0.027 & 0.027 & 0.026 & 0.034 & 0.033 & 0.888 & 0.879 & 0.947 & 100.0\\
\hline
IPW${}^*$ & 0.021 & 0.02 & 0.042 & 0.043 & 0.043 & 0.048 & 0.047 & 0.924 & 0.923 & 0.248 & 100.0\\
\hline
Raking (vanilla)${}^*$ & 0.016 & 0.016 & 0.030 & 0.030 & 0.029 & 0.034 & 0.033 & 0.920 & 0.917 & 0.531 & 99.3\\
\hline
MICE${}^*$ & 0.017 & 0.017 & 0.030 & 0.029 & 0.029 & 0.034 & 0.033 & 0.908 & 0.919 & 0.531 & 100.0\\
\hline
MI-XGB${}^*$ & 0.061 & 0.062 & 0.086 & 0.042 & 0.085 & 0.073 & 0.105 & 0.558 & 0.886 & 0.355 & 100.0\\
\hline
MI-RF${}^*$ & 0.001 & 0 & 0.028 & 0.029 & 0.027 & 0.029 & 0.027 & 0.956 & 0.949 & 0.758 & 100.0\\
\hline
IPCW-TMLE-M & 0.014 & 0.013 & 0.037 & 0.043 & 0.036 & 0.046 & 0.039 & 0.964 & 0.924 & 0.267 & 100.0\\
\hline
IPCW-TMLE-MTO & -0.006 & -0.006 & 0.037 & 0.042 & 0.037 & 0.042 & 0.037 & 0.972 & 0.953 & 0.516 & 100.0\\
\hline
\end{tabular}
\end{table}

\clearpage

\newpage

\subsection{Plasmode Simulation Results Figures}

\subsubsection{ cOR}
\clearpage
\newpage

\begin{figure}[!htb]
\caption[Plasmode Simulation clogOR, census truth, bias]{\textbf{Plasmode Simulation: Census clogOR}. Comparing estimators of the census truth. \textbf{Top graph}: \textbf{\textcolor{black}{Bias}} (median, IQR, min and max of converged simulations); \textbf{Middle graph}: Robust RMSE (rRMSE), using median bias and MAD; \textbf{Bottom graphs}: Nominal and oracle coverage, respectively, with blue confidence bands at $ .95 \pm 1.96 \sqrt{\frac{.05\cdot .95}{1000}}$. True values of clogOR from lightest to darkest scenarios are 0.113, 0.017, -0.192, and -0.069.}

\includegraphics[scale=0.65]{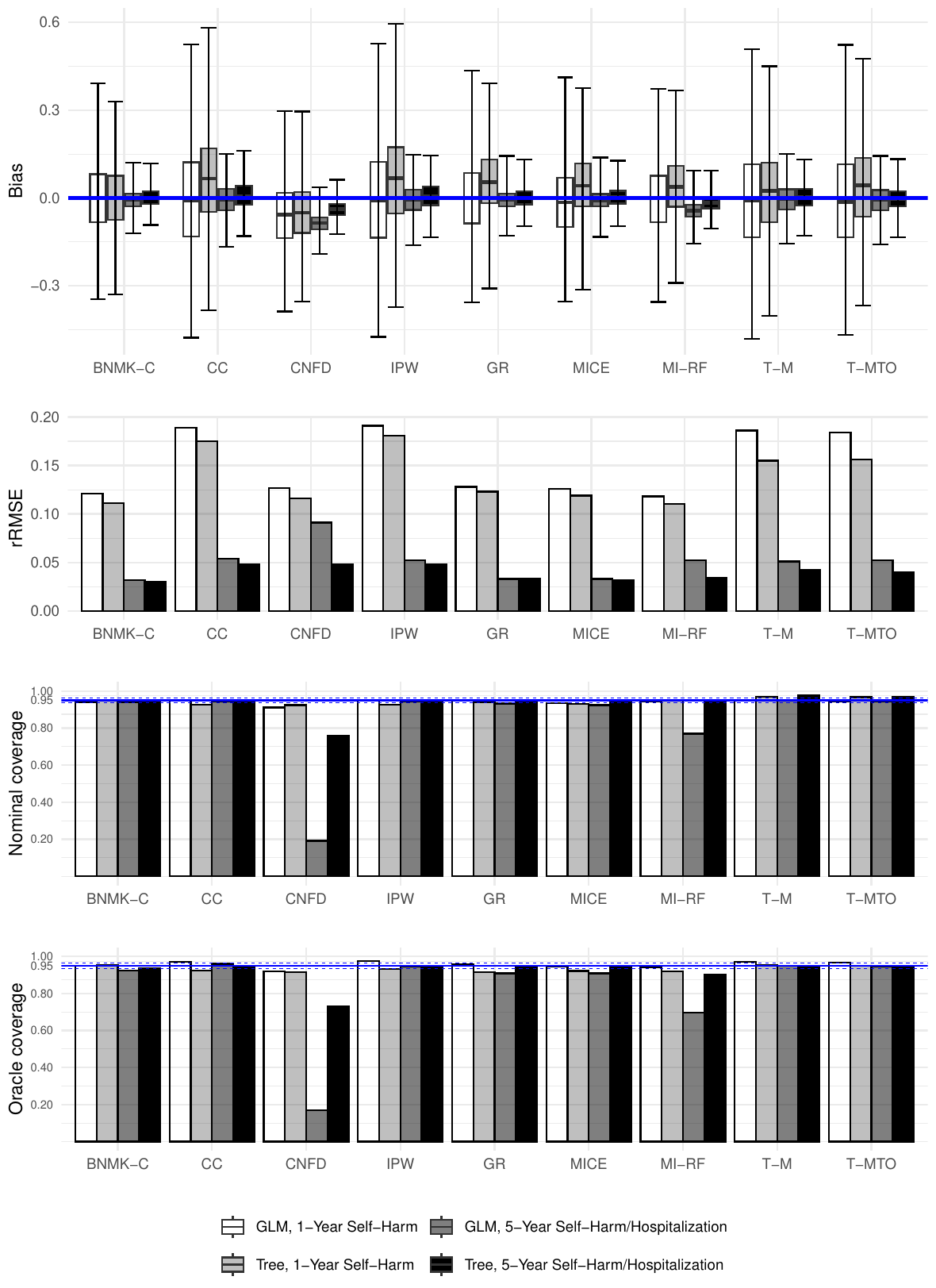}
\label{fig:plasmode_clogOR_census}
\end{figure}

\begin{figure}[!htb]
\caption[clogOR, oracle truth, bias]{\textbf{Plasmode Simulation: Oracle clogOR}. Comparing estimators of the oracle truth. \textbf{Top graph}: \textbf{\textcolor{black}{Bias}} (median, IQR, min and max of converged simulations); \textbf{Middle graph}: Robust RMSE (rRMSE), using median bias and MAD; \textbf{Bottom graphs}: Nominal and oracle coverage, respectively, with blue confidence bands at $ .95 \pm 1.96 \sqrt{\frac{.05\cdot .95}{1000}}$. True values of clogOR from lightest to darkest scenarios are 0.104 and -0.206.}

\includegraphics[scale=0.65]{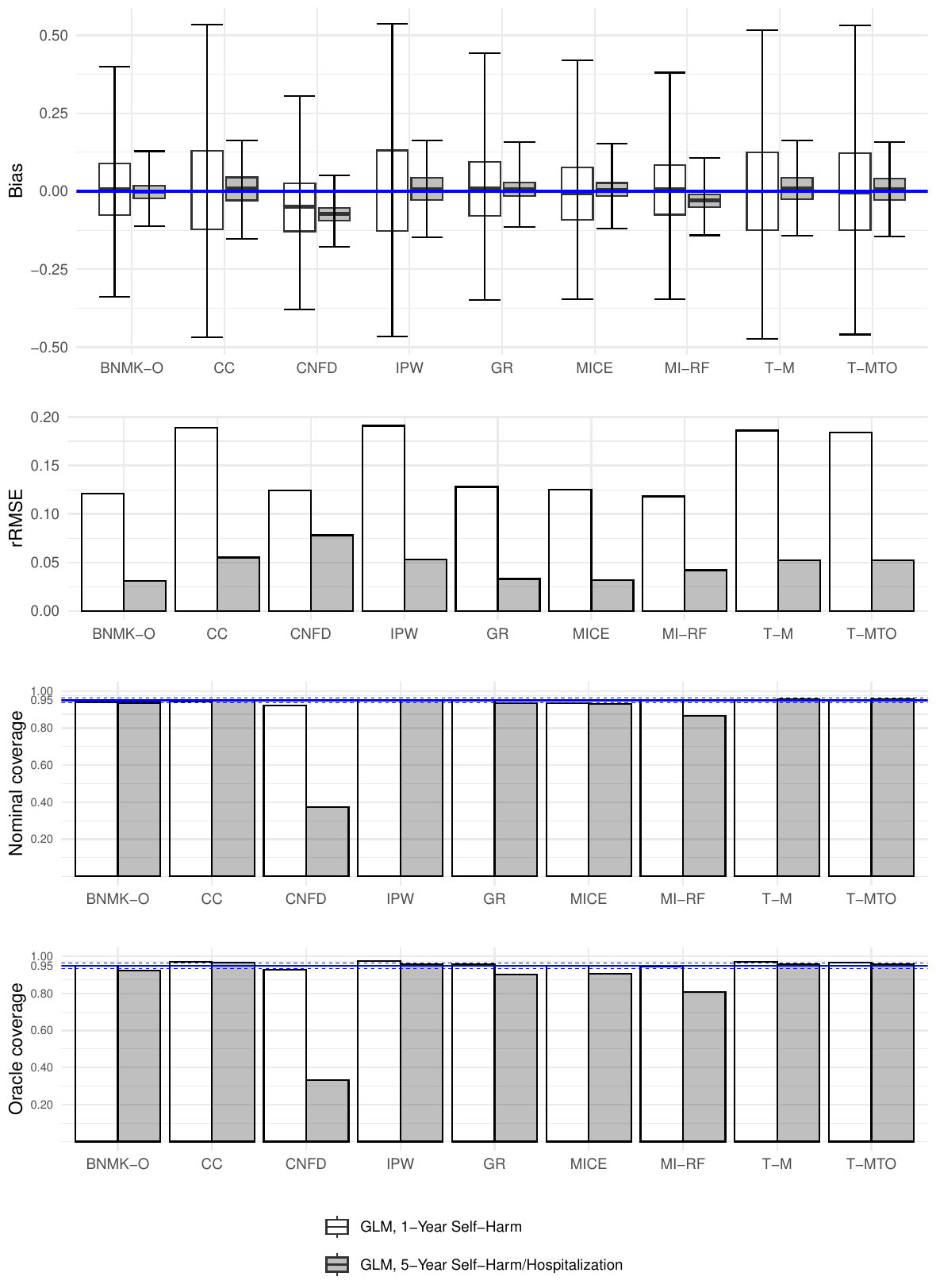}
\label{fig:plasmode_clogOR_oracle}
\end{figure}
\clearpage

\newpage

\subsubsection{ mRR}
\clearpage
\newpage

\begin{figure}[!htb]
\caption[mlogRR, census truth, bias]{\textbf{Plasmode Simulation: Census mlogRR}. Comparing estimators of the census truth. \textbf{Top graph}: \textbf{\textcolor{black}{Bias}} (median, IQR, min and max of converged simulations); \textbf{Middle graph}: Robust RMSE (rRMSE), using median bias and MAD; \textbf{Bottom graphs}: Nominal and oracle coverage, respectively, with blue confidence bands at $ .95 \pm 1.96 \sqrt{\frac{.05\cdot .95}{1000}}$. True values of mlogRR from lightest to darkest scenarios are 0.107, 0.016, -0.159, and -0.058.}

\includegraphics[scale=0.65]{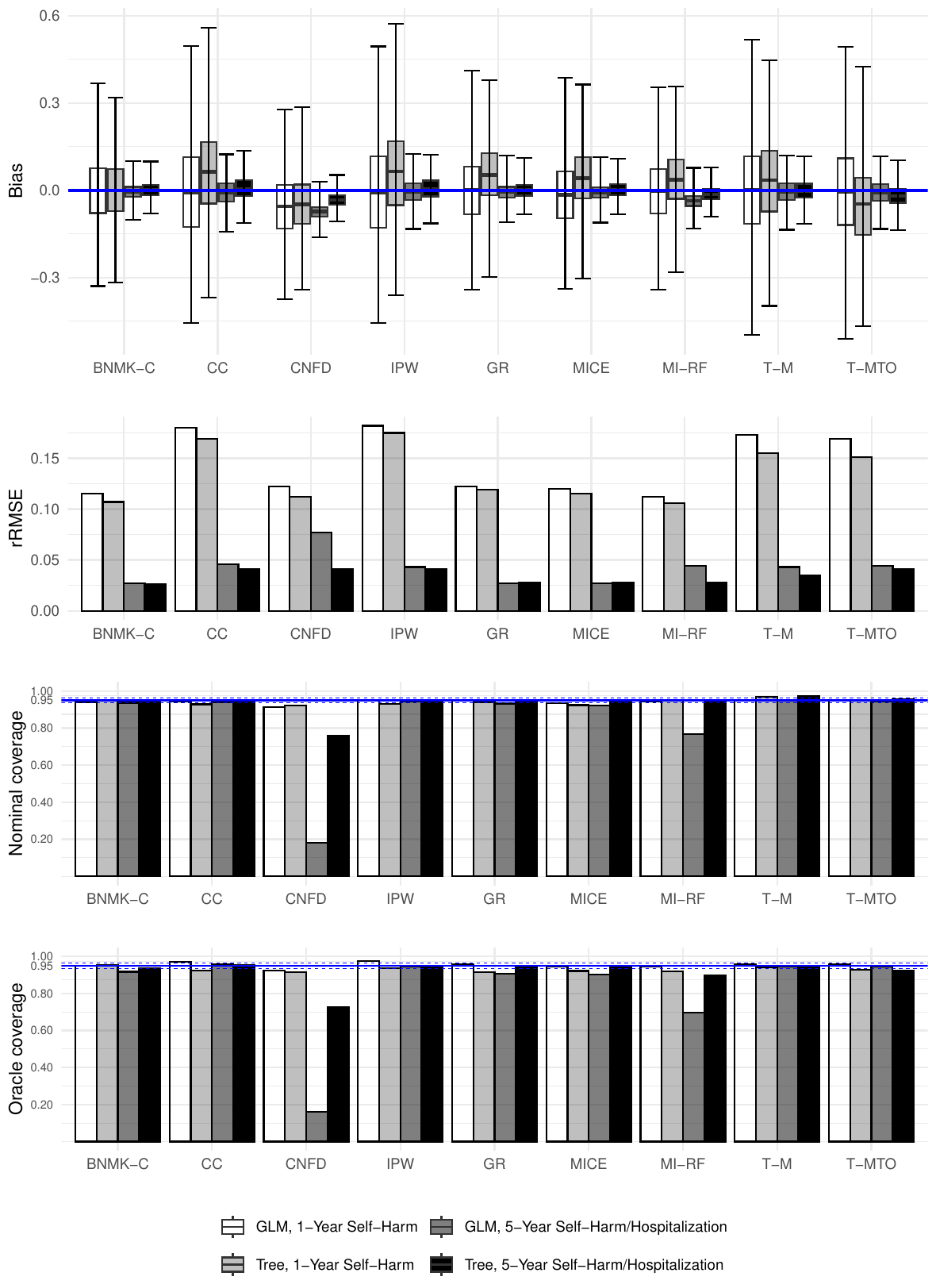}
\label{fig:plasmode_mlogRR_census}
\end{figure}


\begin{figure}[!htb]
\caption[Plasmode Simulation mlogRR, oracle truth, bias]{\textbf{Plasmode Simulation: Oracle mlogRR}. Comparing estimators of the oracle truth. \textbf{Top graph}: \textbf{\textcolor{black}{Bias}} (median, IQR, min and max of converged simulations); \textbf{Middle graph}: Robust RMSE (rRMSE), using median bias and MAD; \textbf{Bottom graphs}: Nominal and oracle coverage, respectively, with blue confidence bands at $ .95 \pm 1.96 \sqrt{\frac{.05\cdot .95}{1000}}$. True values of mlogRR from lightest to darkest scenarios are 0.106, -0.033, -0.170, and -0.02.}

\includegraphics[scale=0.65]{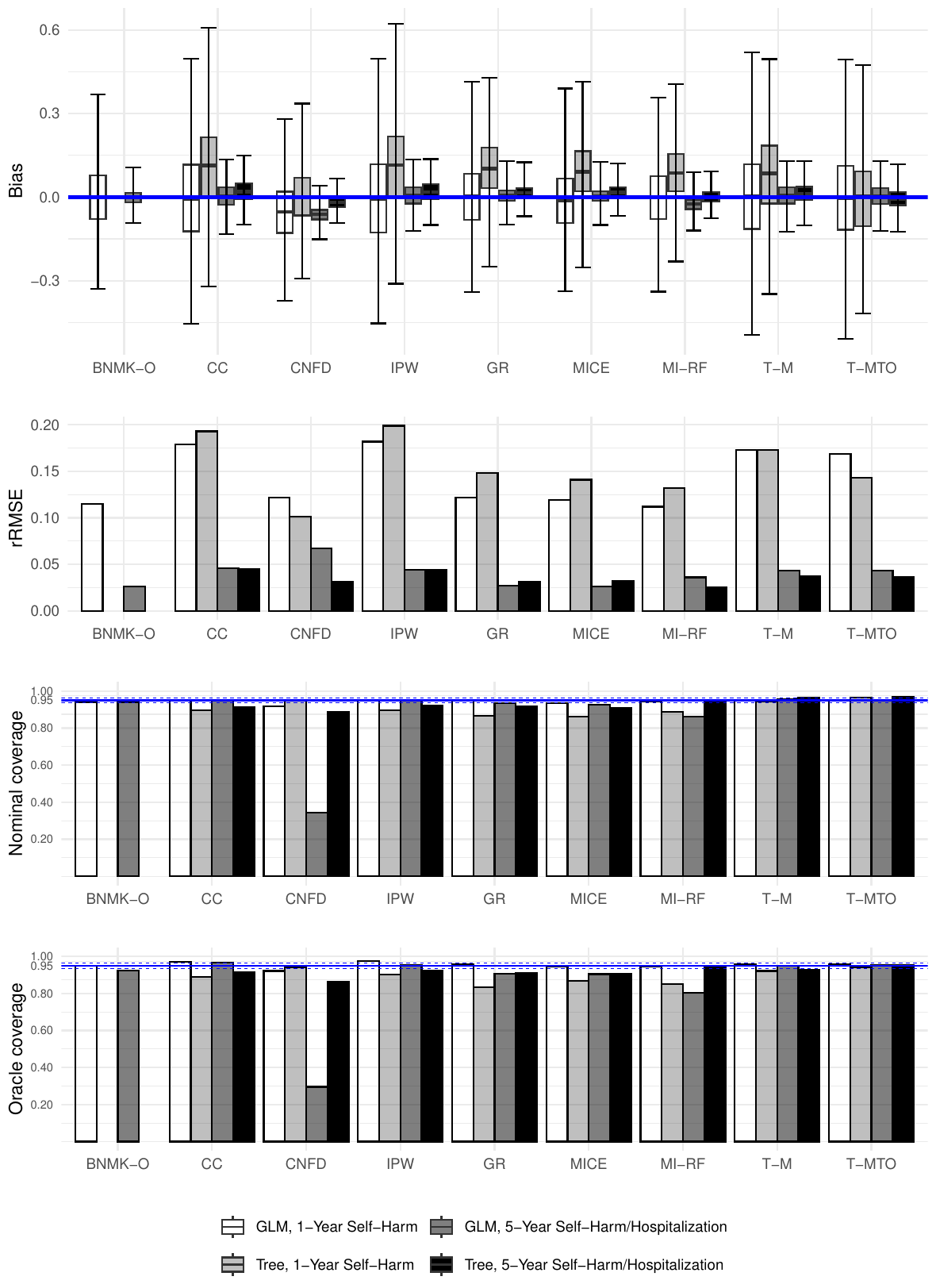}
\label{fig:plasmode_mlogRR_oracle}
\end{figure}
\clearpage

\ifnotarXiv
    \newpage

    \bibliographystyle{chicago}
    \bibliography{references}

    \end{document}
\fi

\fi

\bibliographystyle{chicago}
\bibliography{references}

\end{document}